\documentclass[useAMS,usenatbib,aas_macros]{mn2e}
\usepackage{aas_macros}
\usepackage{amsmath}
\usepackage{graphicx}
\usepackage{color}

\newcommand{\adr}[1]{\textcolor{red}{[#1]}}

\newcommand{\equ}[1]{eq.~(\ref{eq:#1})}
\newcommand{\equs}[1]{eqs.~(\ref{eq:#1})}

\newcommand{\Equ}[1]{Eq.~(\ref{eq:#1})}

\newcommand{\equnp}[1]{eq.~\ref{eq:#1}}
\newcommand{\se}[1]{\S\ref{sec:#1}}
\newcommand{\fig}[1]{Fig.~\ref{fig:#1}}
\newcommand{\figs}[1]{Figs.~\ref{fig:#1}}
\newcommand{\Fig}[1]{Figure~\ref{fig:#1}}

\newcommand{\tab}[1]{Table~\ref{tab:#1}}
\newcommand{\be}{\begin{equation}}
\newcommand{\ee}{\end{equation}}
\newcommand{\ba}{\begin{align}}
\newcommand{\ea}{\end{align}}
\newcommand{\bad}{\begin{equation} \begin{aligned}}
\newcommand{\ead}{\end{aligned} \end{equation}}
\newcommand{\bea}{\begin{eqnarray}}
\newcommand{\eea}{\end{eqnarray}}

\def\la{\langle}

\def\ssim{\!\sim\!}
\def\seq{\!=\!}
\def\ssimeq{\!\simeq\!}

\def\sgt{\!>\!}
\def\slt{\!<\!}
\def\sgsim{\!\gsim\!}
\def\slsim{\!\lsim\!}
\def\sgeq{\!\geq\!}
\def\sleq{\!\leq\!}
\def\sgg{\!\gg\!}
\def\sll{\!\ll\!}
\def\sdash{\!-\!}

\def\sprop{\!\propto\!}

\newcommand{\msun}{M_\odot}
\newcommand{\Msun}{M_\odot}

\newcommand{\ifm}[1]{\relax\ifmmode#1\else$\mathsurround=0pt #1$\fi}
\newcommand{\kms}{\ifmmode\,{\rm km}\,{\rm s}^{-1}\else km$\,$s$^{-1}$\fi}

\newcommand{\kpc}{\,{\rm kpc}}
\newcommand{\pc}{\,{\rm pc}}
\newcommand{\cm}{\,{\rm cm}}
\newcommand{\Gyr}{\,{\rm Gyr}}

\newcommand{\Myr}{\,{\rm Myr}}
\newcommand{\kyr}{\,{\rm kyr}}
\newcommand{\yr}{\,{\rm yr}}
\newcommand{\erg}{\,{\rm erg}}

\newcommand{\cmc}{\,{\rm cm}^{-3}}

\newcommand{\ltsima}{$\; \buildrel < \over \sim \;$}
\newcommand{\lsim}{\lower.5ex\hbox{\ltsima}}
\newcommand{\gtsima}{$\; \buildrel > \over \sim \;$}
\newcommand{\gsim}{\lower.5ex\hbox{\gtsima}}
\newcommand{\prop}{\propto}

\def\Rv{R_{\rm v}}
\def\Vv{V_{\rm v}}

\def\Mg{M_{\rm g}}
\def\Ms{M_{\rm s}}

\def\Sig1{\Sigma_1}

\def\Vsn{V_{\rm SN}}

\def\Mt{M_{\rm tot}}

\def\Rd{R_{\rm d}}
\def\fg{f_{\rm g}}
\def\fgs{f_{\rm gs}}

\def\MT{M_{\rm T}}
\def\RT{R_{\rm T}}
\def\Mc{M_{\rm c}}
\def\Rc{R_{\rm c}}
\def\Rcv{R_{\rm c,v}}
\def\Rci{R_{\rm c,i}}
\def\Vc{V_{\rm c}}
\def\Vcv{V_{\rm c,v}}
\def\Vrot{V_{\rm rot}}
\def\Vrotc{V_{\rm rot,c}}

\def\td{t_{\rm d}}
\def\fdis{f_{\rm dis}}

\def\brho{\bar\rho}

\def\sfr{\dot{M}_{\rm sf}}

\def\Mdotout{\dot{M}_{\rm out}}
\def\Mdotin{\dot{M}_{\rm in}}
\def\Mout{M_{\rm out}}

\def\Vout{V_{\rm out}}
\def\Vkick{V_{\rm kick}}

\def\Mt{M_{\rm tot}}
\def\Rd{R_{\rm d}}
\def\Hd{H_{\rm d}}
\def\fg{f_{\rm g}}

\def\epsf{\epsilon}
\def\eps2{\epsilon_{-2}}

\def\tff{t_{\rm ff}}

\def\Mv{M_{\rm v}}

\def\M11{M_{\rm v,11}}

\def\Md{M_{\rm d}}
 
\def\Vd{V_{\rm d}}

\def\f16b{f_{\rm b,0.16}}
\def\sv25{\phi_{11,-2.5}}

\def\c2h{c_{2{\rm h}}}
\def\s1h{s_{1{\rm h}}}

\def\sigr{\sigma_r}
\def\sigrd{\sigma_{r,{\rm d}}}

\def\mdotsf{\dot{M}_{\rm sf}}
\def\Vsn{V_{\rm sn}}

\def\efade{e_{\rm fade}}
\def\tcool{t_{\rm cool}}
\def\tfade{t_{\rm fade}}

\def\Rfade{R_{\rm fade}}

\def\Esn{E_{\rm sn}}
\def\Egrav{E_{\rm grav}}
\def\Edis{E_{\rm dis}}
\def\Evir{E_{\rm vir}}
\def\Eb{E_{\rm b}}
\def\Eout{E_{\rm out}}
\def\So{S}
\def\Sp{S'}

\def\Mdotout{\dot{M}_{\rm out}}
\def\fsig{f_\sigma}
\def\sigc{\sigma_{\rm c}}
\def\sigcv{\sigma_{\rm c,v}}

\def\fsn{f_{\rm sn}}
\def\fdis{f_{\rm dis}}
\def\fgrav{f_{\rm grav}}
\def\fout{f_{\rm out}}
\def\Vcf{V_{\rm c,50}}
\def\fgc{f_{\rm g,c}}
\def\tenc{t_{\rm enc}}

\def\etag{\eta}
\newcommand{\vela}{\texttt{VELA}~}
\newcommand{\velatwo}{\texttt{VELA-2}~}
\newcommand{\velathree}{\texttt{VELA-3}~}
\newcommand{\velathreecom}{\texttt{VELA-3}}
\newcommand{\velasix}{\texttt{VELA-6}~}
\newcommand{\velasixcom}{\texttt{VELA-6}}

\def\delrho{{\delta}_{\rm {\rho}}}
\def\delmin{{\delta}_{\rm {\rho}}^{\rm min}}

\def\tc{t_{\rm c}}
\def\tauc{\tau_{\rm c}}

\def\taugas{\tau_{\rm gas}}
\def\taug{\tau_{\rm gas}}

\def\SC{S\ }
\def\SCnos{S}
\def\LC{L\ }
\def\LCnos{L}
\def\LS{LS\ }
\def\LSnos{LS}
\def\LL{LL\ }
\def\LLnos{LL}

\def\Sdir{S_{\rm dir}}
\def\Smod{S_{\rm mod}}
\def\fesc{f_{\rm esc}}

\title[Clump Survival and Disruption]
{Conditions for Clump Survival in High-z Disc Galaxies} 

\author[Dekel et al.]
{\parbox[t]{\textwidth}
{Avishai Dekel$^{1,2}$\thanks{E-mail: dekel@huji.ac.il},
Offek Tziperman$^1$, 
Kartick Sarkar$^1$,
Omri Ginzburg$^1$,
Nir Mandelker$^1$,
Daniel Ceverino$^{3,4}$,
Joel Primack$^{2,5}$
}
\\ \\
$^1$Racah Institute of Physics, The Hebrew University, Jerusalem 91904 Israel\\
$^2$SCIPP, University of California, Santa Cruz, CA 95064, USA\\  
$^3$Departamento de Fisica Teorica, Facultad de Ciencias, Universidad Autonoma
de Madrid, Cantoblanco, 28049 Madrid, Spain\\
$^4$CIAFF, Facultad de Ciencias, Universidad Autonoma de Madrid, 28049 Madrid, 
Spain\\ 
$^5$Physics Department, University of California, Santa Cruz,
Santa Cruz, CA 95064, USA\\ 
}

\begin{document}

\large

\pagerange{\pageref{firstpage}--\pageref{lastpage}} \pubyear{2002}

\maketitle

\label{firstpage}

\begin{abstract}
We study the survival versus disruption of the giant clumps in high-redshift 
disc galaxies, short-lived (\SCnos) versus long-lived (\LCnos) clumps 
and two \LC sub-types, via analytic modeling tested against simulations.
We develop a criterion for clump survival, with or without their gas,
based on a predictive survivability parameter $S$. 
It compares the energy sources by supernova feedback and 
gravitational contraction to the clump binding energy and losses by 
outflows and turbulence dissipation.
The clump properties are derived from Toomre instability,
approaching virial and Jeans equilibrium, 
and the supernova energy deposit is based on an up-to-date bubble analysis. 
For moderate feedback levels, we find that \LC clumps exist with 
circular velocities $\sim\!50\kms$ and masses $\sgeq\!10^8\msun$. 
They are likely in galaxies with circular velocities $\sgeq \!200\kms$,
consistent at $z \ssim 2$ with the favored stellar mass for discs,
$\sgeq\!10^{9.3}\msun$. 
\LC clumps favor disc gas fractions $\sgeq\! 0.3$, low-mass bulges 
and redshifts $z\ssim 2$. 
The likelihood of \LC clumps is reduced if the feedback is more
ejective, e.g., if the supernovae are optimally clustered,
if radiative feedback is very strong,
if the stellar initial mass function is top-heavy,
or if the star-formation-rate efficiency is particularly high.
A sub-type of \LC clumps (\LSnos),
which lose their gas in several free-fall times but
retain bound stellar components, may be explained by a smaller contraction 
factor and stronger external gravitational effects, where clump mergers 
increase the SFR efficiency. The more massive \LC clumps (\LLnos) retain most 
of their baryons for tens of free-fall times with a roughly constant 
star-formation rate.
\end{abstract}


\section{Introduction}
\label{sec:intro}

The typical massive star-forming galaxies at the peak epoch of galaxy 
formation, in the redshift range $z\!\sim\!1\!-\!3$, are dominated by  
extended rotating and turbulent gas-rich discs
\citep{genzel06,forster06,genzel08,tacconi10,genzel11,tacconi13,genzel14_rings,
forster18},
each hosting several giant star-forming clumps
\citep{elmegreen05,genzel06,genzel08,guo12,guo15,fisher17,
guo18,huertas20,ginzburg21}.
Most of these clumps are commonly assumed to form by gravitational disc 
instability \citep{toomre64}\footnote{Possibly triggered by compressive modes
of turbulence \citep[][ and work in progress]{inoue16}.}.
This process has been simulated both in isolated discs
\citep{noguchi99,immeli04_a,Immeli04_b,bournaud07c,genzel08}
and in a cosmological setting
\citep{dsc09,agertz09,cdb10,ceverino12},
particularly in \citet{mandelker14}, \citet{mandelker17},
\citet{ginzburg21} and \citet{dekel22_mass}.
Given the high gas fraction in $z\ssim 2$ galaxies 
\citep[e.g.,][]{daddi10,tacconi18},
the typical clump mass predicted by Toomre instability
is on the order of a few percents of the disc mass, 
which makes the clumps play an important dynamical role in the evolution of
these discs, a process dubbed ``violent disc instability" (VDI)
\citep[e.g.,][]{dsc09,db14}.
Gas-rich clumps are predicted to form with a mass near and below the 
characteristic Toomre mass and preferably at relatively large disc radii 
\citep[see][for evidence in simulations and observations]{dekel22_mass}.
Those clumps that survive disruption by stellar and supernova feedback are 
predicted to migrate toward the disc centre due to VDI-driven torques
in a few disc orbital times, corresponding to several hundred Megayears 
at $z\ssim 2$
\citep[e.g.,][]{dsc09,cdb10,krum_burkert10,cacciato12,forbes12,forbes14a,
krum18,dekel20_ring}.

\smallskip 
Simulations reveal that the clumps can be divided into two major types.
One consists of long-lived clumps (hereafter \LC clumps) that remain bound 
and live for tens of
clump free-fall times or more, allowing them enough time
to complete their inward migration.
The other involves short-lived clumps (hereafter \SC clumps), 
which are disrupted after a few free-fall times \citep{mandelker17}.
Using machine learning, the two types of clumps as identified in the 
\velathree cosmological simulations (\se{sim_type})
 were also identified in the large  
catalog of observed clumps from the {\tt CANDELS-HST} survey \citet{guo18},
with consistent relative distributions of clump masses, radial positions in the
discs and host-galaxy masses \citep{ginzburg21}.\footnote{A word of caution is 
that the observational estimates of clump properties 
suffer from large uncertainties, which require careful considerations of 
systematic errors, possibly using machine learning \citep{huertas20}.}

\smallskip 
According to a wide variety of existing simulations, 
the clump type depends on certain clump properties at formation, on certain
host-galaxy properties, and on the assumed subgrid physical models 
for star formation and especially feedback.
\citet{mandelker17}, based on the \velathree cosmological
simulations, found that \LC clumps tend to form above a threshold mass of 
$\sim\!10^8\msun$, while \SC clumps dominate at smaller clump masses.
The \LC clumps were found to be more compact, round and bound,
while the \SC clumps are more diffuse, elongated and they become unbound.
In terms of galaxy properties,  
the \LC clumps in \velathree tend to reside in more massive galaxies,
consistent with the predicted mass threshold for long-lived extended discs at
a halo mass of $\Mv \ssim 2\times 10^{11}\msun$ \citep{dekel20_flip}.
Based on isolated-galaxy simulations, the clump formation, mass and longevity 
are correlated with a high gas fraction in the disc \citep{fensch21,renaud21}.

\smallskip 
Most importantly, the clump type is very sensitive to the subgrid feedback 
model adopted in the simulation.
The \velathree simulations, that assume a moderate supernova
feedback strength with a modest effective radiative momentum driving of 
$2\!-\!3 L/c$ (where $L$ is the clump luminosity),
give rise to clumps of the two types. 
On the other hand, the \velasix simulations
(Ceverino et al. in preparation),
of the same suite of galaxies but with a stronger feedback, 
turn out to produce mostly \SC clumps with a negligible population of \LC 
clumps even at the massive end (see below).
This recovers the results from simulations that put in very strong winds 
\citep{genel12}, or that include enhanced radiative feedback 
of $30\!-\!50 L/c$ due to strong trapping of infra-red photons and additional
elements of feedback
\citep{Hopkins12b,oklopcic17}. Such simulations produced only \SC clumps.

\smallskip 
On the theory side,
\citet{dsc09}, based on \citet{ds86}, crudely estimated that supernova 
feedback by itself may not have enough power to disrupt the massive clumps.
\citet{murray10} argued that momentum-driven radiative
stellar feedback, enhanced by infra-red photon trapping, could disrupt the
clumps on a dynamical timescale, as it does in the local molecular
clouds. However, \citet{kd10} pointed out that such an explosive
disruption would be possible in the high-redshift giant clumps only if the
efficiency of star-formation rate (SFR) in a free-fall time is as high as
$\epsf\!\sim\!0.1$,
significantly larger than what is implied by the observed Kennicutt-Schmidt
relation in different types of galaxies at different redshifts,
namely $\epsf$ of about one to a few percent
\citep{tacconi10,kdm12,freundlich13}.
\citet{dk13} proposed instead that outflows from high-redshift clumps, with
mass-loading factors $\etag$ of order unity to a few, are driven by steady
momentum-driven outflows from stars over many tens of free-fall times. Their
analysis is based on the finding from high-resolution 2D simulations that
radiation trapping is negligible because it destabilizes the wind
\citep{krum_thom12,krum_thom13}.
Each photon can therefore contribute to the wind momentum
only once, so the radiative force is limited to $\sim\!L/c$. 
Combining radiation, protostellar plus main-sequence winds, and supernovae, 
\citet{dk13} estimated the total direct injection rate of momentum into the 
outflow to be $\sim\!2.5L/c$. The adiabatic phase of clustered supernovae 
and main-sequence winds may double this rate \citep{gentry17}. 
The predicted mass loading factors of order unity were 
argued in \citet{dk13} to be consistent with the values deduced from the 
pioneering observations of outflows from giant clumps 
\citep{genzel11,newman12}. 
They concluded that most massive clumps are
expected to complete their migration prior to gas depletion. 
With the additional gas accretion onto the clumps, they argued that the 
clumps are actually expected to grow in mass as they migrate inward.

\smallskip 
\citet{dekel22_mass} studied the evolution of the properties of the massive 
clumps during their migration through the disc.  They developed an idealized 
analytic ``bathtub" model for a single clump, where clump gas turns into stars 
and the clump exchanges gas and stars with the disc through accretion,
feedback-driven outflows and tidal stripping.
The clump evolution is governed by the balance between these processes under
conservation of gas and stellar mass, which lead to an analytic solution.
The model reveals how the clump evolution depends on the assumed
efficiency of accretion, star formation and outflows.
The model predictions were confronted with simulation results,
both of isolated gas-rich galaxies at high resolution and of
high-redshift galaxies in their cosmological framework,
exploring a variety of feedback strengths at moderate levels.
The theoretical predictions were confronted with the {\tt CANDELS} clump 
catalog of \citet{guo18} at $z\!=\!1.5\!-\!3$, indicating that 
the observed gradients of clump properties as a function of distance from the
disc centre are consistent with the predicted migration of massive clumps
and the associated mild evolution of clump properties.

\smallskip 
Our goal in this paper is to understand the conditions for clump survival versus
disruption by feedback, and thus seek the origin of the different clump types
in terms of clump and disc properties as well as feedback strength.  
We do this first via analytic modeling.
For clumps in their early stages of evolution,
we define a survivability parameter $\So$, the ratio of (a fraction of) 
the clump mass to the outflowing mass. The value of this quantity below or 
above unity distinguishes between \SC and \LC clumps. 
This parameter can be derived from basic clump 
properties based on the energy balance in the clump.
We consider the energy sources by supernova feedback and gravitational driving
of turbulence, and energy sinks by dissipation of turbulence and outflows,
in comparison with the clump gravitational binding energy.
%
In this model, the clump properties are deduced from Toomre marginal 
instability \citep[as in][]{dsc09}, with the kinematics of fully collapsed
clumps obeying Jeans equilibrium \citep{ceverino12}.
The input by supernova feedback, inspired by \citet{ds86}, is based on 
up-to-date studies of bubble evolution as in \citet{dekel19_ks}, fine-tuned 
in our ongoing work. 
The gravitational driving of turbulence is estimated from the gravitational
collapse and clump encounters.
The analytic model is then confronted with the clumps in the \velathree and 
\velasix cosmological simulations. 

\smallskip 
Beyond the distinction between \LC and \SC clumps, we address a new sub-division
into two types of \LC clumps that will be identified in the simulations below. 
One sub-type, to be termed \LLnos, consists of clumps that keep a relatively
high gas fraction for many tens of clump free-fall times.
The other sub-type, to be termed \LSnos, is made of clumps that lose most of 
their gas to outflows in $\lsim\!10$ free-fall times, but keep long-lived
bound stellar components that survive for many tens of free-fall times.
Our analytic model provides physical explanations for the basic distinction
between \SC and \LC clumps as well as for the sub-division into 
\LS and \LL clumps.

\smallskip 
The paper is organized as follows.
In \se{survival} to \se{thresholds} we present the analytic model,
and in \se{sim_types} and \se{sim_model} we compare the model to simulations.
In \se{survival} we introduce the survivability parameter $\So$.
In \se{clumps_drains} we model the clump binding energy and the energy
drains by dissipation of turbulence and outflows. 
In \se{sources} we model the energy sources by supernova feedback and by
gravity.
In \se{non-virial} we introduce a correction for clumps that have not reached
virial equilibrium.
In \se{types} we use the model to explain the division to \SC and \LC clumps
and the sub-division to \LS and \LL clumps. 
In \se{thresholds} we identify the dependence of the clump type on the 
different properties of the clump and the host disc as well as the feedback
strength. 
Then, 
In \se{sim_types} we present the division to clump types in the simulations.
Finally,  
in \se{sim_model}, we confront the model predictions with the simulation
results. 
In \se{conc} we summarize our conclusions and discuss them.

\section{Condition for clump survival}
\label{sec:survival}

\subsection{A survivability parameter}

We consider a clump of mass $\Mc(t)$, at time $t$ after its formation,
that has lost mass $\Mout(t)$ to outflows up until that time.
For a clump in virial equilibrium, 
we define a clump survivability parameter $\So$ such that
a criterion for {\it significant mass loss} through outflows by time $t$
is
\be
\So(t) \equiv \frac{0.5\, \Mc(t)}{\Mout(t)} \leq 1 \, .
\label{eq:S} 
\ee
While $\Mout$ is expected to grow steadily in time, 
$\Mc$ may tend to have a rather constant value after the early 
stages of clump buildup, or decline for clumps that lose
a significant fraction of their gas to outflows \citep{dekel22_mass}.
When $\So$ is above unity, the clump has retained most of its mass,
while when $\So$ is below unity, the clump has lost most of its mass to
outflows.  
\Equ{S} is motivated by noticing that for a clump in virial equilibrium, 
an instantaneous removal of half its mass (in a free-fall time)
makes it unbound with zero energy.
We will soften the requirement for virial equilibrium in \se{non-virial}
and generalize the definition of $\So$ accordingly in \equ{S_sim_f}.

\smallskip 
However,
a mass loss with $\So \sleq 1$ does not necessarily imply a total disruption.
If the mass loss is adiabatic, over many free-fall times, 
part of the system may remain bound.
A criterion for destructive instantaneous mass loss is thus
$\dot{M}_{\rm out}(t)\,\tff \sgeq 0.5\Mc(t)$, where $\tff$ is the clump
free-fall time. If we crudely approximate 
$\dot{M}_{\rm out}(t) \ssim \Mout(t)/t$, where $t$ is the time since clump
formation, the criterion for {\it total disruption} becomes 
\be
\tau \So(t) \leq 1 \, , 
\label{eq:St} 
\ee
where $\tau \seq t/\tff$. 
Thus, at $\tau \ssim 1$, a value of $\So \sleq 1$ that implies removal
of most of the gas also indicates a total disruption, $\tau\So \sleq 1$,
namely an \SC clump.
However, if half the mass is lost only by a later time, $\So(\tau) \sleq 1$ at 
$\tau$ well above unity,
the mass-loss would be totally destructive (\SCnos)
only if $\tau\So \sleq 1$ as well,
while a bound (stellar) remnant is expected (\LSnos) if $\tau\So \sgt 1$.

\smallskip 
The quantities $\Mc$ and $\Mout$ that define $\So$ in \equ{S} can be measured 
from simulations,
as we do below in one version of our analysis, though the measured $\Mout$ 
may be rather uncertain, especially at early times when the clump radius is
likely to evolve rapidly.
 
\subsection{The survivability by energy balance}

As an alternative, we wish to predict $\So$ from theoretical considerations,
where we express
$\Mc$ and $\Mout$ by physical quantities that we can estimate using physical
arguments and based on an energy balance.   

\smallskip 
Assuming a spherical clump of mass $\Mc$ within radius $\Rc$,
the potential is characterized by a circular velocity $\Vc^2 \seq G\Mc/\Rc$.
In virial equilibrium, the total energy of the gas per unit mass is
\be
-\Evir = -\frac{1}{2} \Vc^2 = -\frac{1}{2} \frac{G \Mc}{\Rc} \, .
\label{eq:Evir} 
\ee
We tentatively assume that $\Vc$ and the gas mass $\Mg$ are roughly constant
during the early phase of clump evolution,
before a significant amount of the gas is ejected.
This is motivated by analytic modeling and simulations of clump evolution
\citep{dekel22_mass}, and it will be tested further in simulations below.

\smallskip 
The cumulative energy balance by time $t$, per unit gas mass, can be written as
\be
\Esn(t) + \Egrav(t) -\Edis(t) = \Evir + \Eout(t) \, ,
\label{eq:energy}  
\ee
where all the quantities are positive.
Except $\Evir$, they are all integrals of their rates over time.
Each will be evaluated below as a function of certain basic parameters
in terms of $\Vc$ and $\tau$.
The terms on the left-hand side represent sources of energy gain and loss.
The term $\Esn$ is the energy deposited in the gas by stars and supernovae.
The term $\Egrav$ represents the additional mechanical energy sources
largely associated with gravity,
such as the gravity associated with the initial collapse of the clump,
continuous accretion, mergers, tidal effects and shear,
all capable of driving turbulence.
The term $\Edis$ represents the dissipative losses of the turbulence,
cascading down to small scales where it thermalizes and can cool radiatively.
%
The terms on the right-hand side represent the energy required to bring the
gas from virial equilibrium to outflow.
Here $\Evir$ is the energy required for bringing the gas from $-\Evir$
to $E\seq 0$, ready for escape, and
$\Eout$ is the extra kinetic energy carried by the outflow.

\smallskip 
Using \equ{energy}, the survivability parameter $\So$ as defined in \equ{S} 
can be approximated by
\be
\So \ssimeq \frac{\fgc^{-1}\,\Evir}{\Eout} \, ,
\label{eq:S2} 
\ee
where $\fgc$ is the gas fraction in the clump, starting somewhat below unity
and declining in time. 
This arises from \equ{Evir} and the relation of $\Eout$ to $\Vc$,
\be
\Eout = \frac{1}{2} \frac{\Mout}{\Mg} \Vout^2
\simeq \frac{\Mout}{\Mg} \Vc^2 \, ,
\label{eq:Eout0}  
\ee
where the typical outflow velocity is crudely assumed to be near the escape 
velocity,
$\Vout^2 \ssimeq 2\, \Vc^2$.
Thus, $\So$ can be expressed as
\be
\So = \frac{\fgc^{-1}}{\Sp^{-1} -1} \, , 
\quad \Sp \equiv \frac{\Evir}{\Esn+\Egrav-\Edis} \, .
\label{eq:So_Sp}  
\ee
We note that the critical value for losing mass by outflows
that corresponds to $\So\seq 1$ is $\Sp \seq 1/(\fgc^{-1}+1)$. 
This is $\Sp \ssim 0.5$ for $\fgc\ssim 1$ 
and $\Sp \ssim \fgc$ for $\fgc \sll 1$, 
which are valid respectively early and late in the clump evolution.

\smallskip 
The cumulative specific energies that enter $\So$ in \equ{So_Sp}
will be evaluated below
based on theoretical arguments. They all turn out to be proportional
to $\Evir \tau$, and can be written as 
\be 
\Esn = \fsn\, \Vcf^{-1}\, \Evir\, \tau \, , \quad
\Egrav = \fgrav \, \Evir\, \tau \, , 
\label{eq:factors1} 
\ee 
\be
\Edis = \fdis\, \Evir\, \tau \, , \quad
\Eout = \fout\, \Evir\, \tau \, ,
\label{eq:factors2}  
\ee
with the $f$ factors functions of basic clump parameters,
and where $\Vcf \seq \Vc/50\kms$ is the main explicit clump property that
enters.
The clump  $\Vc$ (and $\Evir$) will be predicted assuming Toomre disk 
instability \citep{toomre64}. 
The feedback energy $\Esn$ will be estimated from the theory of supernova 
bubbles as a function of the turbulence velocity dispersion $\sigc$.
The dissipative energy loss
$\Edis$ can also be estimated as a function of $\sigc$. 
In turn, $\sigc$ can be related to $\Vc$ assuming Jeans equilibrium and the
degree of angular-momentum conservation during clump formation. 
The value of $\Egrav$ can be crudely evaluated during clump formation 
and at later times.
In the following estimates we learn that
$\fsn$ is of order unity or larger, $\fdis \ssim 1$,
$\fgrav$ is of order unity or smaller,
and $\fout \sll 1$.

\smallskip 
Inserting \equs{factors1} and (\ref{eq:factors2}) in $\Sp$ from \equ{So_Sp} 
we obtain
\be
\Sp = \frac{1}{(\fsn\, \Vcf^{-1} + \fgrav - \fdis)\, \tau} \, ,
\label{eq:Sp}  
\ee
which serves as our operational expression for $\Sp$ and $\So$.
Interestingly, the key quantity $\Vc$ enters $\Sp$ only through the first 
term.
The dominant factor in determining the value of $\Sp$ is the supernova
feedback represented by the first term in the denominator of \equ{Sp}, 
as $\fsn$ varies with the feedback strength and
$\Vc$ varies between massive and low mass clumps. 
The time dependence enters via $\tau$ as a multiplicative factor for all 
the terms in the denominator,
as well as through possible time dependencies of each $f$ factor.

\def\taus{\tau_{(S\seq 1)}}
\def\tauts{\tau_{(\tau S\seq 1)}}

\smallskip 
According to \equ{So_Sp}, the time $\taus$ when $\So$ equals unity is
when $\Sp^{-1} \seq 1+ \fgc^{-1}$, which from \equ{Sp} is at
\be
\taus = \frac{1+\fgc^{-1}}{\fsn\Vcf^{-1}+\fgrav-\fdis} \, .
\label{eq:tau1}
\ee
If the clump is to lose a significant fraction of its gas, a larger value of 
$\taus$ indicates a later time for gas loss.
Similarly, 
the time that indicates full disruption, when $\tau S \seq 1$, is given by
\be
\tauts = [\fsn\, \Vcf^{-1} +\fgrav -\fdis - \fgc^{-1}]^{-1} \, .
\label{eq:tau2}
\ee

\section{Clump properties \& energy drains}
\label{sec:clumps_drains}

In the coming two sections 
we evaluate the relevant energies and the associated 
$f$ factors that enter $\Sp$ in \equ{Sp}.
In the current section we address the relevant clump properties, using Toomre
instability to derive the binding energy and using Jeans equilibrium to relate
rotation to dispersion velocity, and evaluate the energy 
losses by outflows and dissipation. In the following section we evaluate 
the energy gains by supernovae and gravity.

\subsection{Star formation rate and outflow energy}
\label{sec:Eout}

The outflow energy can be expressed as a function of the SFR, $\sfr$,
via the mass-loading factor
\be
\eta = \frac{\Mdotout}{\sfr} \, .
\label{eq:eta}
\ee
According to the simulations described below, $\eta$ is expected to be
of order unity.
%
Following the standard convention, 
the SFR can be expressed via the free-fall time in the clump, $\tff$, 
and the SFR efficiency parameter, $\epsf$, as
\be
\mdotsf = \epsf\, \frac{\Mg}{\tff} \, .
\label{eq:sfr}
\ee

\smallskip 
The clump free-fall time is
\be
\tff = \left( \frac{3\,\pi}{32\,G\,\rho} \right)^{1/2}
     \simeq  5.2 \Myr \, n_2^{-1/2} \fgc^{1/2} \, ,
\label{eq:tff}
\ee
where $\rho$ is the total density in the clump,
$n \seq 100 \cmc\, n_{2}$ is the corresponding gas number density
and $\fgc\!\lsim\!1$ is the gas fraction in the clump in its early phase.
We very crudely assumed here $\rho \seq \fgc^{-1}\, m_{\rm p}\, n$.
A value of $n \ssim 100 \cmc$ is expected if the clump has collapsed in 3D
by a factor of a few from a gas density of order $1\cmc$ in the disc.

\smallskip 
The SFR efficiency per free-fall time, $\epsf$,  has to be of order a few 
percent in order to match the Kennicutt-Schmidt law in different 
environments and redshifts \citep[e.g.,][]{kdm12}.
The effective value of $\epsf$ as defined for the whole clump mass and
free-fall time may be larger if the star formation actually occurs in dense
sub-regions within the clump, where the free-fall timescale is shorter.
We denote $\epsilon \seq 0.03\,\epsilon_{.03}$.

\smallskip
Assuming a constant SFR and $\Mg$,
using \equs{eta} and (\ref{eq:sfr}) one can write
\be
\Mout = \eta\, \sfr\, t 
      = 0.03\, \epsf_{.03}\, \eta\, \Mg\, \tau \, ,
\label{eq:Mout}
\ee   
and express the outflow energy per unit gas mass from \equ{Eout0} as
\be
\Eout = 0.03\,\epsf_{.03}\, \eta\, \tau\, \Vc^2 \, .
\label{eq:Eout}  
\ee

\smallskip 
Comparing \equ{Eout} to \equ{Evir}, 
assuming the fiducial values of $\eta$
and $\epsf$, we learn that $\Eout \!\ll\! \Evir$ in the first few free-fall 
times, and they become comparable at $\tau \ssim 16$.

\subsection{Toomre clump binding energy}
\label{sec:Evir}

\smallskip
In order to evaluate the clump properties that enter $\Evir$ in \equ{Evir},
we assume that the disc is in marginal Toomre instability \citep{toomre64}
with $Q\seq 1$. We assume that the clump mass is $\mu\,\MT$, 
where $\MT$ is the Toomre mass corresponding to the fastest growing scale.
Following \citet{dsc09}, \citet{cdb10} and \citet{ceverino12},
we define a key quantity, the cold mass fraction,
\be
\delta \equiv \frac{\Md}{\Mt} \, .
\label{eq:delta}
\ee
Here $\Md$ is the cold mass in the disk (gas and young stars), and
$\Mt$ is the total mass encompassed by the sphere of the 
disc radius $\Rd$ (including gas, stars and dark matter).
Assuming a local power law for the disc rotation curve, 
$\Vd(r) \!\prop\! r^\alpha$, with a characteristic value $\Vd$,
and a disc radial velocity dispersion $\sigrd$,
a Toomre parameter of $Q \seq 1$ implies that
\be
\delta = \sqrt{2}(1+\alpha)^{1/2} \,\frac{\sigrd}{\Vd} \, .
\label{eq:deltaQ1}
\ee
Then, the Toomre proto-clump radius relative to the disc radius is
\be
\frac{\RT}{\Rd} = \frac{\pi}{4\,(1+\alpha)}\, \delta \, ,
\label{eq:RT}
\ee
and the Toomre clump mass relative to the disc mass is
\be
\frac{\MT}{\Md} = \left( \frac{\RT}{\Rd} \right)^2
= 0.025\, \delta_{.2}^2 \, ,
\label{eq:MT}
\ee
where $\delta \seq 0.2\,\delta_{.2}$, the density in the proto-clump is
assumed to be similar to the mean density in the disc, 
and $\alpha\seq 0$ is assumed.
We thus have  
\be
\Mc = \mu\MT \quad {\rm and} \quad \Rci\seq\mu^{1/2}\RT \, ,
\label{eq:mu}
\ee
where $\Rci$ is the initial radius of the proto-clump patch in the disc.
The clump mass in terms of the cold-disc mass becomes
\be
\Mc = 0.25\times 10^8\msun\, \mu_{.5}\, \delta_{.2}^2\, M_{\rm d,9.3} \, ,
\label{eq:McMd}
\ee
where $\Md = 10^{9.3} \msun\, M_{\rm d,9.3}$. 

\smallskip
In order to obtain the clump circular velocity,
we assume that the clump has collapsed by a contraction factor $c$
from its initial radius to a final radius $\Rc$, 
\be
\Rci = c \Rc \, .
\label{eq:c}
\ee
Then, combining \equ{RT} and \equ{c},
the clump potential well is characterized by
\be
\Vc^2 = \frac{G\,\Mc}{\Rc}
= \frac{\pi}{4\,(1+\alpha)}\, c\, \mu^{1/2}\, \delta^2\, \Vd^2 \, ,
\label{eq:Vc2}
\ee
recalling that $\Vd^2 \seq G(\delta^{-1}\Md)/\Rd$.
Assuming hereafter a flat rotation curve, $\alpha=0$, 
we obtain for the characteristic clump circular velocity
\be
\Vc \simeq 50\kms \,c_3^{1/2}\,\mu_{.5}^{1/4}\, \delta_{.2}\,
V_{{\rm d},200} \, ,
\label{eq:Vc}
\ee
where $c\seq 3\,c_3$, $\mu \seq 0.5 \mu_{.5}$, $\delta\seq 0.2\,\delta_{.2}$,
and 
$\Vd\seq 200\kms V_{{\rm d},200}$.
This could be inserted in \equ{Evir} for $\Evir$.

\smallskip
The fiducial values of the parameters are crudely justified as follows.
A contraction factor of $c \ssim 3\sdash 5$ is expected for clump virialization
and rotation support \citep{ceverino12}.
A galactic circular velocity of 
$\Vd \ssimeq 1.5\Vv \ssimeq 200\kms$
is expected for a galaxy in a halo of 
$\Mv \ssim 3 \times 10^{11}\msun$
at $z \ssim 2$, based on the virial relation, \equ{MvVv} below.
This is above the threshold for long-lived disks at all redshifts
\citep{dekel20_flip}.
This halo mass implies a stellar mass of 
$\Ms \ssim 3\times 10^{9} \msun$
\citep{behroozi19}, and a comparable gas mass for $\fg\ssimeq 0.5$
(see \se{galaxy_mass}).
A value of $\delta \ssim 0.2$ can be expected for a gas fraction in the disc 
of $\fg \ssim 0.5$, given the contribution of the bulge and the dark matter to
$\Mt$ in the denominator of $\delta$ (see \se{fg}).
%
\Equ{Vc} is in good agreement with the values obtained for clumps in the
\velathree simulations \citep[][Fig.~8]{mandelker17}, where the typical galaxy
masses at $z\ssim 2$ are indeed comparable to the fiducial values mentioned
above, to be discussed further below.

\subsection{Rotation and dispersion: Jeans equilibrium}
\label{sec:sigma}

Certain energy terms in \equ{energy} depend on the clump   
velocity dispersion, $\sigc$,
which characterizes the turbulence in the clumps, and is significantly 
larger than the speed of sound of $\sim\!10\kms$ at a typical temperature
of $\sim\!10^4$K. 
In $\sigc$ we refer to the one-dimensional component of the velocity dispersion,
specifically the radial component.
Following \citet{ceverino12}, assuming that the clump is in Jeans equilibrium,
$\sigc$ can be evaluated from $\Vc$ and the clump contraction factor $c$, 
as follows.

\smallskip
Assuming cylindrical symmetry, the Jeans equation in the equatorial plane reads
\be
\Vc^2 = \Vrotc^2 - \frac{r}{\rho} \frac{\partial(\rho\,\sigc^2)}{\partial r}
\simeq \Vrotc^2 + 2\,\sigc^2 \, ,
\label{eq:Jeans}
\ee     
where $\Vrotc$ is the rotation velocity in the clump,
and $\sigc$ is the radial velocity dispersion, which is assumed to be constant
in the second equality.
The factor 2 is for an isothermal-sphere-like density profile,
$\rho \!\prop\! r^{-2}$. In practice, simulations show that for entire discs 
this factor can vary  
between unity near the effective radius of the gas and 4 at five effective 
radii, well outside the main body of the disc \citep{kretschmer21_kvir},
indicating that variations may be expected within the clumps as well.
We note that with no rotation, $\sigc^2 \ssimeq 0.5 \Vc^2$.

\smallskip
For a disc in marginal Toomre stability with $Q \seq 1$, and a clump
contraction factor $c$, we obtain from \equ{deltaQ1} to \equ{c}
that the clump is related to the disc via
\be
\Vc^2 = \frac{\pi}{2}\, c\, \mu^{1/2}\, \sigrd^2 \, ,
\label{eq:Vc_sigrd}
\ee
independent of $\alpha$.
Assuming that angular momentum is conserved during the collapse of the clump,
one obtains \citep[as in eq.~19 of][]{ceverino12} a second relation between
clump and disc,
\be
\Vrotc^2 = \frac{\pi^2}{32}\, (1+\alpha) \,\mu\, c^2 \sigrd^2 \, . 
\label{eq:Vrot1}
\ee 
Using \equs{Vc_sigrd} and (\ref{eq:Vrot1}), we obtain within the clump
\be
\Vrotc^2 = \frac{\pi}{16}\, (1+\alpha)\, c\, \mu^{1/2}\, \Vc^2 
\simeq 0.2\, c\, \mu^{1/2}\, \Vc^2\, ,
\label{eq:Vrot2}
\ee
where the second equality is for a flat rotation curve, $\alpha\seq 0$.
Note that full rotation support, $\Vrotc \seq \Vc$, is obtained for 
a maximum contraction factor $c \seq 5\, \mu^{-1/2}$.
Inserting \equ{Vrot2} in \equ{Jeans} we obtain
\be
\frac{\sigc^2}{\Vc^2} \simeq 0.5 - 0.1\, c\, \mu^{1/2}
\equiv 0.5\, \fsig \, .
\label{eq:sigma_Vc}
\ee 
The factor $\fsig$ defined here can range from $\fsig\seq 1$
in the case of no rotation to $\fsig\seq 0.58$ 
when angular-momentum is conserved during clump collapse, 
with the fiducial values $c\seq 3$ and $\mu \seq 0.5$.
%
This gives
\be
\sigc \simeq 35 \kms \, \fsig^{1/2}\, V_{\rm c,50} \, .
\label{eq:sigma_35}
\ee
For $\Vc\seq 50 \kms$, a value of $\sigc \ssimeq 35\kms$ is expected
if there is no angular momentum,
and 
$\sigc \ssimeq 27\kms$ when angular momentum is conserved,
so we denote $\sigc \seq 30\kms\, \sigma_{30}$.

\subsection{Dissipation of Turbulence}
\label{sec:Edis}

\smallskip 
\Equ{sigma_Vc} allows us to express $\Edis$, the dissipative loss of the 
turbulence per unit mass, as a function of $\Vc$.
Assuming that the timescale for turbulence decay is $\gamma\, \tff$, 
where $\gamma$ is a factor of order unity, the dissipated energy by time $t$ is
\be
\Edis = \frac{3}{2} \sigc^2\, \gamma^{-1}\, \tau
= \frac{3}{4} \fsig\, \gamma^{-1}\, \Vc^2\, \tau \, ,
\label{eq:Edis}
\ee
where the turbulence is assumed to be isotropic with $\sigc$ the one-dimensional
velocity dispersion.
The corresponding dimensionless factor in \equ{factors2} is
\be
\fdis = 1.5\, \fsig\, \gamma^{-1} \, .
\label{eq:fdis0}
\ee
A comparison to \equ{Evir} indicates that $\Edis$ becomes comparable to $\Evir$
at $\tau$ of order unity or a few, when $\Eout$ in \equ{Eout}
is still much smaller.

\section{Sources: supernovae and gravity}
\label{sec:sources}

Complementing the previous section, we evaluate in this section
the energy input to the clump gas by supernovae and by gravity.

\subsection{Energy deposit by Supernova Feedback}
\label{sec:Esn}

\def\esn{e}

Here, we evaluate $\Esn(t)$, the energy per unit mass
deposited in the clump gas by supernovae bubbles till time $t$. 
While the feedback also contains contributions from stellar winds and radiative
pressure, we tentatively limit the analysis here to the energy deposited in
the ISM by supernovae.
The calculation is inspired by \citet{ds86}, following the revised treatment
of supernova bubbles in \citet{dekel19_ks} and ongoing work (Tziperman, Sarkar,
Dekel, in preparation).
The energy is evaluated here as the sum of the energies deposited by 
individual supernova bubbles.
We assume a constant SFR during the time relevant for clump evolution,
motivated by \citet{dekel22_mass}.
The calculation is simplified by the realization that the relevant time of 
one or more free-fall times is much larger than the fading time of each 
bubble, $\tfade$, and that overlap of active bubbles younger than $\tfade$ is 
negligible as long as they occur in random positions within the clump.
An alternative calculation could address super-bubbles generated by supernovae 
that are clustered in subclumps, e.g., in the spirit of \citet{gentry17}
and \citet{dekel19_ks}.

\subsubsection{An individual supernova bubble}
\label{sec:individual}

According to the standard model for the evolution of a supernova bubble in a
uniform medium, 
the first important stage is the Sedov-Taylor adiabatic phase, during which the
shock radius grows as $R\!\prop\!t^{2/5}$ and its velocity slows down as
$V\!\prop\!t^{-3/5}$. The end of this stage is commonly defined at 
the cooling time, $\tcool$, when the bubble has lost one third of its initial
energy to radiation\footnote{This is when the radiated energy is comparable to
the kinetic energy and to the thermal energy, each contributing one third of
the supernova energy at $\tcool$ \citep[][Figure A1]{kim15,sarkar21}.}. 
For an assumed cooling rate \citep[][eq.~9]{dekel19_ks}, 
which is a fair approximation for a solar-metallicity gas at temperatures in
the range $10^{5-7.3}K$, the cooling time
is estimated to be \citep[][eq.~12]{dekel19_ks}
\be
\tcool \simeq 3.95 \kyr \, n_2^{-0.55} e_{51}^{0.22} \, ,
\label{eq:tc}
\ee 
where $e_{51}$ is the initial supernova energy in units of $10^{51}\erg$.
The shell velocity and radius just prior to the cooling time are
$V_{\rm cool} \ssimeq 340\kms\, n_2^{0.13}\, e_{51}^{0.066}$
and $R_{\rm cool} \ssimeq 3.53\pc\, n_2^{-0.42}\, e_{51}^{0.29}$.

\smallskip
The second stage is the snow-plow phase, where a dense shell of cold gas is
mainly pushed by the pressure of the enclosed hot central volume.
The shell mass increases as it sweeps up the ambient gas, and it slows down
as $R \!\prop\! t^{2/7}$ and $V \!\prop\! t^{-5/7}$.
The velocity of the shock radius in the snow-plow phase is 
$V_{\rm s} \seq (2/7)\, R_{\rm cool}/t$ 
\citep[note the discontinuity in the velocity at $\tcool$, e.g.,][]{sarkar21}.
In the snow-plow phase, the total energy in the bubble is commonly assumed 
to decline as $\prop t^{-4/7}$, but we adopt a slightly steeper decline
of $\prop t^{-0.7}$, based on more involved theoretical arguments and 
spherical simulations \citep{dekel19_ks}.

\smallskip
The bubble is regarded as faded away at time $\tfade$,
when its outward velocity becomes comparable to the velocity dispersion 
outside it, $\sigc$. 
The energy is deposited until that time, and soon after the bubble
melts into the general clump medium. 
In analogy to eq.~18 of \citet{dekel19_ks}, the fading time is evaluated to be 
\be
\tfade = 73\kyr\, n_2^{-0.37}\, e_{51}^{0.32}\, \sigma_{30}^{-7/5} \, ,
\label{eq:tfade}
\ee
where 
$\sigma_{30} = 30\kms\,\sigc$.
This fading time is significantly smaller than the clump free-fall time of 
$\sim 5\Myr$ (\equnp{tff}), over which the clump evolves.
The shell radius at that time is
\be
\Rfade = 8.1 \pc \, n_2^{-0.37}\, e_{51}^{0.32}\, \sigma_{30}^{-2/5} \, ,
\label{eq:Rfade}
\ee
which is significantly smaller than the clump radius of $\gsim\! 200\pc$.
The energy deposited in the medium by a single bubble until $\tfade$ is
\be
\efade \simeq 8.7\times 10^{49} \erg\, n_2^{-0.13}\, e_{51}^{0.93}\,
\sigma_{30}^{0.98} \, .
\label{eq:efade} 
\ee

\subsubsection{Multiple supernova bubbles}
\label{sec:multiple}

Following \citet{ds86}, we write the total deposited energy by time $t$ as
\be
\Mg\,\Esn(t) = \int_0^{N(t)} e(t-t_*)\, dN(t_*) \, ,
\ee
where $e(t')$ is the energy deposited by a single supernova from its birth
to time $t'$, 
and $N(t)$ is the cumulative number of supernovae by time $t$. 
Assuming a constant SFR during the early phases of the clump lifetime,
\be
N(t) = \nu\, \sfr\, t \, ,
\label{eq:N}
\ee
with $\nu = 0.01 \Msun^{-1} \nu_{.01}$ the number of supernovae per
solar mass of forming stars, determined by the stellar initial mass function.
A change of variables gives
\be
\Mg\,\Esn(t) = \nu\, \sfr\, \int_0^t e(t')\, dt' \, .
\ee
For $t \sgg \tfade$, this simplifies to 
\be
\Mg\,\Esn(t) = \efade\, \nu\, \sfr\, t \, .
\ee
Expressing the SFR via \equ{sfr} we obtain
\be
\Esn(t)  
= \frac{1}{2} (51\kms)^2\, n_2^{-0.13}\, \nu_{.01}\, \epsilon_{.03}\, 
e_{51}^{0.93}\, \sigma_{30}^{0.98}\, \tau .
\label{eq:Esn_sigma}
\ee
Expressing $\sigc$ by $\Vc$ using \equ{sigma_Vc} and \equ{sigma_35} we obtain
\be
\Esn(t) = \frac{1}{2} (55\kms)^2 
n_2^{-0.13} \nu_{.01}\, \epsilon_{.03} e_{51}^{0.93} \fsig^{0.49} 
V_{\rm c,50}^{0.98} \tau .
\label{eq:Esn}
\ee
The corresponding dimensionless factor in \equ{factors1}, ignoring the weak $n$
dependence and rounding the power $0.98$ to $1$, is
\be
\fsn = 1.22\, \nu_{.01}\, \epsilon_{.03}\, e_{51}^{0.93}\, \fsig^{0.49}  \, .
\label{eq:fsn0}
\ee
The characteristic velocity for supernova feedback, $\Vsn$, is defined by
$\Esn \seq 0.5\,\Vsn^2$. For the fiducial values of the parameters it is
$\Vsn \ssim 50\kms\, \tau^{1/2}$.

\smallskip
Comparing to \equ{Edis},
we find that $\Esn(t)$ is comparable to $\Edis(t)$ (if $\Vcf \ssim 1$), or
larger (if $\Vcf$ is significantly smaller than unity). 
They are always larger than $\Eout(t)$ of \equ{Eout}. 
They both become comparable to $\Evir$ of \equ{Evir} at $t \ssim \tff$. 
Note that $\Esn$ is roughly proportional to $\Vc$, while the other energies
scale with $\Vc^2$.

\smallskip 
In order to test the validity of neglecting bubble overlap,
we estimate the fraction of the clump volume that is occupied by
active bubbles at a given time. By inserting 
$\tfade \ssim 70\kyr$ from \equ{tfade} in \equ{N}, 
with the fiducial values for $\nu$ and $\epsf$, 
and using 
$\Rfade \ssim 8\pc$ 
from \equ{Rfade},  
the volume in active bubbles is estimated to be 
$\sim\! 10^6 \pc^3$. 
This is while the clump volume, assuming a radius of $\Rc \ssim 300\pc$,
is $\sim\! 10^8\pc^3$.
This implies that only a small fraction of the clump  
is occupied by active bubbles, which 
justifies the neglect of bubble overlap as long as the supernovae are spread
over the bubble volume.

\subsection{Gravitational Driving of Turbulence}
\label{sec:Egrav}

During the formation of the clump, in the first one to a few free-fall times,
the contraction by a factor $c$ is associated with a deepening of the potential 
from $-G\Mc/(c\Rc)$ to $-G\Mc/\Rc$, namely a gravitational energy gain of
\be
\Egrav = (1-c^{-1})\, \Vc^2 \, .
\label{eq:Egrav}
\ee 
This is comparable to $\Evir$, and to $\Edis$ and $\Esn$ at $t\ssim \tff$.
Thus, at $\tau \ssim 1$, we expect in \equ{factors1} 
$\fgrav \seq 2\,(1-c^{-1})/\tau$, of order unity.

\smallskip
At later times, we expect the gravitational energy input rate to become
smaller, such that $\Egrav(t)$ becomes smaller than $\Edis(t)$ and
$\Esn(t)$. This implies that, with $\Egrav \seq \fgrav\,\Evir\, \tau$,
we expect $\fgrav \ssim 1$ at $\tau \ssim 1$, and $\fgrav \slt 1$ later.
It could in general be much smaller than unity, except during episodes of
mergers, intense accretion, or excessive tidal effects and shear, when 
it could grow to $\fgrav \slsim 1$. 

\smallskip 
To get an estimate for $\fgrav$ long after $\tau \seq 1$, we appeal to
gravitational clump encounters.
For discs that are self-regulated at marginal Toomre instability,
\citet{dsc09} estimated the timescale for stirring turbulence by 
gravitational encounters between clumps to be on the order of the disc 
dynamical time $\td$ or larger (their eq. 18 in \S 3).
This is the timescale for the encounters to generate a specific energy change 
of order $\sigrd^2$, and it is estimated to be 
\be
\tenc \simeq 2\,\beta_{.2}^{-1}\, Q_{.67}^4\,\td
\sim 10\,c_3^{3/2}\,\tff \, . 
\ee
Here $\beta \ssim 0.2$ is the instantaneous disc mass fraction in clumps,
and $Q \ssimeq 0.67$ is the value of the Toomre parameter
for marginal instability in a thick disc.
The second equality stems from the clump free-fall time being 
$\tff \ssim c^{-3/2} \td$.
This indicates that at $\tau$ significantly larger than unity, typically,
\be
\fgrav \lsim 0.1\, c_3^{-3/2} \, ,
\label{eq:fgrav_enc}
\ee
namely significantly smaller than unity, on the order of $0.1$.
The value of $\fgrav$ could be larger during episodes of excessive 
gravitational interactions. 
This is especially true during clump mergers, and possibly also 
in episodes of intense accretion, strong tidal effects and shear. 
While only a fraction of the clump encounters lead to actual mergers,
$\fgrav$ inside the merging clumps is expected to be larger than estimated
in \equ{fgrav_enc}.

\smallskip
Based on \equ{Egrav} and \equ{fgrav_enc}, we crudely adopt
\be
\fgrav \sim
\begin{cases}
2(1-c^{-1}) \ \ {\rm if}\ \tau \sim 1\\
0.3 \ \ \ \ \ {\rm if}\ \tau > 1, \ {\rm for\ low\ } c\\
0.1 \ \ \ \ \ {\rm if}\ \tau > 1, \ {\rm for\ high\ } c\, .\\
\end{cases}
\label{eq:fgrav0}
\ee



\section{Non-virialized clumps}
\label{sec:non-virial}

\def\f{p}

We present here a modification to $\So$ of \equ{S}, that is especially
relevant for the \SC clumps that disrupt on a free-fall timescale.
Such clumps may not have reached an equilibrium before being disrupted, 
such that the binding energy per unit mass in them is smaller than $\Evir$ 
at virial equilibrium. 
This introduces corrections to the expressions in the previous sections, 
in particular the survivability parameter $\So$.
The binding energy per unit mass, the gravitational energy gained during the
collapse of the clump and the energy needed for disrupting the clump,
can be written as
\be
\Eb = \frac{G\Mc}{\Rc} - \frac{G\Mc}{\Rci} 
= (1-\f) \Vc^2 \, ,
\label{eq:Eb}
\ee
where $\Vc^2 = G\Mc/\Rc$, with $\Vc$ and $\Rc$ the clump circular velocity and
radius but not necessarily their virial values.
The factor $\f$ should be $\f \seq c^{-1}$ where $c=\Rci/\Rc$, but we note that 
for $\f\seq 0.5$ the binding energy equals the virial energy, and we are back 
to the analysis of virialized clumps. We therefore define
\be
\f =  
\begin{cases}
c^{-1}\ \ {\rm if}\ c \sleq 2\\
0.5 \ \ \ {\rm if}\ c \sgt 2 \, .\\
\end{cases}
\label{eq:f(c)}
\ee
We note that $\Rc$ and $\Vc$ relate to their virial analogs as
$\Rc \seq 2\f \Rcv$ and $\Vc^2 \seq (2\f)^{-1} \Vcv^2$.

\smallskip
The factor $\f$ could be interpreted as the fraction of $\Mc$ that is left 
bound after the instantaneous mass removal. This is because, if we equate the 
energy change by the removal, 
\be
\Delta E = \frac{G\Mc}{\Rc} - \frac{\f G\Mc}{\Rc} = (1-\f) \Vc^2 \, , 
\ee
with $\Eb$ of \equ{Eb}, we see that this $\f$ is the same as the $\f$
in \equ{f(c)}. 
This implies that in order to unbind the clump, one needs to remove a fraction 
$(1-\f)$ of its mass.  We therefore generalize \equ{S} to
\be
\So = \frac{(1-\f)\,\Mc}{\Mout} \, .
\label{eq:S_sim_f}
\ee

\smallskip
Now \equ{S2} remains the same but with $\Evir$ replaced by $\Eb$,
namely $\Vcv^2$ replaced by $\Vc^2 \seq (2\f)^{-1} \Vcv^2$.
Accordingly, as in \equ{Eout0}, we assume escape with $\Vout^2 \seq 2 \Vc^2$. 
Then, in \equ{So_Sp}, the expression for $\So$ remains the same,
and $\Evir$ is replaced by $\Eb$ in the expression for $\Sp$.
Defining the ``$f$" factors as in \equ{factors1} and \equ{factors2},
the expression for $\So$, replacing $\Sp$ of \equ{Sp}, becomes
\be
\So =\frac{\fgc^{-1}}{\Sp^{-1} -1} , \ \
\Sp = \frac{2(1-\f)}{(\fsn\Vcf^{-1} + \fgrav - \fdis)\tau} \, .
\label{eq:S_mod_f}
\ee
The only difference is the factor $2(1-\f)$ in $\Sp$. One should also note that
$\Vc$ here is the circular velocity of the clump as is, even when it is not 
in virial equilibrium.

\smallskip
We next evaluate the potential modifications 
to the different energies in terms of the clump parameters.
The expression for $\fgrav$ remains the same as in \equ{fgrav0}.
The expressions for $\Edis$ and $\Esn$, in \equ{Edis} and \equ{Esn_sigma}, 
depend on $\sigc$. In order to relate $\sigc$ to $\Vc$, we used \equ{sigma_Vc}, 
derived assuming Jeans equilibrium (\equnp{Jeans}), clump properties in Toomre 
instability with $Q\seq 1$ (\equnp{Vc_sigrd}), 
and conservation of angular momentum (\equnp{Vrot1}). 
If the clump is not in equilibrium, the relation between $\sigc$ and $\Vc$ 
becomes uncertain. We found in \citet{ceverino12} that the 
final value of the clump velocity dispersion, $\sigcv$, is comparable to 
$\sigrd$ of the disc. We thus make the assumption that $\sigc$ is constant 
during the clump collapse. 
As in \equ{sigma_Vc}, we parameterize the relation between $\sigc$ and $\Vc$ 
in terms of the parameter $\fsig$, of order unity,
$\sigc^2/\Vc^2 \!\equiv\! 0.5\fsig$.
Then, 
the expressions for $\Edis$ and $\Esn$ remain as in \equ {Edis} and \equ{Esn},
and so are $\fdis$ and $\fsn$ in \equ{fdis0} and \equ{fsn0}.
The value of $\fsig$ can be determined either from $\sigc \seq \sigcv$
(version 1) or from $\sigc \seq \sigrd$ (version 2), 
with comparable results for typical clumps.
Based on \equ{sigma_Vc} and \equ{Vc_sigrd}, we can have, respectively,
\be
\fsig = 
\begin{cases}
1 - 0.2\, c\, \mu^{1/2}\, , \ \ \sigc \seq \sigcv\\ 
(4/\pi)\, c^{-1} \mu^{-1/2}\, , \ \ \sigc \seq \sigrd\, .
\end{cases}
\label{eq:fsig_f2}
\ee 
We adopt hereafter version 2, $\sigc \seq \sigrd$,
with version 1 yielding similar results.

\section{Survivability of three clump types}
\label{sec:types}

Based on the previous sections,
the ``$f$" factors that enter $\Sp$ in \equ{Sp} or \equ{S_mod_f}
can be summarized as
\be
\fsn \simeq 1.22\, \nu_{.01}\, \epsilon_{.03}\, e_{51}^{.93}\, \fsig^{.49} \ ,
\label{eq:fsn}
\ee
\be
\fdis \simeq 1.5\, \fsig\, \gamma^{-1} \sim 1 \, ,
\label{eq:fdis}
\ee
\be
\fgrav \sim 
\begin{cases}
2(1-c^{-1}) \ \ {\rm if}\ \tau \sim 1\\
0.3 \ \ \ \ \ {\rm if}\ \tau > 1, \ {\rm for\ low\ } c\\
0.1 \ \ \ \ \ {\rm if}\ \tau > 1, \ {\rm for\ high\ } c\, .\\
\end{cases}
\label{eq:fgrav}
\ee
The $\Vc$ that enters the first term in \equ{Sp} is given by \equ{Vc},
and $\fsig$ is given by \equ{fsig_f2}.
For completeness,
\be
\fout \simeq 0.06\, \epsilon_{.03}\, \eta \, .
\label{eq:fout}
\ee
We note that for $\Vc \ssimeq 50 \kms$ and $\fsn \ssimeq 1$,       
we have $\Esn(t) \ssimeq \Edis(t)$, 
and they are both comparable to $\Evir$ at $t \ssimeq \tff$,
while $\Eout(t) \ssimeq \Edis(t)/16$ at any time.

\subsection{Short-lived versus long-lived clumps}
\label{sec:types_SC_LC}


\smallskip
At the end of the formation of the clump, near $\tau \ssim 1$, we expect 
$\fgrav \ssim 1$ based on \equ{Egrav}. 
Then, for the fiducial $\fdis \ssim 1$, \equ{Sp} simplifies to  
\be
\Sp \ssimeq \frac{\Vcf}{\fsn\, \tau} \, .
\label{eq:Sp_ff}
\ee
This is to be multiplied by $2(1-\f)$ if correcting for non-virialized clumps 
as in \equ{S_mod_f}.
This implies that the clump survival is mainly determined by the competition 
between the energy deposited by feedback and the binding energy of the 
clump.
For moderate feedback strength, $\fsn \ssim 1$, and for massive clumps of 
fairly deep potential wells, $\Vcf\ssim 1$,
we obtain $\Sp \ssimeq 1$. For $\fgc \ssim 1$ this yields $\So \sgg 1$.
These clumps are thus expected to keep most of their gas during the early
phase of clump evolution. 
We term these {\it long-lived} clumps (\LCnos).
If, on the other hand, $\fsn \Vcf^{-1} \sgt 2$, namely the feedback is
stronger and/or the clump $\Vc$ is lower, 
we get $\Sp \slt 0.5$, which implies $\So \slt 1$.
In this case $\tau \So$ is also smaller than unity at the early times,
so these clumps are expected to disrupt on a free-fall timescale.  
We term these {\it short-lived} clumps (\SCnos)

\smallskip
The conditions for \LC clumps
can arise from a large $\Vc$ with the fiducial values for
supernova feedback, or from a weaker feedback.
Based on \equ{Vc}, a large $\Vc$ can be obtained for either
a massive disc with a large $\Vd$ or
a large-$\delta$ disc due to a high gas fraction 
or a low-mass bulge and low central dark-matter mass.
A high clump contraction factor and a clump mass that is comparable to
the Toomre 
mass can also contribute to a large $\Vc$, but in a weaker way.
Based on \equ{S2}, a low $\fgc$ will increase $\So$ 
and help the survivability.
On the other hand,
the conditions for \SC clumps 
can be due to a smaller $\Vc$, namely lower-mass clumps,
or to stronger feedback that yields a larger $\fsn$.

\subsection{Two types of long-lived clumps}
\label{sec:types_LS_LL}

Among the \LC clumps that keep most of their gas until after the first 
few free-fall times, we envision two sub-types.
Certain \LC clumps, to be termed \LL clumps, may lose only a small
fraction of their gas mass to outflows for tens of free-fall times, 
keeping a non-negligible gas mass and SFR until late in their inward migration 
at $\tau \sim 50 \sdash 100$. 
Other \LC clumps, to be termed \LS clumps,
may lose most of their gas mass within the first 
few or $\sim$ten free-fall times, but unlike the \SC clumps that lose
their gas more rapidly, the \LS clumps manage to 
form stellar components and keep them bound for long lifetimes, showing only
little gas and low SFR. 
These are the clumps that, despite having $\So$ drop to below unity
by $\tau \ssim 10$, have $\tau \So \sgt 1$ in \equ{St}, 
indicating that they do not fully disrupt due to the rather slow outflow rate.
The properties of the \LS clumps are expected to be in certain ways between 
those of the \SC clumps and the \LL clumps.
The existence of \LS clumps was hinted in simulations already in 
\citet{mandelker17}, where they were not analyzed in as much detail as
the \SC and \LL clumps.
In the simulation results described below we analyze this population of 
clumps as well.
Next, we try to consider a possible way to understand the origin of this
intermediate population of \LS clumps.

\smallskip 
A while after $\tau \ssim 1$, for \LC clumps that have survived the
early phase of clump formation, one may inquire
what values of the parameters in \equ{Sp} may distinguish between \LS and \LL
clumps. It seems that the value of $\fsn \Vcf^{-1}$ has a crucial role.
Lets assume $\fdis \ssim 1$.
Since at $\tau$ of a few or greater we expect $\fgrav$ to become smaller than 
unity, $\Sp$ in \equ{Sp} is larger than in \equ{Sp_ff}, which was roughly 
valid at $\tau \ssim 1$.
If $\fgrav \sll 1$, and $\fsn \Vcf^{-1} \ssim \fdis \ssim 1$,
then in \equ{Sp} $\Sp$ is close to unity and in \equ{tau1} $\taus \sgg 1$.
At $\tau$ of a few this corresponds to $\So \sgg 1$, namely an \LL clump.
If, on the other hand, $\fsn \Vcf^{-1}$ is larger than unity,
and $\fgrav$ is not too small,
we expect $\Sp$ to decline to $\sim 1/\tau$, so as long as $\fgc$
has not dropped yet by a large factor (e.g. if it is declining slower than 
$1/\tau$), we obtain $\So \slt 1$, namely significant mass loss. 
This would be a non-disrupted \LS clump with $\tau \So \sgt 1$
because roughly $\tau \So \prop \fgc^{-1}$, a growing function of time. 

\smallskip 
For a numerical example, if the \LS clumps have $\fsn \Vcf^{-1} \ssimeq 2$, 
$\fgrav \ssimeq 0.3$ and $\fdis \ssimeq 1.3$ (because of the higher $\fsig$),
with $\fgc \sim 0.67$, we get $\So \ssimeq 1.5/(\tau-1)$, which would cross 
the critical value at $\taus \ssimeq 2.5$.
This is compared to much larger values of $\So$ and $\taus$ for \LL clumps,
where typical values may be $\fsn \Vcf^{-1} \ssimeq 1$, 
$\fdis \ssimeq 1$ and $\fgrav \sll 1$.
The bimodality is emphasized by the vanishing values obtained in the
denominator of \equ{So_Sp} for \LL clumps, which could yield large values
of $\Sp$ near unity and therefore very large values of $\So$.   
For $\tau$ significantly larger than unity, these clumps have $\tau\So \sgt
1$, namely no total disruption, as expected for \LS clumps.

\smallskip 
Motivated by the simulation results described below,
it is possible that the distinction between \LS and \LL clumps may be 
associated with a difference in collapse factor, with the \LS clumps 
contracting less than the \LL clumps. 
For a given mass, a smaller $c$ would correspond to a larger $\Rc$ and
therefore a smaller $\Vc$ for \LS clumps.
Based on \equ{Vrot2}, this will be associated with a weaker rotation support.
The larger $\Rc$ in the \LS clumps may be associated with a larger $\fgrav$ 
due to more intense clump mergers and stronger tidal effects.
These may increase the SFR efficiency $\epsf$ (e.g., via shocks),
and thus increase $\fsn$.
The combination of a larger $\fsn$, a smaller $\Vc$ and a larger $\fgrav$ 
can significantly lower $\Sp$ in \equ{Sp} for \LS compared to \LL clumps, 
thus generating a bimodal distribution of \LC clumps.
This possible explanation will be tested in the simulations.

\smallskip 
The distinction between the \LS clumps and \SC clumps could be addressed
in terms of $\tauts$, given in \equ{tau2}.
Assuming for simplicity $\fgrav \seq \fdis \seq 1$, namely \equ{Sp_ff},
and $\fgc \seq 1$, we obtain
\be
\tauts \ssim [\fsn\,\Vcf^{-1} - 1 ]^{-1} \, .
\ee
For $\fsn\,\Vcf^{-1}$ significantly larger than unity, we get a small
$\tauts$, namely an \SC clump.
For $\fsn\,\Vcf^{-1}$ only slightly above unity, we get
$\tauts \sgg 1$, namely an \LS clump.

\section{Thresholds for clump survival}
\label{sec:thresholds}

We now analyze the distinction between \SC and \LC clumps in terms of
the implied thresholds for the relevant physical quantities. 
Appealing to the critical value $\So \ssim 1$ at $\tau \ssim 1$, 
we investigate the critical condition $\Sp \ssim \Vcf/(\fsn\,\tau) \ssim 0.5$, 
based on \equ{Sp_ff}, in terms of the relevant
disc and clump properties.
These predictions will be compared to simulations in the following sections.

\subsection{Galaxy velocity and  mass}
\label{sec:galaxy_mass}

Assuming tentatively a moderate feedback, $\fsn \ssimeq 1$,
one of the strong dependences of $\Vc$ in \equ{Vc}  
is on $\Vd$, meaning that \LC clumps below and near the Toomre mass
are expected in galaxies above a threshold disc rotation velocity.
With the fiducial values of $c_3 \ssim \delta_{.2} \ssim \mu_{.5} \ssim 1$ 
in \equ{Vc} and \equ{Sp_ff}, we typically expect \LC clumps of 
$\Vc \sgt 25\kms$ in discs of $\Vd \sgt 100\kms$.
Note that these velocities could double if $\fsn \ssim 2$ or if $\So$ is 
considered at $\tau \ssim 2$, implying that our model predictions should be 
considered as semi-qualitative estimates only.


\smallskip
We recall that discs tend to be long lived, not disrupted by mergers in an 
orbital time, when they reside in halos more massive
than a threshold mass of $\Mv \ssim 2\!\times\!10^{11} \msun$
\citep{dekel20_flip}.
This is also the mass range where extended rings survive instability-induced
inward mass transport due to a massive central bulge (or a central cusp of 
dark matter) \citep{dekel20_ring}. Such a bulge forms by a wet compaction
event \citep{zolotov15}, which typically occurs near a similar critical mass
\citep{tomassetti16}.
This mass is slightly smaller but in the ball park of the ``golden mass" of 
most effective galaxy formation \citep{dekel19_gold}.

\smallskip
Above this critical mass, the disc is supported by rotation,
so $\Vd$ can be assumed to be comparable to and slightly larger than the 
halo virial velocity $\Vv$, e.g., $\Vd \ssimeq 1.5\, \Vv$.
Then $\Vd$ that enters \equ{Vc} can be related to the halo mass
via the halo virial relation \citep[e.g.,][appendix]{db06},
\be
M_{{\rm v},11.3} \simeq V_{{\rm v},120}^{3} (1+z)_3^{-3/2}
\simeq 1.37\, V_{{\rm d},200}^{3} (1+z)_3^{-3/2} \, . 
\label{eq:MvVv}
\ee
Here, the virial quantities $\Mv$ and $\Vv$
are measured in units of $10^{11.3}\msun$ and $120\kms$ respectively, 
and $(1+z)$ is normalized to $3$ ($z\seq 2$).
At $z \ssim 2$, the typical stellar mass is related to the halo mass as
\be
\Ms \simeq 2\times 10^{9}\msun \, M_{{\rm v},11.3} \,f_{\rm sv} \, ,
\label{eq:MsMv}
\ee
with $f_{\rm sv} \ssimeq 1$ assumed hereafter 
\citep[][from galaxy-halo abundance matching]{moster18,behroozi19}.
The corresponding disc gas mass is
\be 
\Md \ssim 2\times 10^9\msun\, M_{{\rm v},11.3} \,\fgs \, ,
\label{eq:Md}
\ee 
where $\fgs \ssim 1$ is the gas-to-stellar mass ratio, such that the
gas-to-baryonic fraction is 
$\fg \seq (\fgs^{-1} +1)^{-1} \ssim 0.5$. 
Then, from \equ{MvVv},
\be
V_{{\rm d},200} \simeq 0.9\, M_{{\rm v},11.3}^{1/3}\, (1+z)_3^{1/2}
           \simeq 0.9\, M_{{\rm d},9.3}^{1/3}\, \fgs^{-1/3}\, (1+z)_3^{1/2} .
\label{eq:VdMd}
\ee
Thus, for masses above the threshold mass for discs, we expect 
$\Vd \sgt 100\kms$ at all redshifts, so $\Sp \sgt 0.5$ in
\equ{Sp}, namely \LC clumps.

\smallskip
In galaxies below the critical mass for discs, the velocity dispersion is 
significant, with $\Vd/\sigrd \ssim 1$, such that in Jeans equilibrium the 
rotation velocity $\Vd$ is smaller than $\Vv$ by a factor of a few
\citep{dekel20_flip,dekel20_ring,kretschmer21_kvir}.
This reduces $\Vd$ further, well beyond the $\Md^{1/3}$ dependence of 
\equ{VdMd}, such that $\Vd$ drops below $100\kms$ immediately below the
threshold mass for discs even at high redshifts. 
This brings $\Sp$ to below 0.5, thus not allowing \LC clumps in galaxies of 
$\Mv \slt 10^{11}\msun$, namely $\Md \slt 10^9\msun$. 
Indeed, in the \velathree simulations there are hardly any long-lived clumps in
disks of $\Md \!<\!2\times 10^{9}\msun$ \citep[][section 6.2.1]{mandelker17}, 
associated with $\Mv \!<\!2\times 10^{11}\msun$ haloes at $z\ssim 2$.

\subsection{Clump mass}
\label{sec:clump_mass}

The clump circular velocity $\Vc$ that enters $\Sp$ can be expressed in terms 
of the clump mass $\Mc$.
Using \equ{VdMd} and \equ{McMd} in \equ{Vc}, we obtain
\be
\Vcf \simeq c_3^{1/2}\, \mu_{.5}^{-1/12}\, \delta_{.2}^{1/3}\, \fgs^{-1/3}\,
(1+z)_3^{1/2}\, M_{\rm c,7.5}^{1/3} \, ,
\label{eq:VcMc}
\ee
where $\Mc = 10^{7.5}\msun\, M_{\rm c,7.5}$.
The value of $\delta/\fgs$, both referring to the galaxy gas disc,
is expected to be similar for different galaxies
(\se{fg}), so we are left with no explicit dependence on the galaxy mass.
The dependence on $\mu$ became negligible, so for a fixed $c$,
$\Vc$, and therefore $\Sp$, is determined by the absolute value of $\Mc$.
It implies that when sampling clumps in galaxies of different masses, 
one expects a threshold for survival at a critical clump mass,
with $\Vc \ssim 50\kms$ corresponding to $\Mc \slsim 10^8\msun$. 
On the other hand, we expect no clear threshold as a function of the clump 
mass relative to the disc mass, 
via $\mu$ or $\Mc/\Md$.

\smallskip  
This is indeed comparable to the transition clump mass from short-lived to 
long-lived clumps in the cosmological \velathree simulations at 
$\Mc\ssim 10^8\msun$ \citep[][Fig.~8]{mandelker17}, 
where the massive galaxies with clumps have 
$\Mv \ssim (2\sdash 9)\times 10^{11}\msun$ at $z \seq 2$, when 
most clumps are detected.
Indeed, in these simulations there is only a minor distinction between 
long-lived and short-lived clumps as a function of $\Mc/\Md$.

\subsection{Cold fraction and gas fraction}
\label{sec:fg}

The clump circular velocity $\Vc$ of \equ{Vc}
that enters $\Sp$ also depends strongly on the cold fraction
$\delta \seq \Md/\Mt$,
with a preference for \LC clumps at higher values of $\delta$.


\smallskip
The threshold in $\delta$ can be translated to a threshold in gas fraction
with respect to the baryons, for given stellar (disc plus bulge) and dark
matter components, and for the typical clumps of a given fraction $\mu$
of the Toomre mass.  
We assume four components within the disc radius: a gas disc, a stellar disc, 
a bulge and dark matter, and describe the relevant relations between them by 
three structure parameters as follows.
Let $f_{\rm db}$ be the ratio of dark-matter to baryonic mass (and adopt
a fiducial value of
$f_{\rm db} \seq 1$, but it could vanish in dark-matter-deficient cores). 
Let $f_{\rm bd}$ be the stellar bulge to disc ratio (fiducial value 
$f_{\rm bd} \seq 1$, but it could vanish in a bulge-less disc).
Let $f_{\rm cs}$ (between zero and unity)
be the fraction of the stellar disc that adds to the
gas in constituting the cold disc mass $\Md$, which participates in the
instability and therefore enters $\delta$ (fiducial $f_{\rm cs} \seq 0$, 
ranging from zero to unity).
The gas fraction is then
\be
\fg \frac{\Mg}{\Mg + \Ms} =
\frac{(1+f_{\rm db})(1+f_{\rm bd})\,\delta -f_{\rm cs}}
{1+ f_{\rm bd} -f_{\rm cs}} 
\lsim 0.4\, \delta_{.2}\, ,
\label{eq:delta_fg}
\ee
in which the last equality is 
for the fiducial values $f_{\rm db}\seq f_{\rm bd}\seq 1$ and $f_{\rm cs}=0$.
We then learn from the condition $\Sp \sgt 0.5$ that,
with the other parameters at their fiducial values,
the threshold for clump survival is $\fg \slsim 0.4$, 
which is in the ball-park of the observed values at $z\ssim2$
\citep{tacconi10,daddi10,tacconi18}.

\smallskip 
Relevant possible deviations from the fiducial values of the structure
parameters are likely to decrease the threshold value of $\fg$ and thus 
lead to increased survival.
As a first example, 
if stars contribute to the cold disc, say $f_{\rm cs} \seq 0.5$, 
for a survival threshold at $\delta\seq 0.2$ the corresponding gas-fraction 
threshold is at $\fg \ssim 0.2$.
Secondly,
a smaller bulge-to-disc ratio, $f_{\rm bd} \!<\! 1$, with the other parameters
fixed, leaves the threshold at $\fg \ssimeq 0.4\,\delta_{.2}$.
Thirdly, for a smaller dark-matter fraction, $f_{\rm db} \!<\! 1$,
with the other parameters fixed, the threshold gas fraction becomes smaller.
For instance, with no dark matter, the threshold is at 
$\fg \ssimeq 0.2\,\delta_{.2}$. 
In these three cases, for a given $\delta \ssim 0.2$ threshold for survival,
the gas fraction threshold would be equal or lower than for the fiducial 
structural cases $f_{\rm cs}\seq 0$ and $f_{\rm dn} \seq f_{\rm bd} \seq 1$, 
with a value 
\be
\fg \ssim (0.2\sdash 0.4)\, \delta_{.2} \, .
\ee

\smallskip 
We note that
the clump circular velocity $\Vc$ has an additional dependence on gas fraction
through $\Vd \!\prop\! \fgs^{-1/3}$, which somewhat weakens the overall 
$\fg$ dependence of $\Vc$ to roughly $\Vc \!\prop\! \fg^{2/3}$. 

\smallskip 
It is possible that the clump contraction factor $c$ 
is somewhat larger for a higher gas fraction in the disc, because more
dissipation is permitted.
If so, then the $\delta$ threshold for survival, based on \equ{Vc}, 
would become lower by a factor $c_3^{-1/2}$, and the $\fg$ threshold 
would decrease accordingly, thus increasing the clump survivability.

\smallskip 
Indeed, a comparable gas-fraction
threshold for the survival of massive clumps at $\fg\ssim 0.3$ has been
found in simulations of isolated galaxies of a given mass and a small bulge
\citep{fensch21}.
We note that the $\fg$ dependence in \equ{Vc}, stemming from the
$\delta$ dependence, is stronger than the mass
dependencies and slightly stronger than the dependence on the energy per 
supernovae in $\Sp \sprop \fsn^{-1} \sprop e^{-0.93}$.
This is consistent with the finding of \citet{fensch21}.

\smallskip 
One can note in \equ{VcMc} that when averaged over galaxies of different
masses, and allowing $\Mc$ to be at it's critical value for survival, 
the dependence on the gas fraction that stems from the product 
$\delta^{1/3}\, f_{{\rm gs}}^{-1/3}$, becomes significantly weaker than in the 
discussion above for a fixed $\Vd$ and a typical clump of a given $\mu$.

\subsection{Feedback strength}
\label{sec:feedback}

The supernova energy in $\Sp$, via $\fsn$, is almost linear with $e_{51}$. 
This is about unity for a typical supernova, 
but the energies from supernovae that cluster strongly in a star-forming cloud
add up in a super-linear way \citep{gentry17}, which may lead to a larger
effective value of $e_{51}$, representing both thermal and kinetic feedback.
This would lead to clump disruption when all other parameters are at their
above fiducial values, including massive clumps of about a Toomre mass.

\smallskip 
On top of the supernova feedback, there is feedback from stellar winds
and radiative feedback from massive stars, which may or may not be
boosted by infrared photon trapping at high gas densities.
These additional feedback mechanisms should be modeled in detail.
Here, for very qualitative results, we tentatively model them very crudely
by an increase in the energy per supernova $e_{51}$.
This would translate to a value of $\fsn$ somewhat above unity.

\smallskip 
It turns out that in a cosmological simulation suite, \velasixcom, 
similar to \velathree but with a stronger feedback, even
the clumps more massive than $\sim\!10^8\msun$ tend to disrupt after one to 
several disc dynamical times (see below, and Ceverino et al. in preparation), 
while clumps of similar masses tend to survive in \velathree 
for many free-fall times \citep{mandelker17}.
The feedback in \velathree included thermal feedback from 
supernovae, stellar winds, and radiative feedback from massive stars with no 
photon trapping, with a total injected momentum of $\sim\! 3\,L/c$.
In comparison,
the feedback in \velasix includes additional  
kinetic feedback from supernovae to represent strongly clustered supernovae, 
and stronger radiative feedback form 
massive stars due to infrared photon trapping at high gas densities.
In our current simplistic model based on energetics,
the associated enhanced feedback strength may very crudely be referred to as 
an increase in the effective $e_{51}$, bringing it to a value of a few, 
thus increasing
the disruptive term $\fsn$ that enters $\Sp$ in \equ{So_Sp}.
The same may be qualitatively true for other analyses and simulations that 
incorporated very strong winds or enhanced radiative feedback due to strong 
IR trapping, at a momentum level of tens of $L/c$ 
\citep{murray10,genel12,hopkins12,oklopcic17},  
in which all clumps were indeed disrupted on a few free-all timescales.
Some of these simulations also had low gas fractions, which contributed to
the poor survivability as discussed in \se{fg}.
We note that several arguments have been put forward against the plausibility 
of such extremely strong radiative feedback
\citep{kd10,krum_thom12,krum_thom13,dk13}.

\subsection{Star-formation efficiency}
\label{sec:sfr}

The disruptive factor $\fsn$ is linear with $\epsf$,
the SFR efficiency per free-fall time in the star-forming
region. This efficiency is assumed to have a given fixed value under normal
conditions, at the level of a few percent, based on observations and theory 
\citep[e.g.,][]{kdm12}. The actual value for our giant clumps would depend
on whether the star-forming region is regarded as the whole clump or smaller
and denser sub-clumps that may cluster together to form the giant clump or be
fragments of the giant clump.
A value of $\epsilon \ssim 0.1$, which may be valid during mergers
and strong tidal effects and therefore be associated with a high $\fgrav$,
would strengthen the disruption power, 
while a value of $\epsilon \ssim 0.01$ would increase the survivability. 

\smallskip 
The parameter $\nu$, the number of supernovae per one solar mass of forming
stars, depends on the stellar initial mass function (IMF).
For a Chabrier IMF the common value is $\nu \!\simeq\! 0.01\Msun^{-1}$
or lower
\citep{botticella17}.
A top-heavy IMF will make $\nu$ larger, which may tip the balance 
in favor of disruption.

\subsection{Redshift dependence}
\label{sec:redshift}

A redshift dependence enters $\Vc$ in \equ{Vc} through $\delta$ and $\Vd$.
If $\delta$ is proportional to $\fg$ (see \se{fg}), then the dependence of
$\Vc$ on gas fraction at a given disc mass
is roughly $\prop\!\fg^{2/3}$ ($\prop\!\fg$ at a given halo or baryonic mass). 
The average gas fraction is rising from $\fg\!\lsim \!0.1$ at $z\seq 0$ 
to $\fg\ssim0.5$ at $z\ssim 2$ \citep{tacconi10,daddi10,tacconi18}.
This can crudely be described as a power low steeper than 
$\fg \!\prop\! (1+z)^{2}$ in this range.
For given disc mass and gas fraction (or a given halo mass), 
$\Vd \!\prop\! (1+z)^{1/2}$.
Together, the redshift dependence is crudely $\Vc \!\prop\! (1+z)^{2}$.
This strong redshift dependence implies that long-lived clumps are expected 
at $z \ssim 2$ and above and not at much lower redshifts.
The growth of $\fg$ with redshift becomes much milder above $z \sim 2$,
so the preference for surviving clumps increases with redshift at a  
slower pace at higher redshifts.
Naturally, a necessary condition for \LC clumps is being hosted in a 
disc within a halo
above the threshold mass for long-lived discs \citep{dekel20_flip},
which become rare at very high redshifts, when the Press-Schechter mass of
typical haloes is much smaller than the threshold mass for discs.
Thus, while the clump frequency per disc should keep increasing with
redshift, the total number of clumps should peak near $z \ssim 2$.

\smallskip
In the case of a clump mass at it's critical value $\Mc \!\simeq\! 10^8\msun$
in a mixture of galaxy masses,
based on \equ{VcMc}, the dependence of $\Vc$ on gas fraction becomes weaker,
so crudely $\Vc\!\prop\!(1+z)^{1/2}$.
This yields a slightly increased survivability near the critical clump mass
at higher redshift, 
as long as the galaxy is sufficiently massive to allow a long-lived disc.

\section{Clump types in simulations}
\label{sec:sim_types}

\subsection{The VELA simulations}
\label{sec:vela}

\subsubsection{The simulations}

\begin{figure*} 
\centering
\includegraphics[width=0.49\textwidth,trim={0.0cm 0.8cm 0.0cm 0.8cm},clip]
{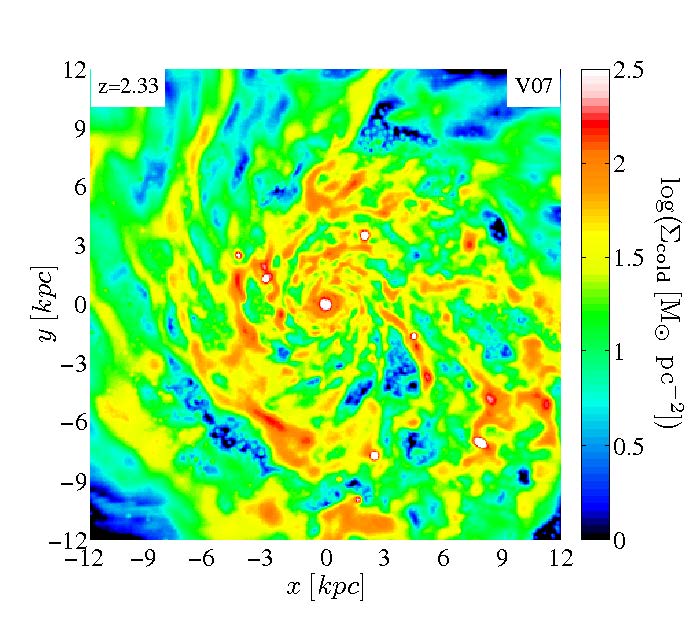} 
\includegraphics[width=0.49\textwidth,trim={0.0cm 0.8cm 0.0cm 0.8cm},clip]
{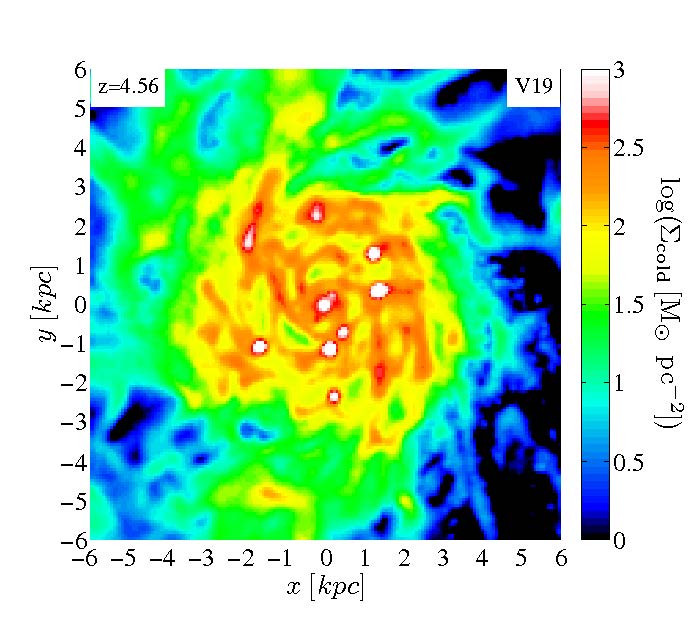} 
\caption{
Face-on views of the surface density of cold mass
in \velathree simulated discs showing giant clumps.
Shown are V07 at $z\!=\!2.33$ (left) and V19 at $z\!=\!4.56$ (right).
The cold mass is comprised of cold gas ($T\!<\!1.5\!\times\!10^4~{\rm K}$)
and young stars ($<\!100~\Myr$).
The disc radii are $\Rd \!=\! 11.8$ and $3.2\kpc$ respectively.
In both cases the disc half-thickness is $\Hd \!\sim\! \Rd / 6$.
The baryonic masses of the discs are
$\Md/(10^{10}\msun) \!\sim\! 3.50$ and $1.35$, and the fractions of cold-mass
are $28\%$ and $33\%$ respectively. Both discs show giant clumps with
masses $\Mc\!>\!10^8\msun$.
}
\label{fig:images_discs}
\end{figure*}

We use here the \vela suite of zoom-in cosmological simulations, 
34 galaxies with halo masses $10^{11}-10^{12}\msun$ at $z \!\sim\! 2$,
which have been used to explore many aspects of galaxy formation at high
redshift
\citep[e.g.,][]{ceverino14,moody14,zolotov15,ceverino15_shape,
inoue16,tacchella16_prof,tacchella16_ms,tomassetti16,
ceverino16_outflow,ceverino16_drops,mandelker17,
dekel20_flip,dekel20_ring,ginzburg21,dekel22_mass}.
The simulations utilize the ART code \citep{krav97,krav03,ceverino09},
which follows the evolution of a gravitating $N$-body system and the
Eulerian gas dynamics with an AMR maximum resolution of $17.5-35\pc$
in physical units at all times.  The dark-matter particle mass is $8.3\times
10^4 \msun$ and the minimum mass of stellar particles is $10^3 \msun$.
The code incorporates gas and metal cooling,
UV-background photoionization and self-shielding in dense gas,
stochastic star formation,
stellar winds and metal enrichment, and feedback from supernovae.

\smallskip 
The thermal feedback model assumes that each stellar particle represents
a single stellar population and injects the luminosity from supernovae and 
stellar winds as thermal heating, at a constant heating rate over
$40\Myr$, the lifetime of the lightest star that explodes as a
core-collapse supernova.
\citep{ceverino09}.
The model adds non-thermal radiative pressure in regions where ionizing
photons from massive stars are produced and trapped. 
In the \velathree version of the feedback model 
\citep[{\tt RadPre} in][]{ceverino14}, 
the radiative pressure is acting in cells that
contain stellar particles younger than $5\Myr$ and whose gas column density
exceeds $10^{21} \cm^{-2}$, as well as in these cell's closest neighbors.
In \velathree, the momentum driving efficiency is $\sim 3 L/c$,
which is comparable to the supernova contribution,
but lower than in certain other simulations where the feedback is very strong
\citep[e.g.,][]{genel12,oklopcic17}.
Further details regarding the feedback and other sub-grid models in \velathree
can be found in \citet{ceverino14} and \citet[][M17]{mandelker17}.

\smallskip 
The \velasix version of the radiative feedback
\citep[based on {\tt RadPre\_IR} in][]{ceverino14} 
includes moderate trapping of infrared photons once that gas density in the
cell exceeds $300\cm^{-3}$.
In addition, the
\velasix feedback model includes momentum injection from the
expanding supernova shells and stellar winds \citep{ostriker11}.
A momentum of $10^6 \msun\kms$ per star more massive than $8\msun$ is injected
at a constant rate over $40\Myr$.
The \velasix model also includes a factor of three boost in the injected
momentum due to clustering of supernovae \citep{gentry17}, which is implemented
in the form of non-thermal pressure as described in \citet{ceverino17}.
Further details regarding the feedback and other sub-grid models in \velasix
can be found in \citet{ceverino14}, \citet{ceverino17} and
Ceverino et al. (in preparation).

\smallskip 
M17 have analyzed the clumps in the \velathree galaxies as
well as in counterparts that lacked the radiation pressure feedback
(\velathree and \velatwo respectively).
They found that with radiation pressure a
significant fraction of the clumps of $\Mc \!\leq\! 10^8\msun$ are
disrupted on timescales of a few free-fall times,
while most of the more massive clumps survive.
The migration inward of these long-lived clumps
produced galacto-centric radial gradients in their properties, such as mass,
stellar age, gas fraction and sSFR, which were found in
\citet{dekel22_mass} to be consistent with an 
analytic model and with observed clumps, distinguishing them from
the short-lived clumps.
However, in M17 the analyzed output times where separated coarsely by 
$\sim\! 100\Myr$,
which prevented a detailed study of the evolution of individual clumps.
In the current study, we have re-simulated several of the \velathree galaxies, 
saving about ten snapshots per disc crossing time, with fine timesteps
that can be as short as $\sim\!5\Myr$.
Among these galaxies, five have \LC clumps (V07, V08, V19, V26, V27),
and four others have only \SC clumps (V11, V12, V14, V25). 

\smallskip  
As in M17, the disc is defined as a cylinder with radius
$\Rd$ and half height $\Hd$, containing 85\% of the cold gas
($T\!<\!1.5\!\times\! 10^4~{\rm K}$) and the young stars (ages $<\!100\Myr$)
within $0.15\Rv$.
The disc stars are also required to obey a kinematic criterion,
that their specific angular momentum parallel to that of the disc, $j_{\rm z}$,
is at least $70\%$ of the maximal possible value it could have given its
galacto-centric distance, $r$, and orbital velocity, $v$, namely
$j_{\rm max}\!=\!v r$.

\smallskip 
The VDI phase typically follows a dramatic event of wet compaction into a 
star-forming ``blue nugget", that tends to occur when the halo mass is
above $\Mv \ssim 10^{11.3}\msun$, and could happen at different redshifts,
typically above $z \seq 1$
\citep[][figure 2]{zolotov15,tomassetti16,tacchella16_prof,tacchella16_ms}. 
An extended, long-lived clumpy disc, evolving to a clumpy ring, 
typically develops after the major compaction event, above the threshold mass 
where the disc is not disrupted by frequent mergers 
\citep{dekel20_flip,dekel20_ring}.

\smallskip 
\Fig{images_discs} shows face on views of two simulated disks,
V07 at $z\!=\!2.33$ and V19 at $z\!=\!4.56$,
shortly after the start of their VDI phase. The figure shows the surface
density of the cold mass, integrated over $\pm\!\Rd$ perpendicular to the disc,
where $\Rd\!\simeq\! 12$ and $3.2\kpc$ for V07 and V19 respectively.
While these two discs are at very different redshifts, with different masses
and sizes, both show giant star-forming clumps with masses
$\Mc\!>\!10^8 \msun$.

\smallskip 
The gas fractions in the \velathree discs are somewhat lower
than estimated in typical observed galaxies at similar redshifts. 
This has been discussed in detail in several papers which used these
simulations \citep[e.g.,][]{zolotov15,tacchella16_ms,mandelker17}.
While the \vela simulations are state-of-the-art in terms of high-resolution
AMR hydrodynamics and the treatment of key physical processes at the subgrid
level, they are not perfect in terms of their treatment of
star-formation and feedback, much like other simulations. Star-formation tends
to occur too early, leading to lower gas fractions later on. The stellar masses
at $z\!\sim\! 2$ are a factor of $\sim\! 1.5\!-\!2$ higher than inferred for
haloes of similar masses from abundance matching 
\citep[e.g.,][]{rodriguez17,moster18,behroozi19}.
However, for the purposes of
the present study, the relatively low gas fractions during the peak VDI phase
would only underestimate the actual accretion of fresh gas onto clumps during
their migration, providing a lower limit on clump survival. The effect of gas
fraction on clump properties and survival in simulated isolated discs
is further discussed in \citet{fensch21}.
In \velasix with the stronger feedback, the agreement with the stellar-to-halo
mass ratio deduced from observations becomes better, but this comes 
at the expense of more dynamical destruction on the clump scales, which may or
may not be more realistic.
We do not attempt in this paper to decide between the different feedback models,
but rather to study the effect of each on the survival and disruption of the 
giant clumps.

\subsubsection{Clump analysis}

Clumps are identified in 3D and followed through time following the method
detailed in M17. Here we briefly summarize the main features.
%
Clumps are searched for within a box of sides $4\Rd$ in the disc plane
and height $4\Hd$ centered on the galaxy centre.
Via a cloud-in-cell interpolation, the mass is deposited in a uniform
grid with a cell size of $\Delta\!=\!70\pc$, two-to-three times the maximum AMR
resolution. We then smooth the cell's density, $\rho$, into a
smoothed density, $\rho_{\rm w}$, using
a spherical Gaussian filter of FWHM $\seq {\rm min}(2.5\kpc, 0.5\Rd)$,
defining a density residual
$\delrho \!=\! (\rho\! -\! \rho_{\rm w})/\rho_{\rm w}$.
Performed separately for the cold mass and the stellar mass,
we adopt at each point the maximum of the two residual values.
Clumps are defined as connected regions containing at least 8
grid cells with a density residual above $\delmin\!=\!10$,
making no attempt to remove unbound mass from the clump.
We define the clump centre as the baryonic density
peak, and the clump radius, $\Rc$, as the radius of a sphere with the same
volume as the clump.
The clump mass, however, is the mass contained in the cells within the
connected region.
%
\textit{Ex-situ} clumps, which joined the disc as minor
mergers, are identified by their dark matter content and the birth place of
their stellar particles. They are not considered further here, where we focus
on the \textit{in-situ} clumps.

\smallskip 
The SFR in the clumps is derived from the mass in stars younger
than $30\Myr$, which is sufficiently long for fair statistics
and sufficiently short for ignoring the stellar mass loss.
Outflow rates from the clumps are measured through shells of radii
$(\Rc, \Rc+100\pc)$.
The gas outflow rate is computed by 
$\dot{M}_{\rm out}\!=\!\Delta^{-1} \sum_i V_r\, m_i$, 
with $\Delta \seq 100\pc$,
where the sum is over cells
within the shell with $V_r\!>\!0$ and a 3D velocity larger than the escape
velocity from the clump, $V^2\!>\!2G\Mc/\Rc$, and where $m_i$ is the gas mass 
in the cell.
The gas accretion rate is computed in analogy to the outflow rates but with
$V_r \slt 0$ and no constraint on $V^2$. We note that this calculation involves
large uncertainties.
%

\smallskip 
Individual clumps, that contain at least 10 stellar particles,
are traced through time based on their stellar particles.
For each such clump at a given snapshot, we search for all
``progenitor clumps" in the preceding snapshot,
defined as clumps that contributed at least $25\%$ of
their stellar particles to the current clump. If a given clump has more
than one progenitor, we consider the most massive one as
the main progenitor and the others as having merged, thus creating a clump
merger tree.
If a clump in snapshot $i$ has no progenitors in snapshot $i\!-\!1$,
we search the previous snapshots back to two disc crossing times before
snapshot $i$. If no progenitor is found in this period, snapshot $i$ is
declared the initial, formation time of the clump, and for that clump $t$ is
set to zero at that time.\footnote{If the mass weighted mean stellar age of the
clump at its initial snapshot is less than the timestep since the previous
snapshot, we set the initial clump time to this age rather than to zero. This
introduces an uncertainty of a few Megayears in the clump age.}
When tracing the evolution of a clump we refer to the main progenitor
and consider the mergers to be part of the accretion onto the clump.
The clump lifetime $\tauc$ is the age of the clump at the last snapshot 
when the clump is still identified.

\begin{figure*} 
\centering
\includegraphics[width=0.33\textwidth]
{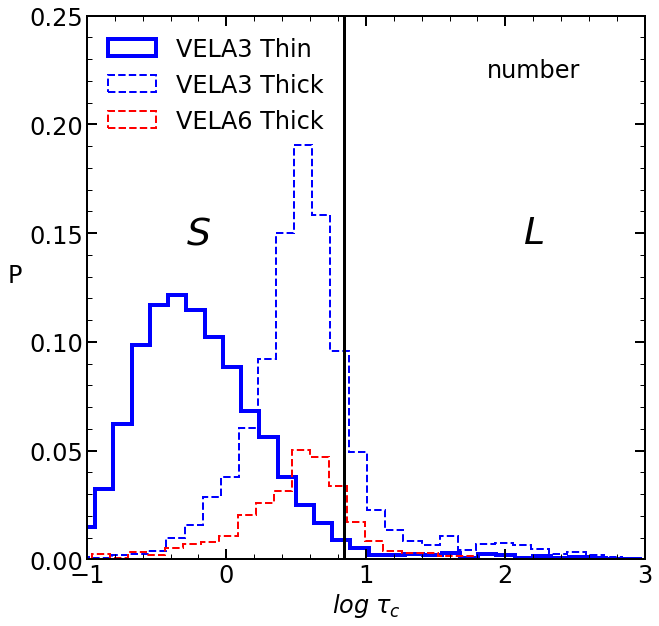} 
\includegraphics[width=0.33\textwidth]
{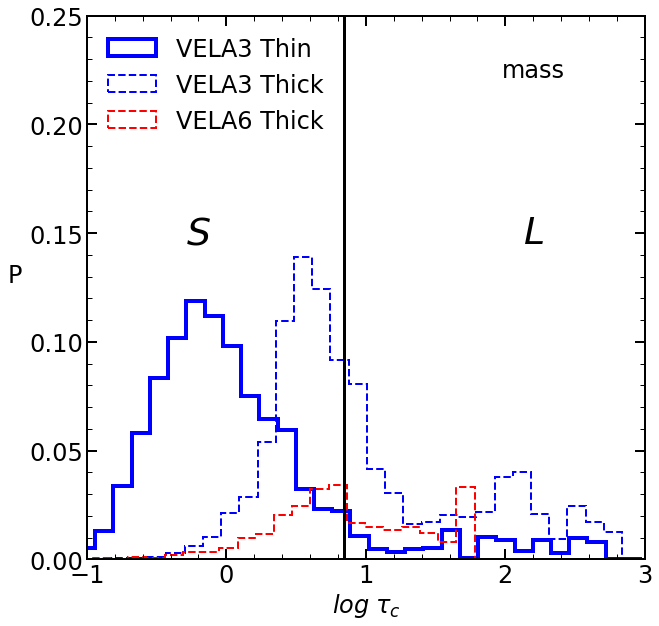} 
\includegraphics[width=0.33\textwidth]
{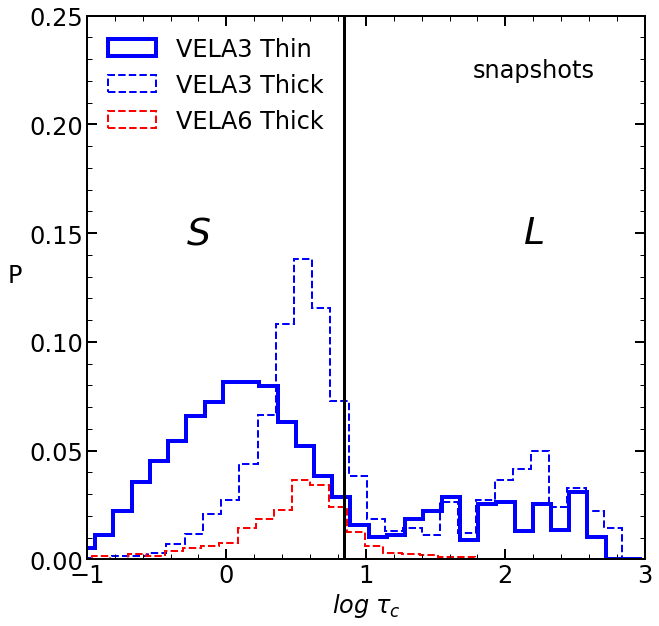} 
\caption{
\SC vs \LC clumps.
The distribution of lifetimes in units of free-fall times,
$\tauc \seq \tc/\tff$, for clumps in \velathree (blue solid line,
normalized to unity).
The minimum clump mass is set in this figure to $2\times 10^7\msun$.
{\bf Left:} Number of clumps.
{\bf Middle:} Weighted by clump mass.
{\bf Right:} Number at each snapshot, to be compared to observations.
We see in the right panel a natural bimodality into \SC and \LC clumps,
which we separate at $\tauc \ssim 7$.
Also shown is the distribution of lifetimes in the
\velasix simulations with the stronger feedback (dashed red),
which is derived from the available coarser output timesteps of
$\sim\! 100 \Myr$.
The comparison with \velathree is performed using the same coarse
timesteps in \velathree (dashed blue).
The \velathree coarse-timestep histogram is independently normalized to unity,
and the corresponding \velasix histogram is based on the same normalization as
\velathree (the number of clumps from the coarse timesteps is
lower than it is for the fine timesteps).
We see that
the coarse-timestep distributions are offset toward larger $\tauc$ (for
reasons that are briefly discussed in the text), but the comparison of
\velathree and \velasix using the coarse timesteps is fair.
The \velasix simulations show less clumps than \velathreecom,
which are predominantly \SC clumps.
}
\label{fig:hist_tc_SC_LC}
\end{figure*}

\begin{figure*} 
\centering
\includegraphics[width=0.33\textwidth]
{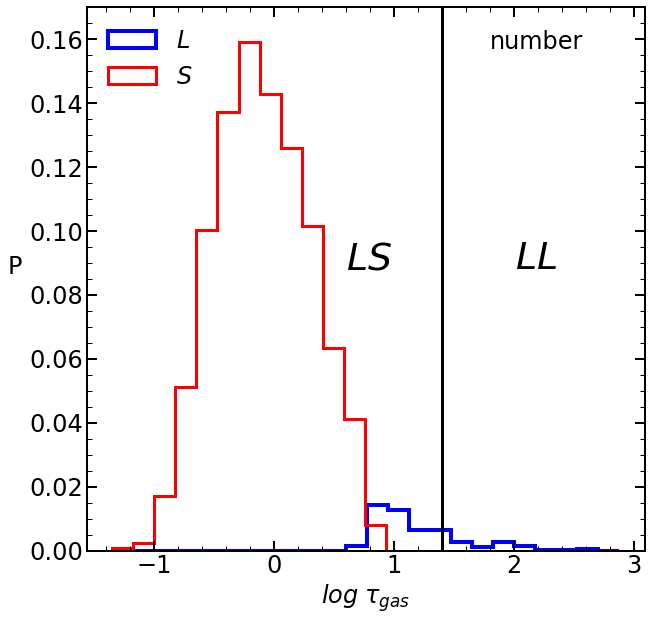} 
\includegraphics[width=0.33\textwidth]
{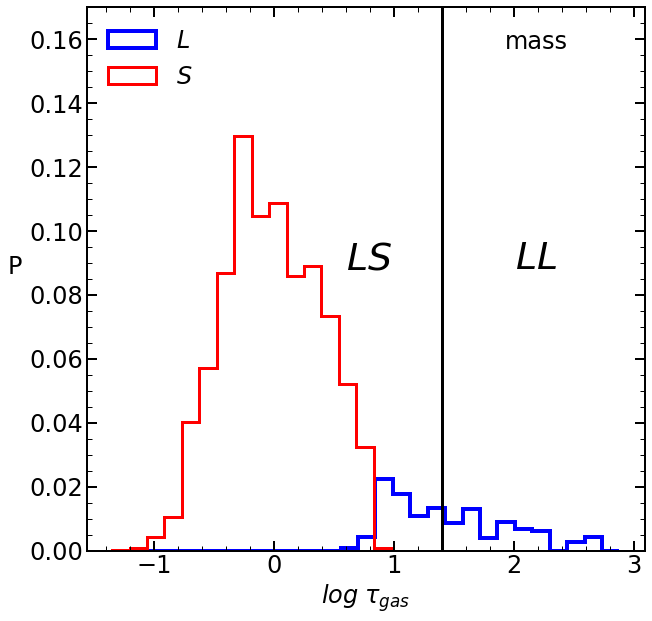} 
\includegraphics[width=0.33\textwidth]
{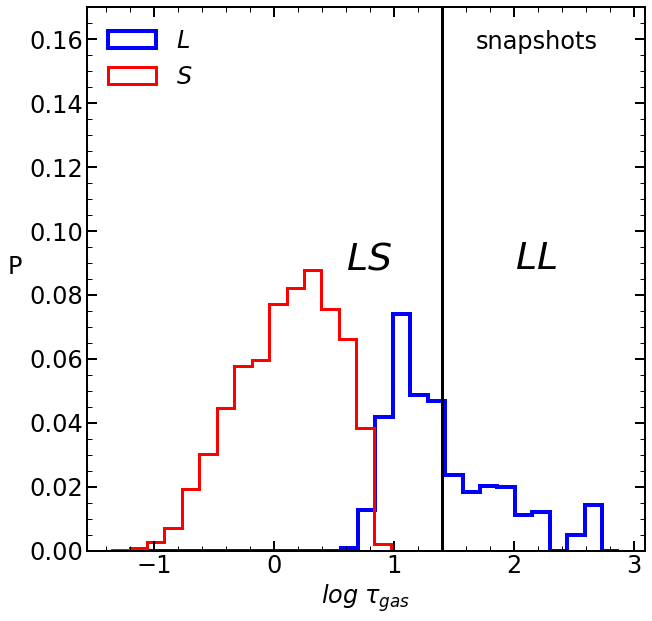} 
\caption{
\LS versus \LL long-lived clumps in \velathreecom.
The distribution of $\taugas$, the time when the clump have lost 90\% of the
gas it had at $\tau \seq 4$ (if $\tauc \slt 4$ then $\taugas$ is set to
equal $\tauc$).
The minimum clump mass is set in this figure to $2\times 10^7\msun$.
{\bf Left:} Number of clumps.
{\bf Middle:} Weighted by clump mass.
{\bf Right:} Number at each snapshot, as observed.
For the \SC clumps ($\tauc \slt 7$) we have $\taugas \ssimeq \tauc$.
There is a natural division of the \LC clumps ($\tauc \sgt 7$)
at $\taugas \ssimeq 25$ to \LS clumps that lose most of their gas moderately
early and \LL clumps that keep most of their gas for tens of free-fall times.
}
\label{fig:hist_taugas_LS_LL}
\end{figure*}

\begin{figure*} 
\centering
\includegraphics[width=0.44\textwidth]
{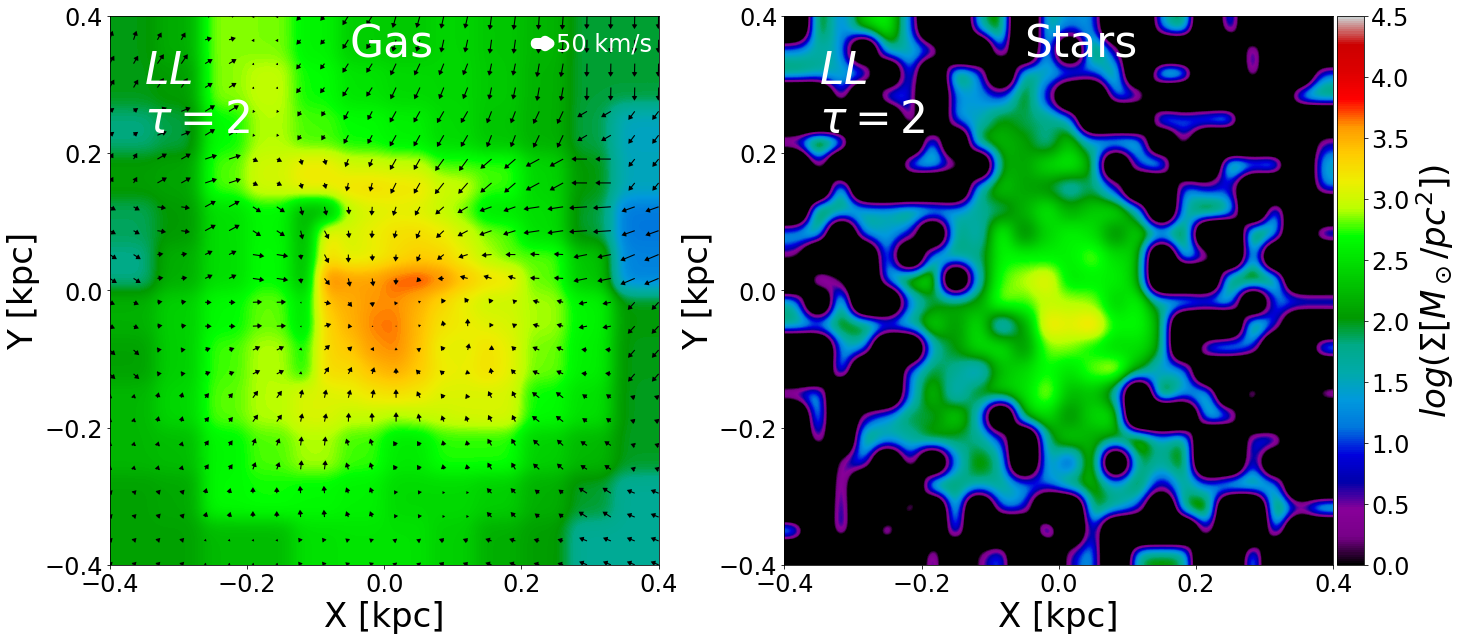} 
\includegraphics[width=0.44\textwidth]
{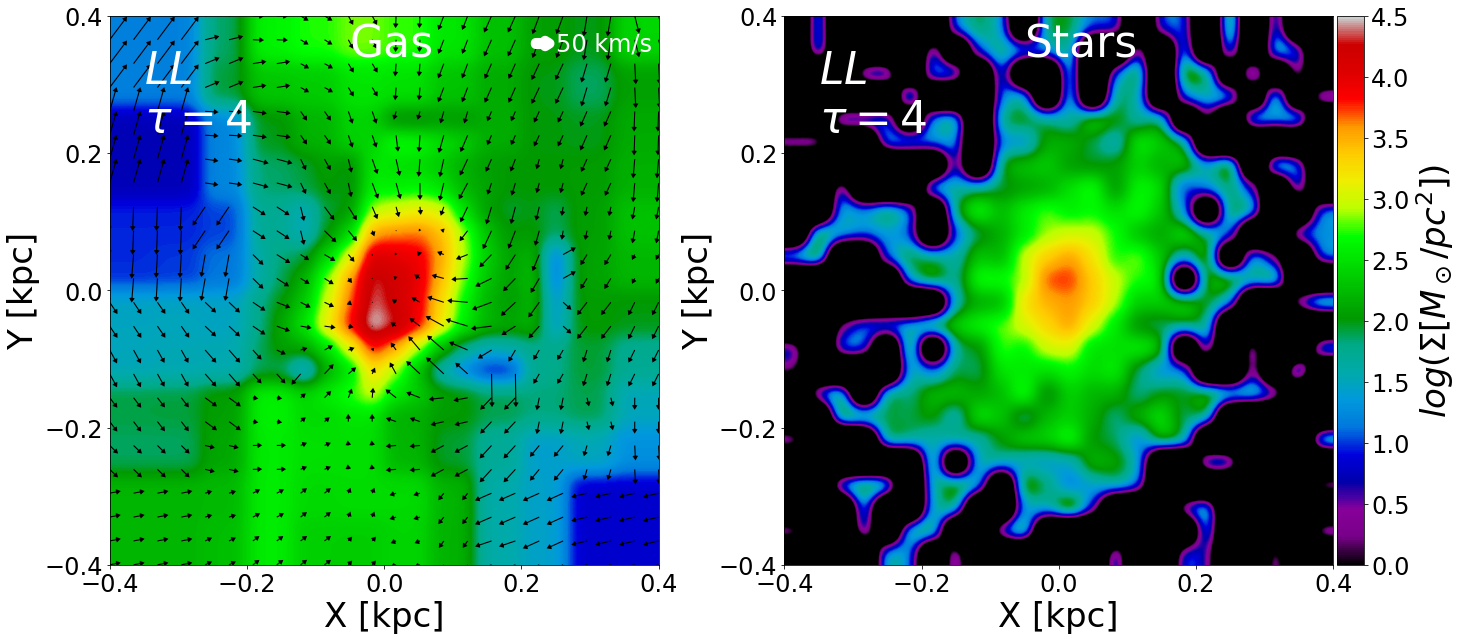} 
\includegraphics[width=0.44\textwidth]
{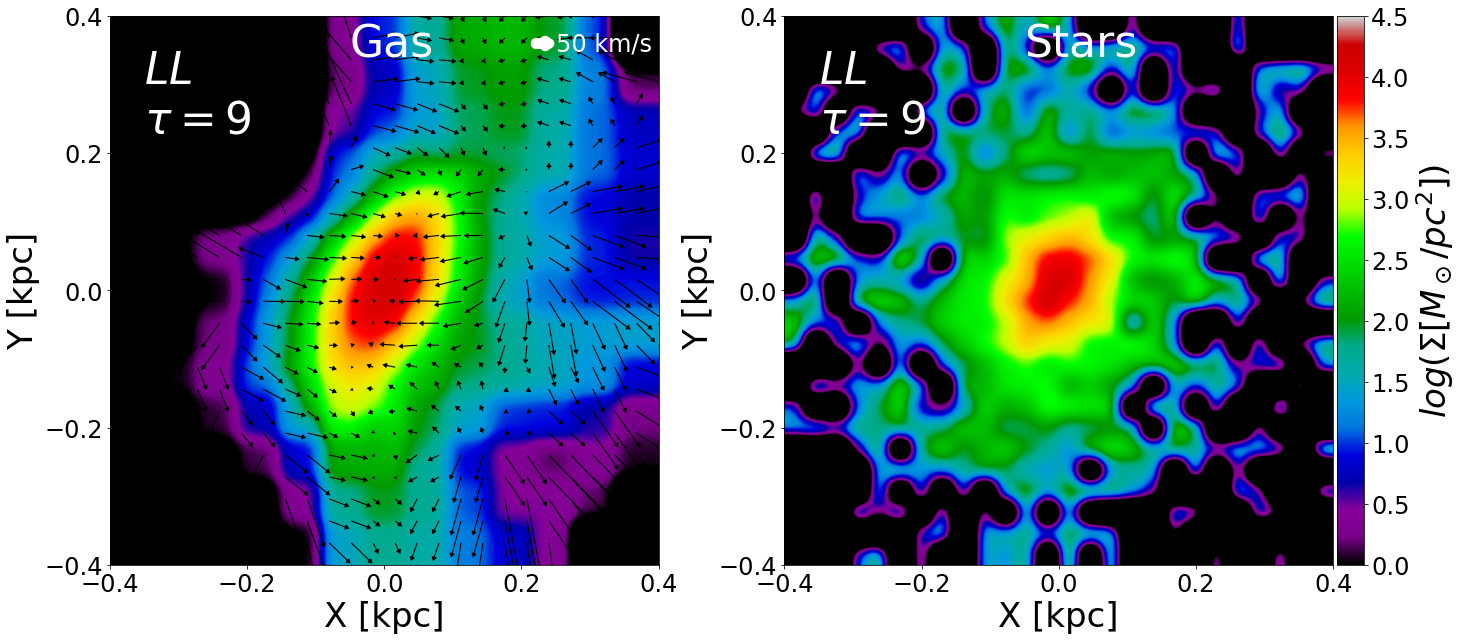} 
\includegraphics[width=0.44\textwidth]
{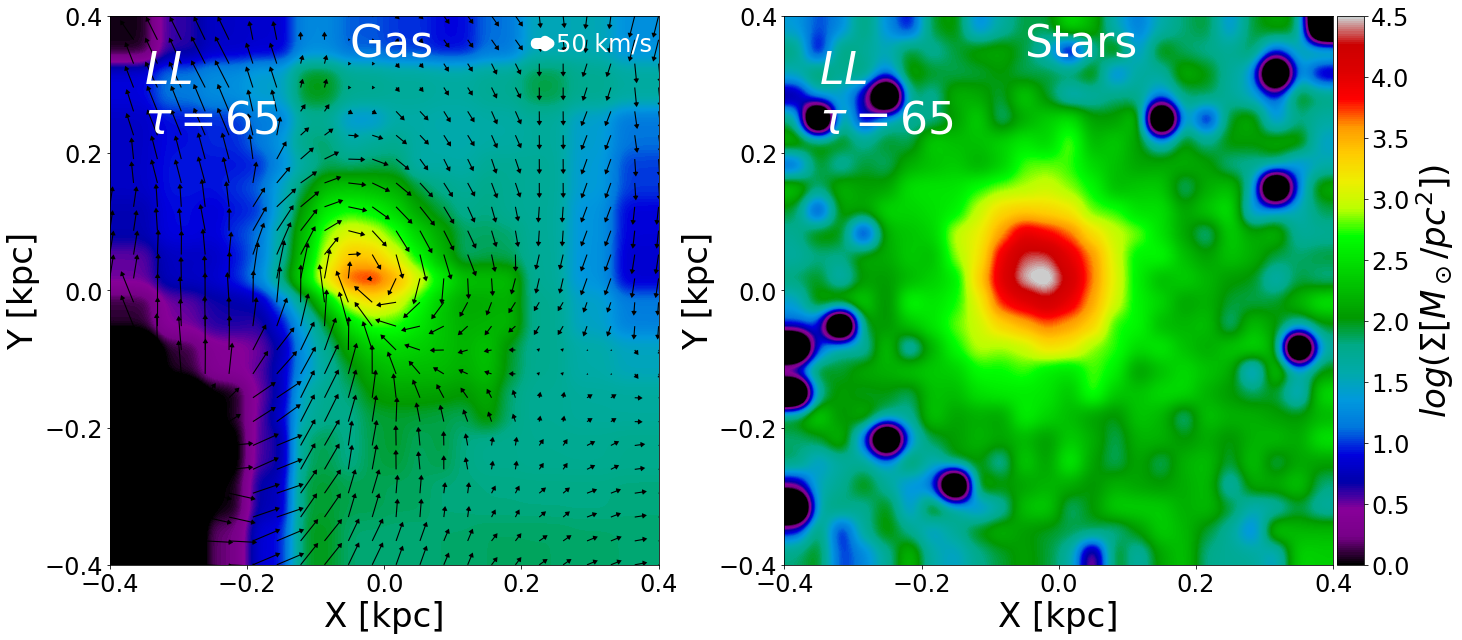} 
\vskip 0.3cm
\includegraphics[width=0.44\textwidth]
{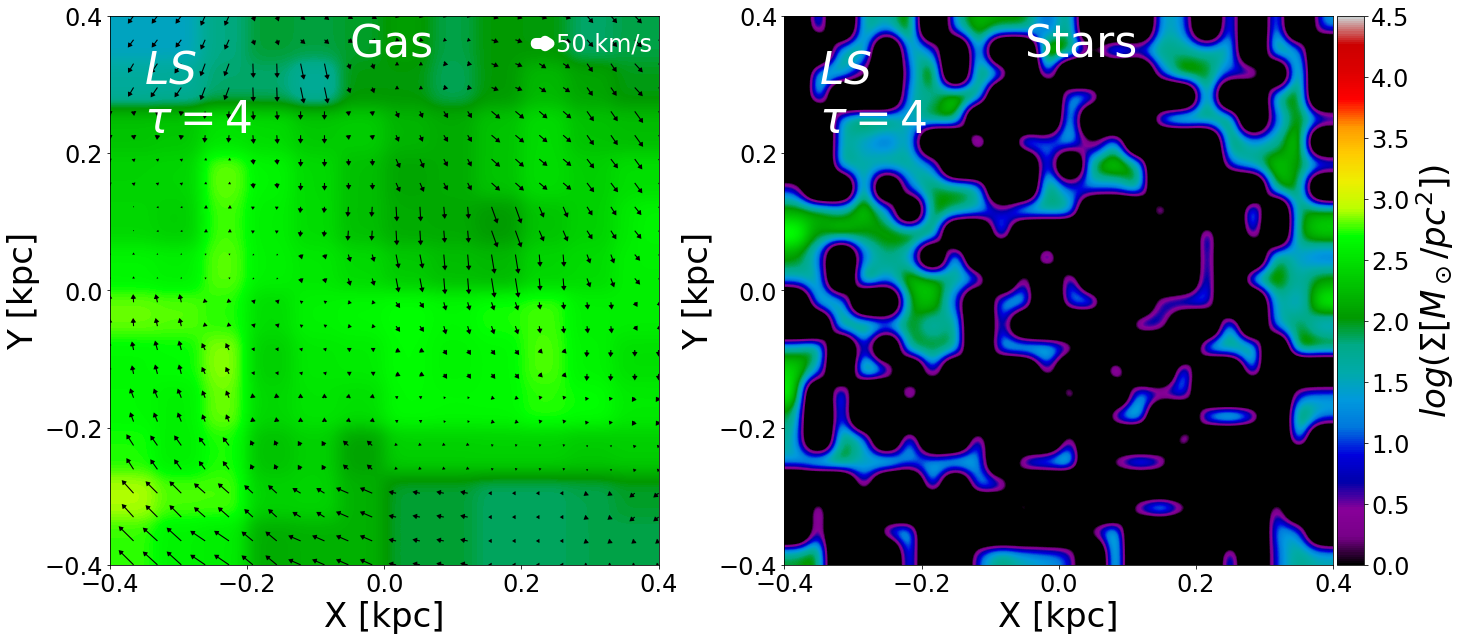} 
\includegraphics[width=0.44\textwidth]
{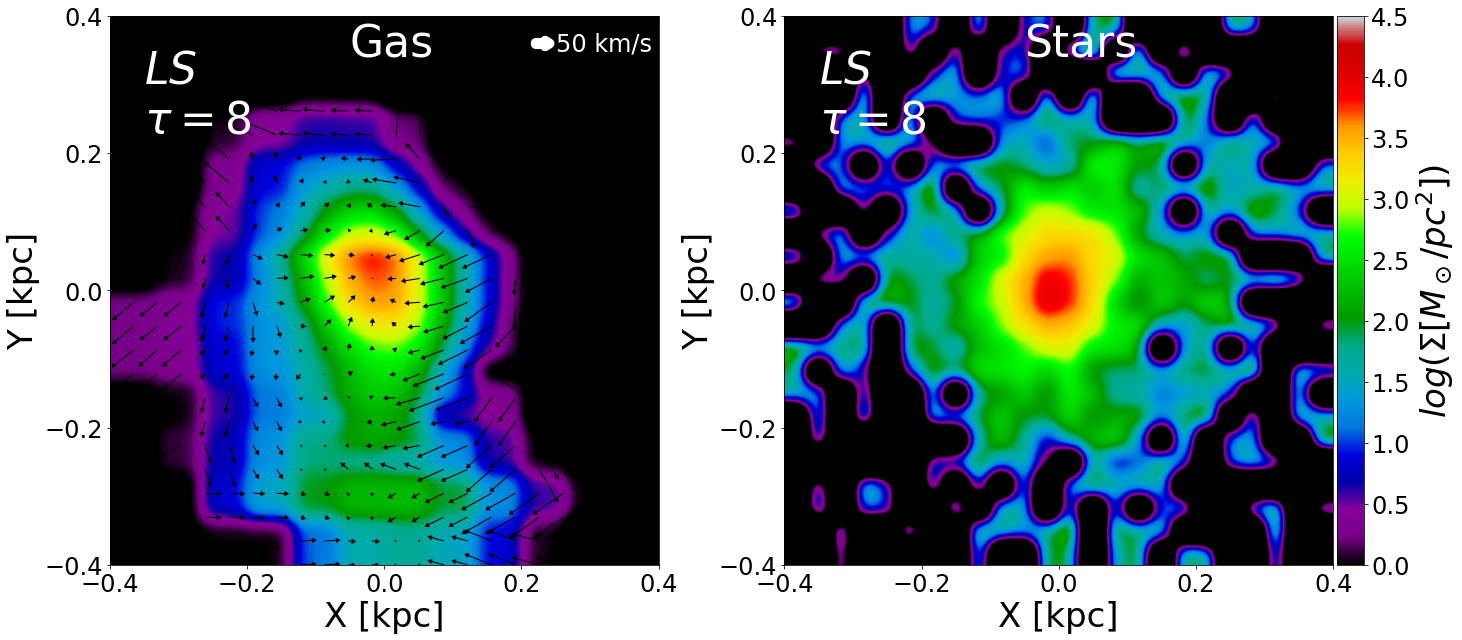} 
\includegraphics[width=0.44\textwidth]
{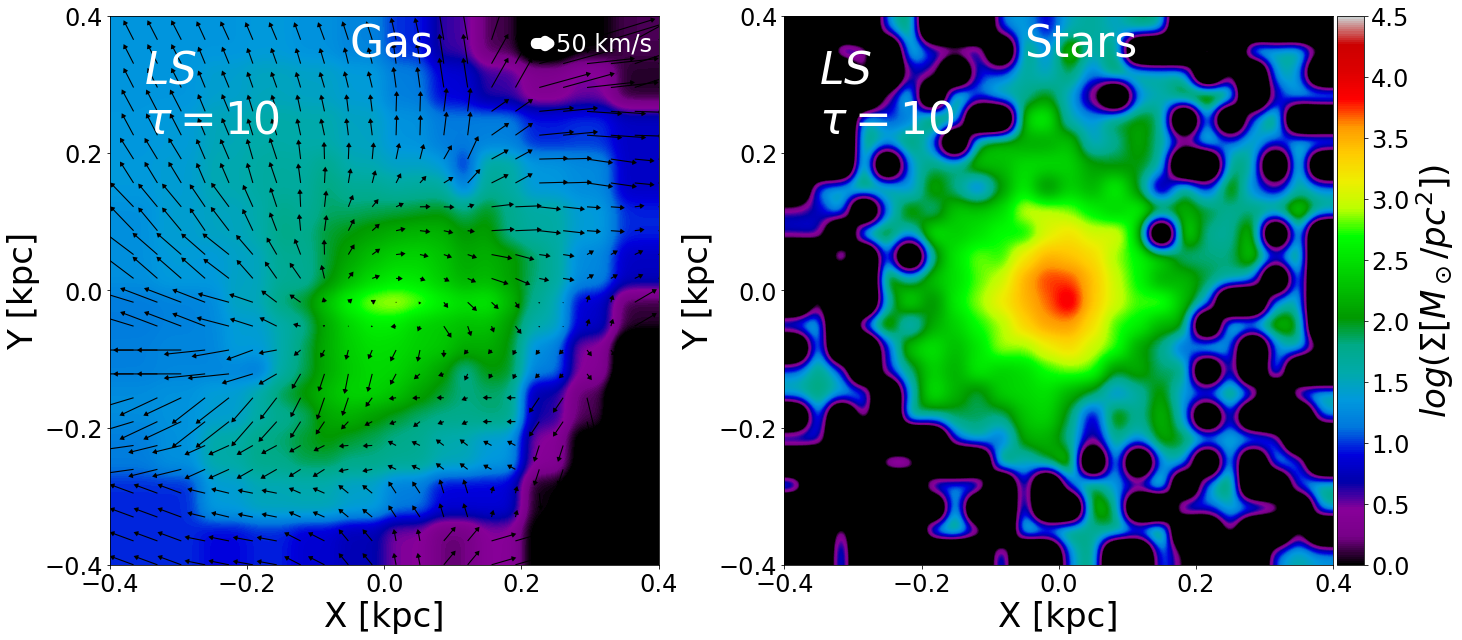} 
\includegraphics[width=0.44\textwidth]
{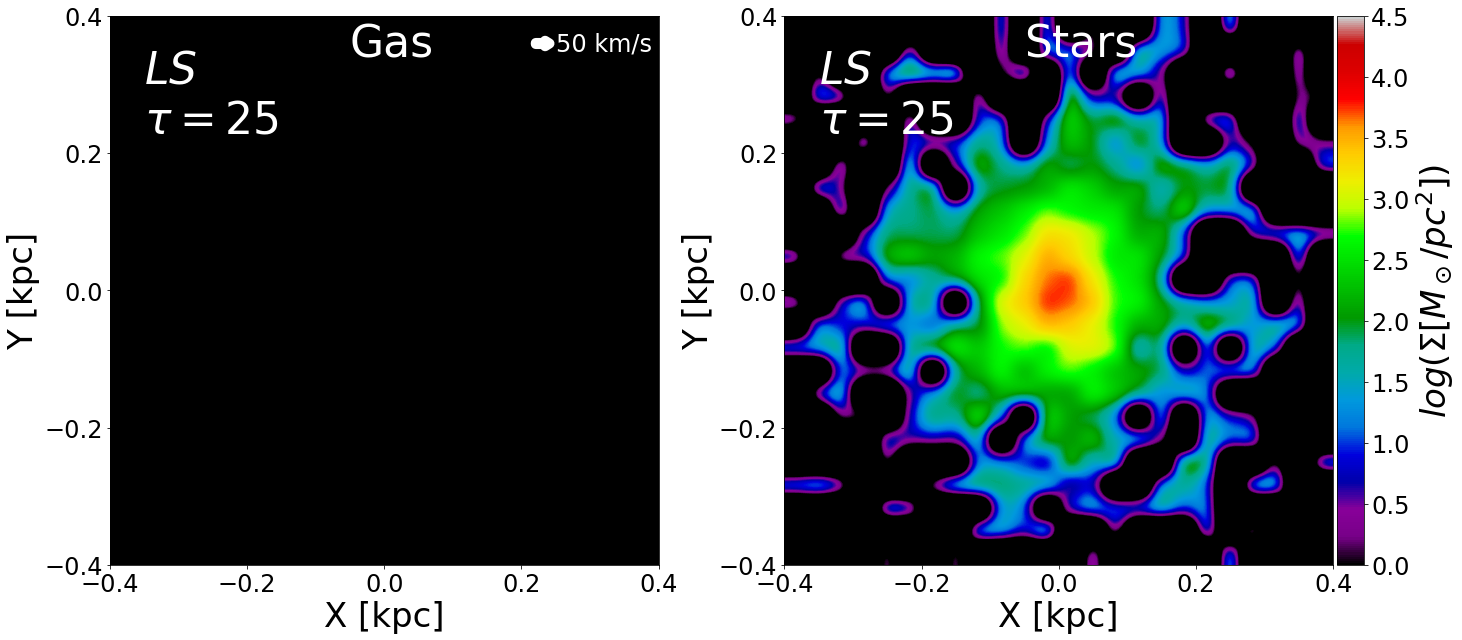} 
\vskip 0.3cm
\includegraphics[width=0.44\textwidth]
{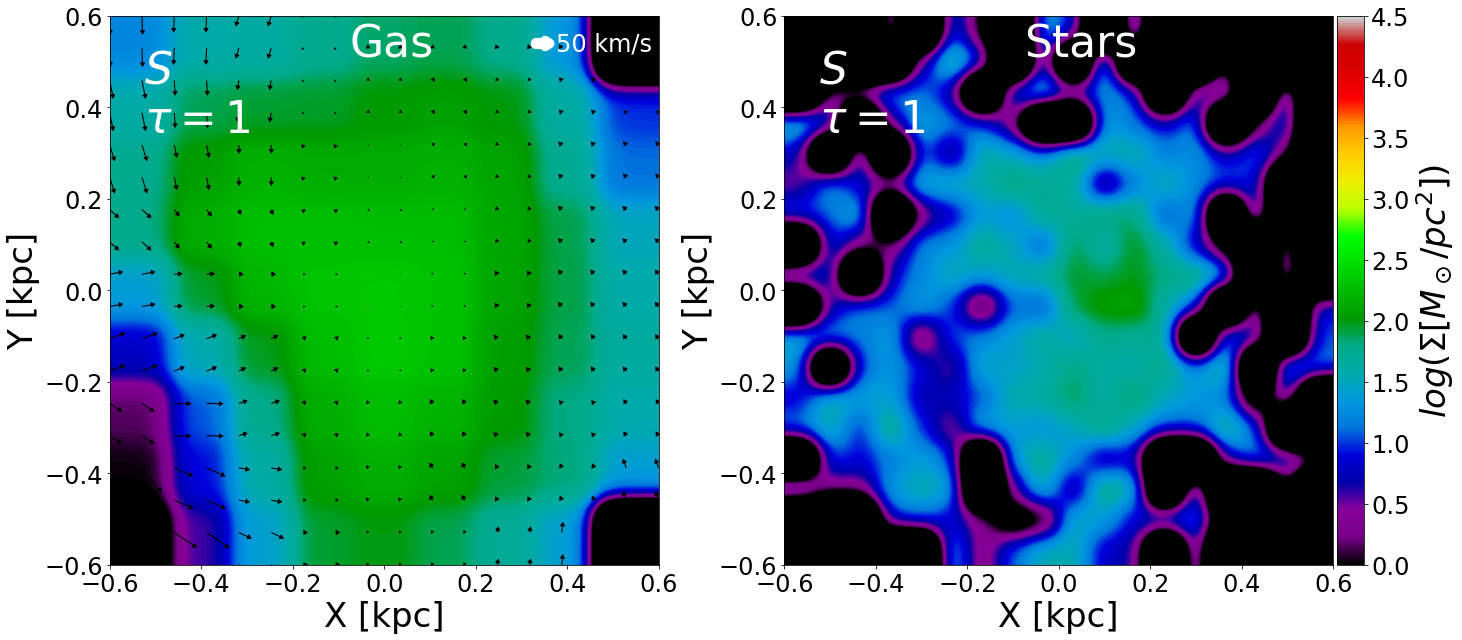} 
\includegraphics[width=0.44\textwidth]
{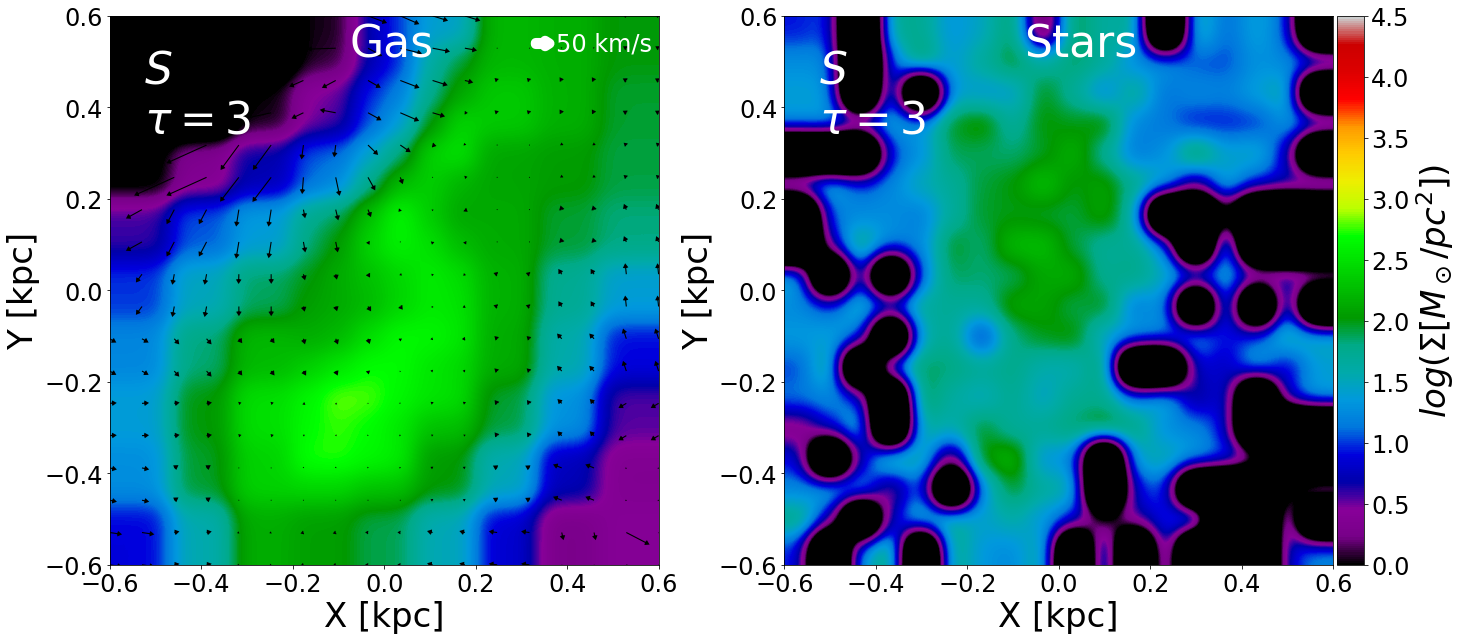} 
\includegraphics[width=0.44\textwidth]
{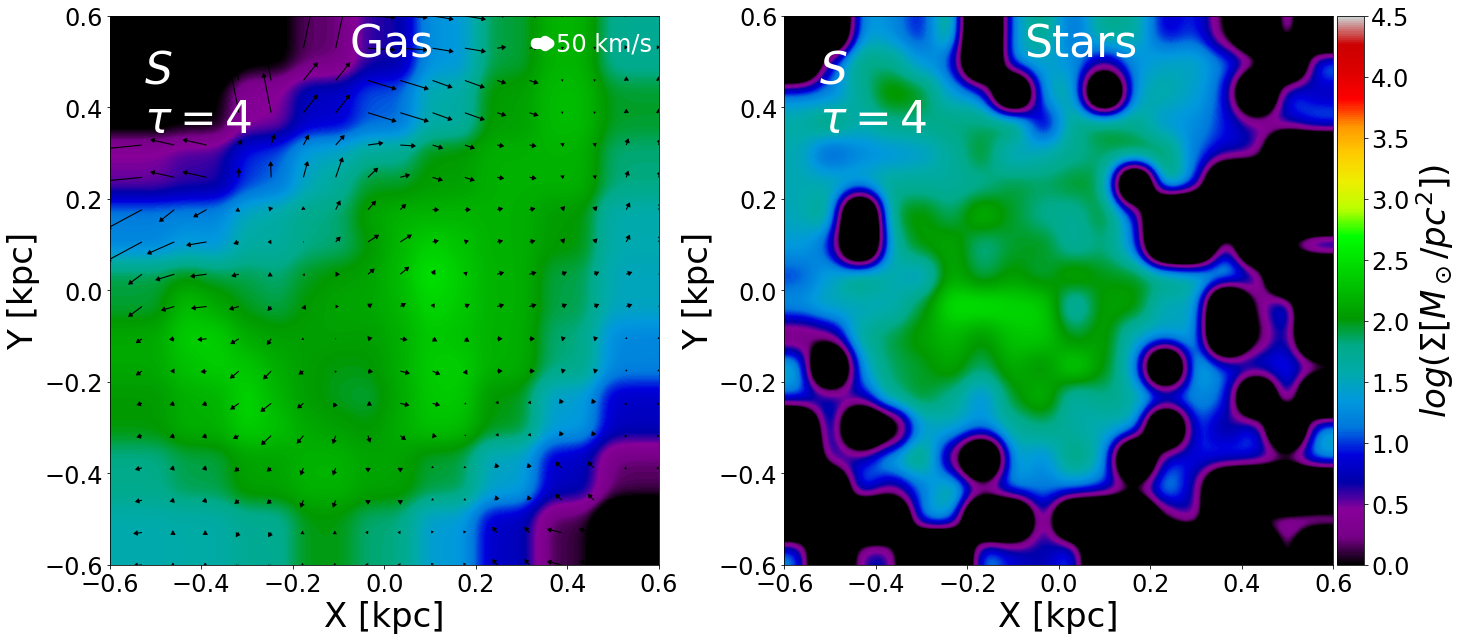} 
\includegraphics[width=0.44\textwidth]
{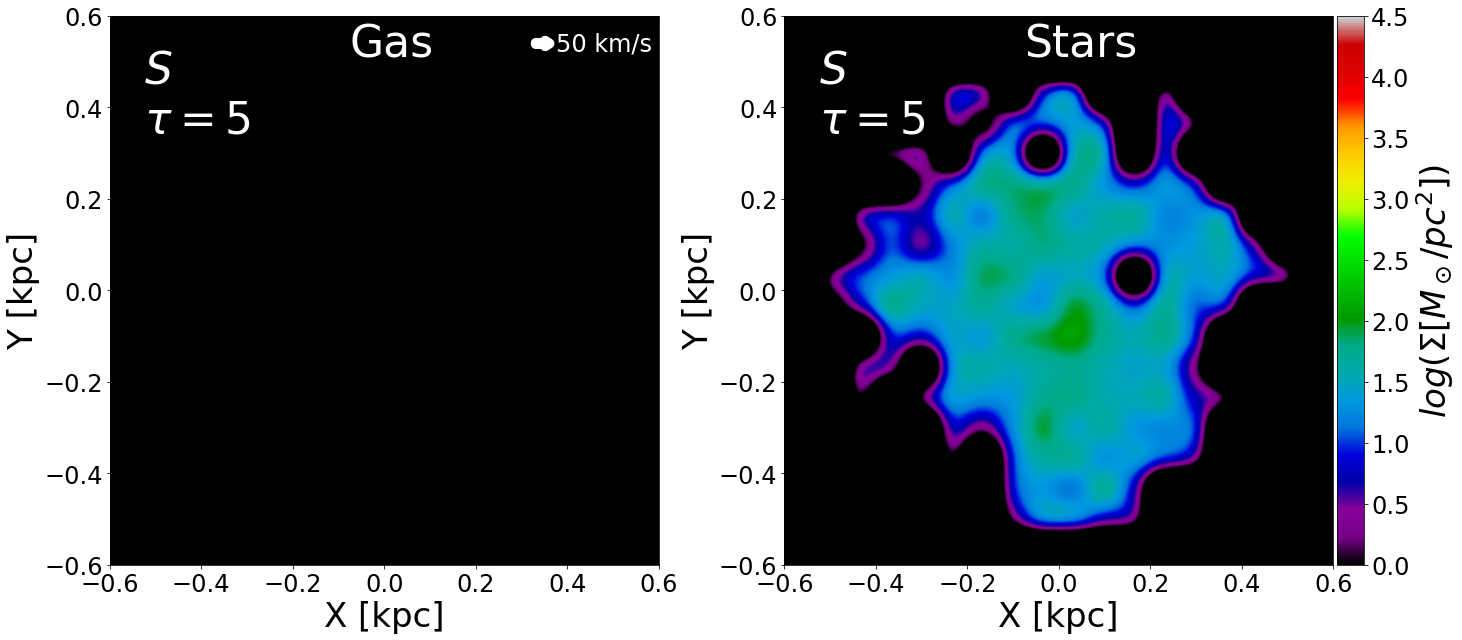} 
\caption{
Images of gas (left) and stars (right) in examples of the 3 clump types at
four times.
{\bf Top:} An \LL clump at $\tau \seq 2,4,9,65$, surviving both in stars and
gas for tens of free-fall times.
{\bf Middle:} An \LS clump at $\tau \seq 4,8,10,25$, surviving in stars but
losing all gas by $\tau \ssim 10$.
{\bf Bottom:} An \SC clump at $\tau \seq 1,3,4,5$, losing all its gas and
dispersing its stars by $\tau \ssim 5$.
}
\label{fig:images}
\end{figure*}

\subsection{Three clump types}
\label{sec:clump_types}






The clumps for analysis were selected as follows.
From the nine galaxies that were re-simulated with fine output timesteps,
output snapshots were selected where the galaxies are discs with an axial ratio
$\Rd/\Hd \sgt 3$.
Clumps were selected for analysis only in snapshots when their baryonic mass
is above a threshold mass $\Mc \sgt 10^7\msun$,
with the additional requirement that their initial gas mass was
$\Mg \sgt 10^6\msun$ at the snapshot when they were first identified.

\smallskip 
As found in M17,
the clumps in the \velathree simulations can be divided into two major
classes based on their lifetimes. This
is demonstrated in \fig{hist_tc_SC_LC}, which shows (in solid blue)
the probability
distribution of clump lifetimes in units of clump free-fall times, the latter
being the mass-weighted average of $\tff$ over the clump lifetime.
The distributions are shown in three different ways, corresponding to
(a) clump number where each clump is counted once (left),
(b) clump mass where each clump is counted once and the mass is the average
over the clump lifetime (middle),
and (c) clump number where each clump is counted at every snapshot in which
it appears (right).  The latter allows direct comparison to the distribution of
clumps in observations.
Here, the lifetimes were determined using the fine timesteps as opposed to the
coarse timesteps used in M17.
We see a bimodality into short-lived and long-lived clumps, which is emphasized
in the right panel. We term them
\SC and \LC clumps respectively, and separate here near
a clump lifetime of $\tauc \seq \tc/\tff \ssim 7$.\footnote{Any choice on the
order $\tauc \ssim 10$ would make sense. For example, M17 used
$\tauc \seq 20$.}
The difference between the medians of the normalized $\tauc$ for \SC and \LC
clumps, $\sim\! 1.9$dex, reflects a large difference of
$\sim\!1.4$dex in $\tc$, aided by a smaller difference of $\sim\!0.5$dex
in $\tff$ (see \fig{size_t}).

\smallskip 
For a quick comparison of the clumps in the \velathree and \velasix
simulations, with different feedback strengths, we appeal to a clump analysis
using the coarser output timesteps of $\sim\!100\Myr$, which are the data
currently available for \velasixcom.
\Fig{hist_tc_SC_LC} shows in dashed lines the distributions of clump lifetimes
using the coarse timesteps.
The dashed \velathree histogram (blue) is normalized to unity, and the dashed
\velasix histogram (red) is normalized based on \velathreecom.
We note in passing that for \velathree the distribution using the coarse
timesteps is different from the fine-timestep distribution, for several reasons.
First, the overall number of clumps is lower in the coarse-timestep analysis,
despite having more galaxies in this analysis.
This is because the fine timesteps allow more clumps to be identified,
and because more zero-lifetime clumps are excluded in the analysis using
coarse timesteps (termed ZLC in M17).
Second, around $\tauc \ssim 10$ there are more clumps at a given $\tauc$ in the
coarse-timestep analysis. This indicates that the corresponding histogram is
shifted towards larger $\tauc$ values, despite the
natural tenancy to measure larger lifetimes with finer timesteps.
This shift is partly due to the way ages are assigned to \SC clumps at the
first snapshot, especially if they are identified in one snapshot only,
or because $\tff$ is somehow underestimated when averaged over less snapshots.
The above comparison between the the fine and coarse timesteps
is not relevant for our current aim to qualitatively compare
between the clumps in \velathree and \velasix using the same coarse timesteps.
We find that in \velasixcom, with the stronger feedback, there are
significantly fewer clumps, and a significantly smaller fraction of \LC clumps.
This indicates that clump survival is a strong
function of the feedback strength, as predicted by our model.
A more detailed comparison between the clumps in the \velathree versus the
\velasix simulations is deferred to Ceverino et al. (in preparation).

\smallskip 
We next realize that the \LC clumps in \velathree
can be divided into two sub-classes based on
the way they lose their gas to outflows. This is demonstrated in
\fig{hist_taugas_LS_LL}, which shows for the \SC and \LC clumps in \velathree
the distribution of $\taug$, the
time when the clump has lost 90\% of the gas mass it had at $\tau\seq 4$.
The latter is typically the time when the gas mass reaches a maximum.
If $\tauc \slt 4$ we set $\taug\seq\tauc$.
Among the \LC clumps, one can identify
a bi-modality into two types, separated
near $\taug \ssim 25$, which we term \LS and \LL clumps respectively
(the first L for the stars, the second S or L for the gas).
The \LL clumps are the long-lived clumps analyzed in M17 (sometimes together
with the \LS clumps) and in
\citet{dekel22_mass}, which keep a significant fraction of their gas
and a non-negligible SFR for many tens of free-fall times, till they complete
their inward migration.
The \LS clumps, which we analyze here in detail,
are in certain ways
between the \SC clumps and the \LL clumps. They lose most of their gas on a
timescale of $\sim\!10$ free-fall times, while keeping a long-lived bound
stellar component with only little SFR.

\begin{figure*} 
\centering
\includegraphics[width=0.33\textwidth]
{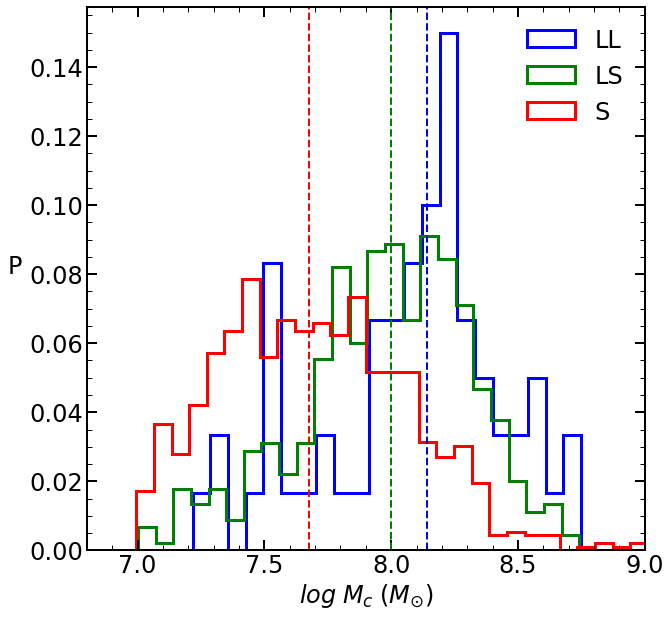} 
\includegraphics[width=0.33\textwidth]
{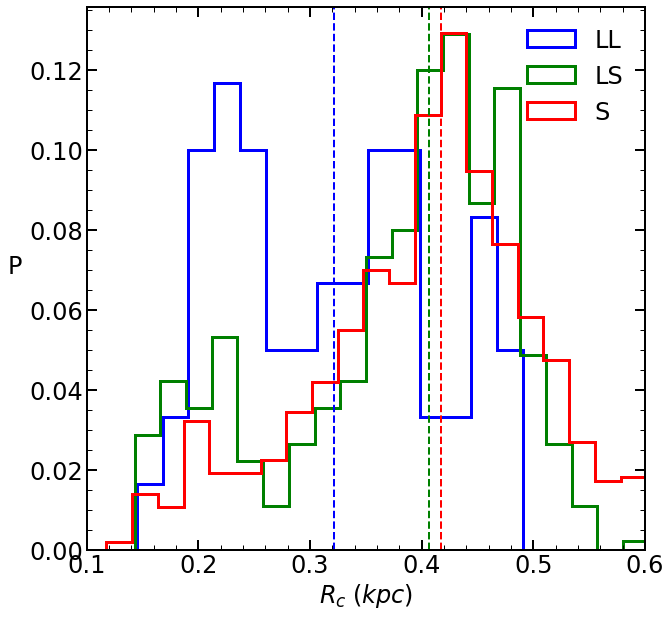} 
\includegraphics[width=0.33\textwidth]
{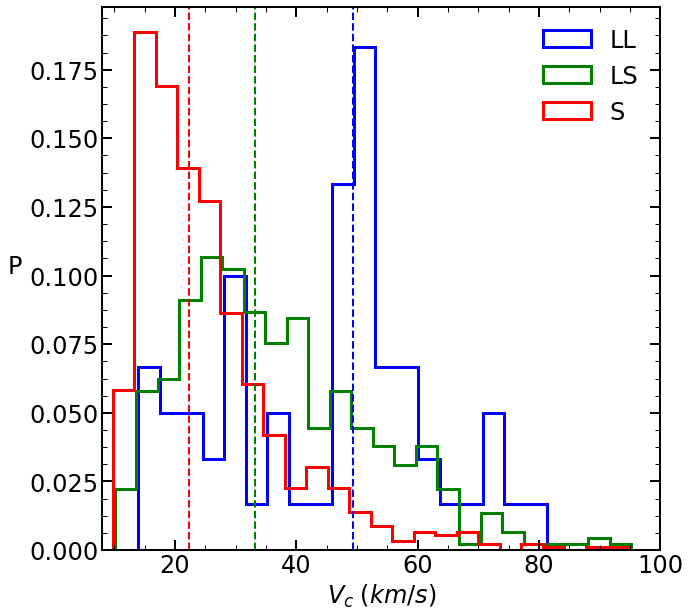} 
\caption{
The probability distribution of the clump properties $\Mc$, $\Rc$ and $\Vc$,
averaged over the early phase $\tau \slt 4$,
for the three clump types.
Each clump is counted once.
The medians are marked.
There is a trend with mass from \SC through \LS to \LLnos, with the \LL clumps
typically above $10^8\msun$.
The \LL clumps tend to have smaller radii.
As a result there is a clear trend with binding energy, via $\Vc$,
from the loosely bound \SC clumps of $\Vc \ssim 20\kms$, through \LSnos,
to the strongly bound \LL clumps of $\Vc \ssim 50\kms$.
}
\label{fig:hist_props}
\end{figure*}

\begin{figure*} 
\centering
\includegraphics[width=0.32\textwidth]
{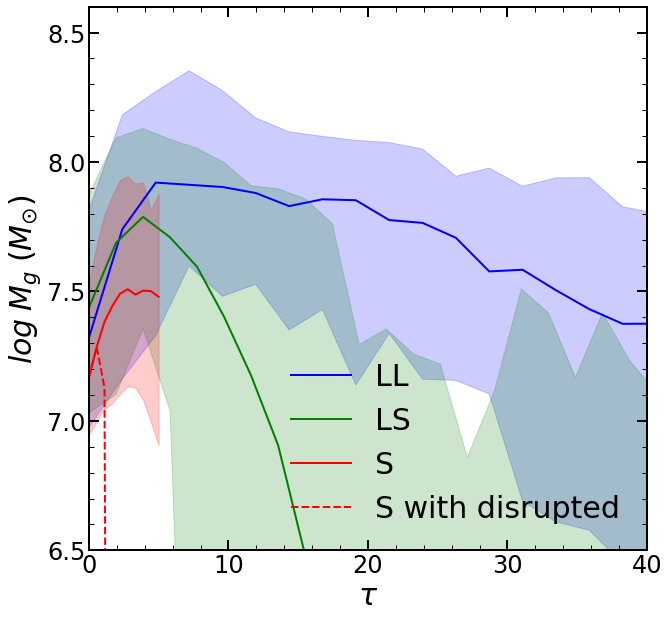} 
\includegraphics[width=0.32\textwidth]
{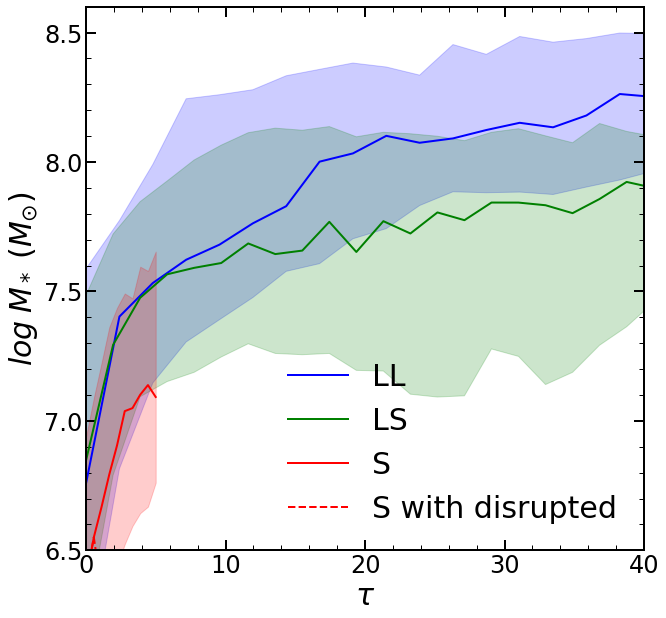} 
\includegraphics[width=0.32\textwidth]
{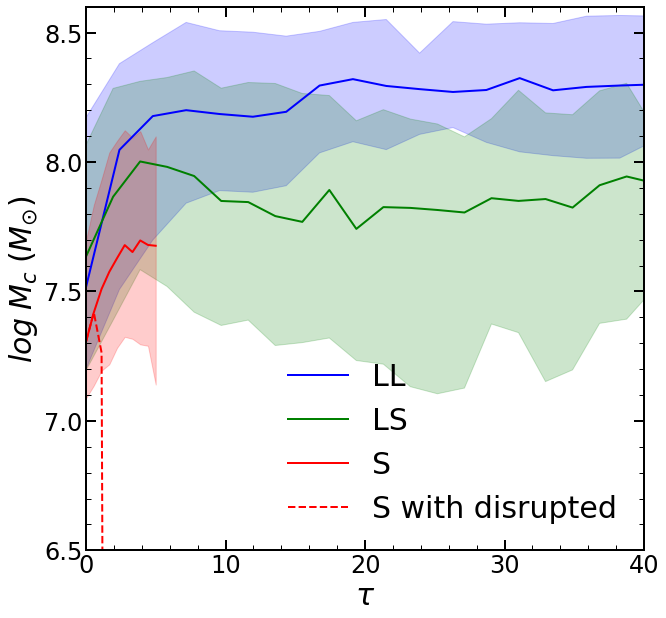} 
\includegraphics[width=0.33\textwidth]
{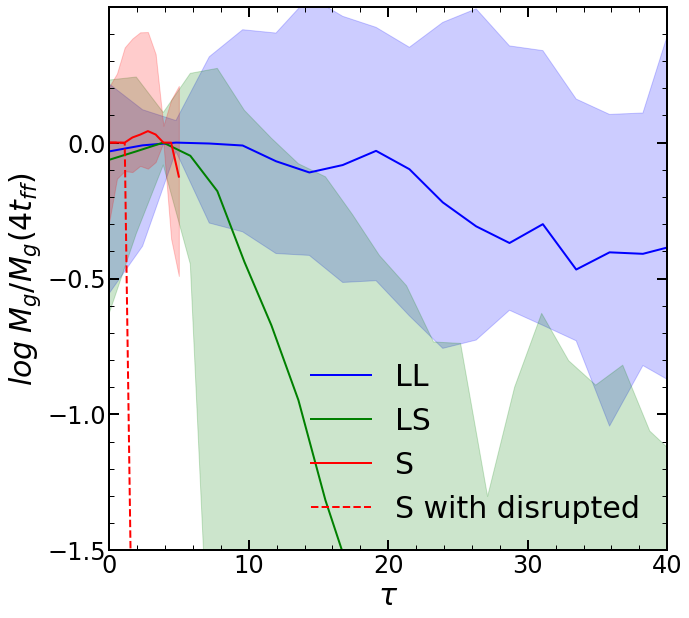} 
\includegraphics[width=0.32\textwidth]
{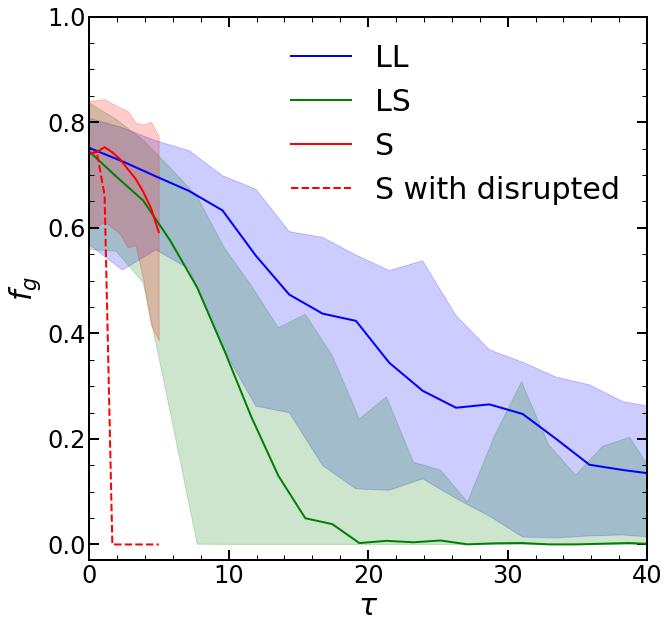} 
\includegraphics[width=0.32\textwidth]
{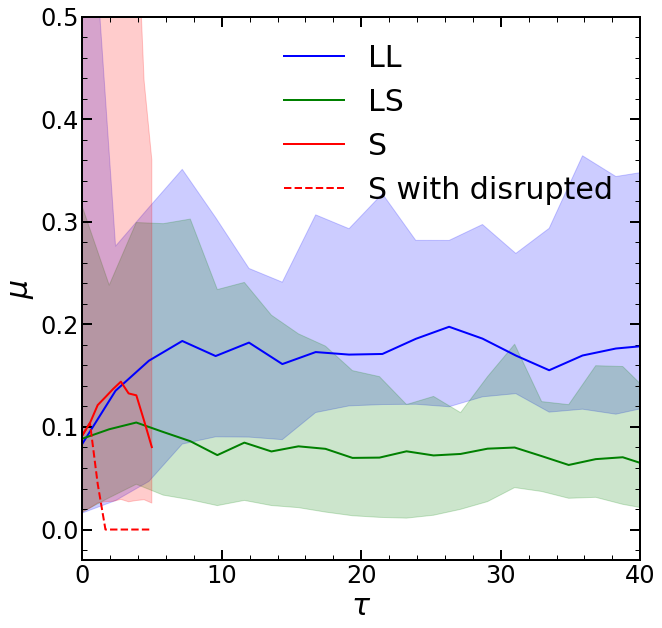} 
\caption{ 
Time evolution of clump mass, $\Mg$, $\Ms$ and $\Mc\seq\Mg+\Ms$, 
for gas, stars and total, as a function of $\tau \seq t/\tff$.
Shown are the medians and $1\sigma$ scatter for each of the three types. 
For the \SC clumps (red), the solid curve and shaded area refer to the small
fraction of clumps that survive until the given time, while in the dashed curve
the disrupted clumps are included with zero mass.
The gas mass is also shown relative to its value at $\tau\seq 4$, near its
peak, and via the gas fraction $\fg\seq\Mg/\Mc$. 
Also shown is $\mu$, the clump mass relative to the Toomre mass.
The \SC clumps lose their gas in the first few free-fall times.
The \LS clumps lose most of their gas to outflows by $\tau \ssim 10$, 
but keep a bound stellar component.
The \LL clumps keep their gas and stars for tens of free-fall times.
}
\label{fig:mass_t}
\end{figure*}

\begin{figure*} 
\centering
\includegraphics[width=0.33\textwidth]
{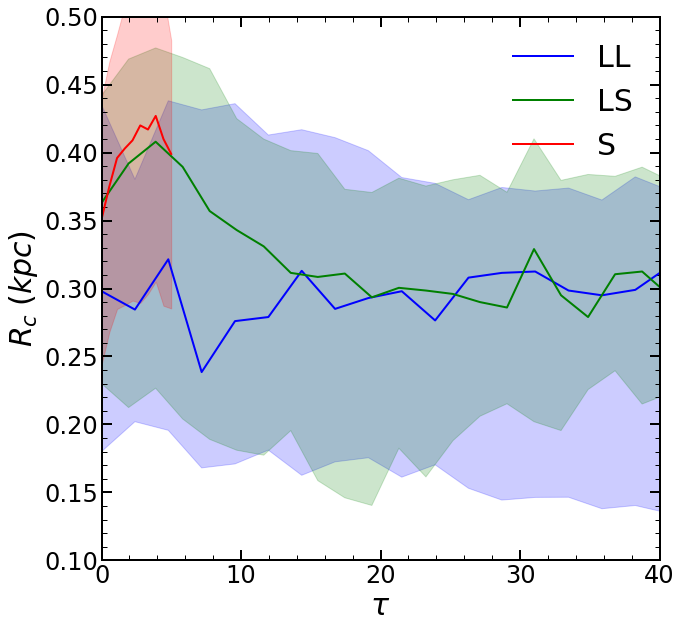} 
\includegraphics[width=0.32\textwidth]
{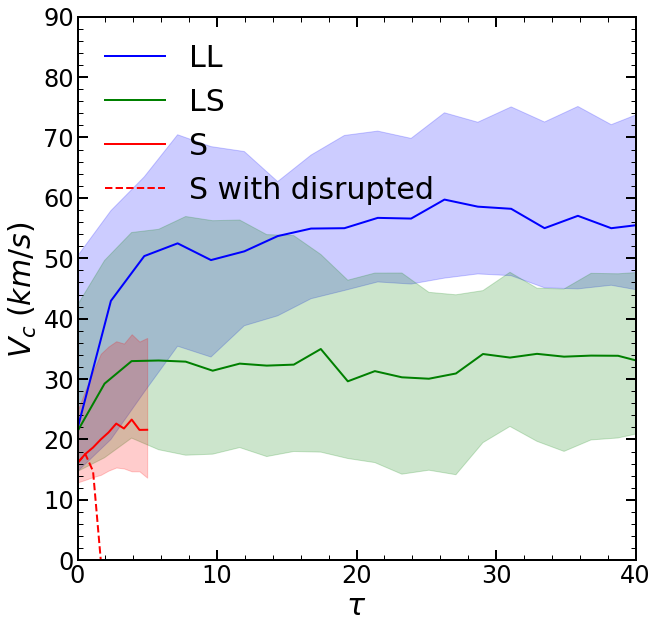} 
\includegraphics[width=0.32\textwidth]
{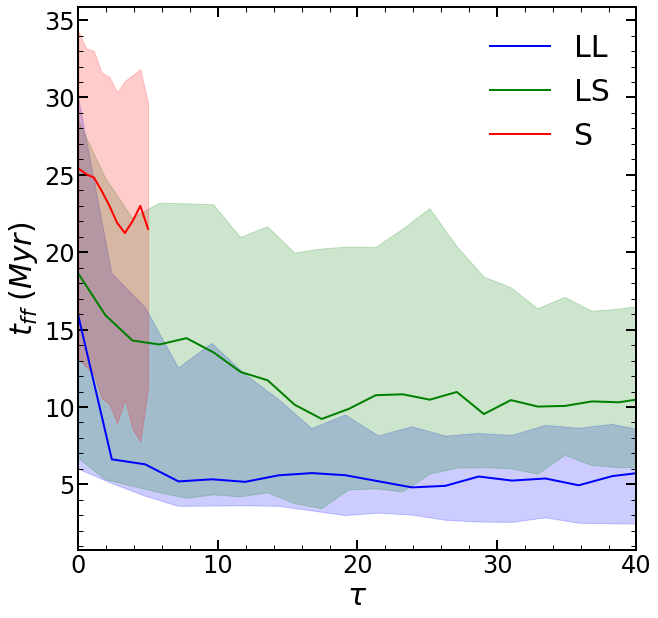} 
\includegraphics[width=0.32\textwidth]
{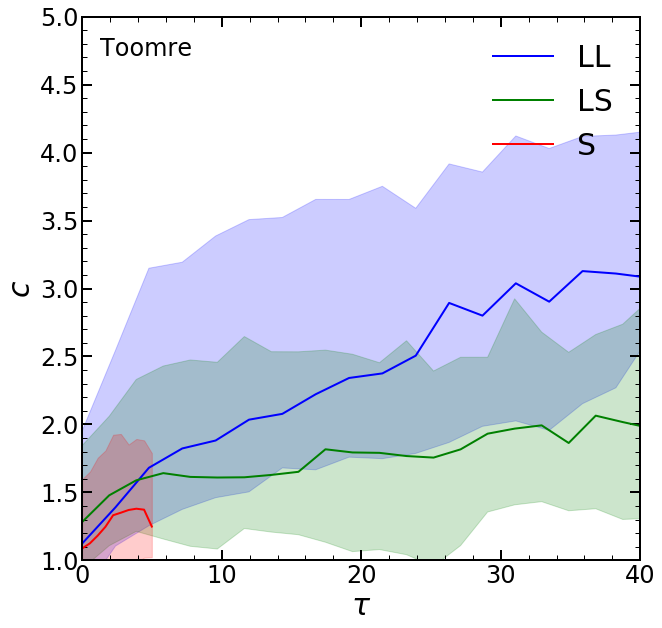} 
\includegraphics[width=0.32\textwidth]
{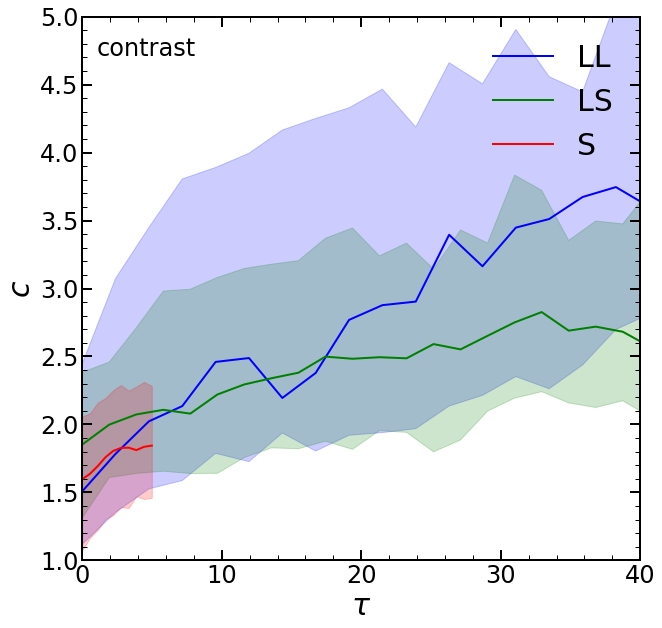} 
\includegraphics[width=0.32\textwidth]
{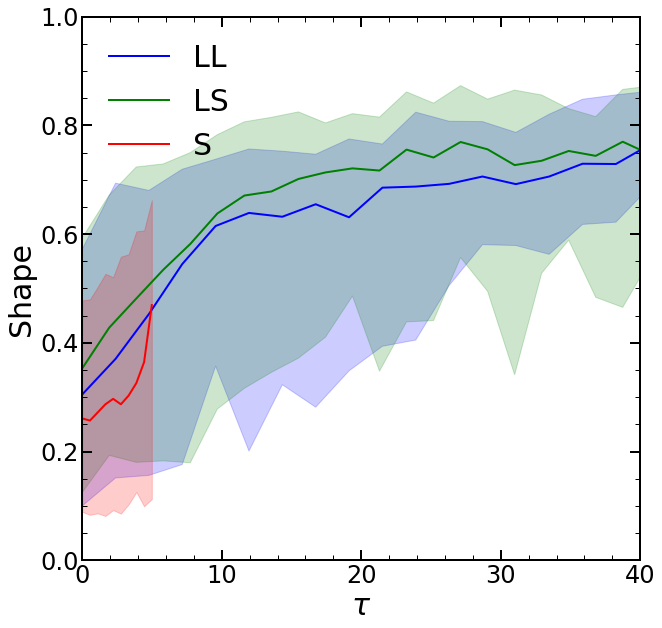} 
\caption{ 
Evolution of clump size, velocity, and free-fall time at $\tau$. 
The density contrast $c$ is measured in two alternative ways, first by 
estimating the initial clump radius from the Toomre analysis (left), 
and second from the 3D density contrast between clump and disk (middle).
The \LL clumps tend to be smaller ($\Rc\ssim 300\pc$), 
more tightly bound $\Vc \ssim 50\kms$, and with short 
free-fall times ($\tff\ssim 5\Myr$).
The \SC clumps are more diffuse and loosely bound ($\Rc\ssim 400\pc$,
$\Vc \ssim 20\kms$, $\tff \ssim 23\Myr$).
The contraction factor is somewhat higher for the \LL clumps.
The difference in clump shape between the clump types is marginal, with the
\SC clumps slightly more elongated. 
}
\label{fig:size_t}
\end{figure*}

\smallskip 
\Fig{images} presents images of gas and stellar surface density in typical 
clumps of the three types in \velathreecom. 
The \LL clump shown at the top at times $\tau\seq 2,4,9,65$
reveals a bound clump both in stars and gas all the way to $\tau \seq 65$.
On the other hand, 
in the \LS clump shown in the middle at $\tau\seq 4,8,10,25$,
the gas has disappeared by $\tau \seq 25$, while the significant stellar
component remains largely unchanged till later times.
The \SC clump shown at $\tau\seq 1,3,4,5$ have lost all its gas already  by 
$\tau \seq 5$, and the stellar component becomes dilute at that time, to
disappear soon thereafter. 

\smallskip 
The three clump types differ by many of their properties beyond the difference
in life-time and gas depletion time that have been used for their 
classification.
A few of these properties are shown in 
\fig{hist_props}, through their distributions in \velathree during the first 4
free-fall times of each clump.
In terms of clump mass, we see a weak systematic decrease of mass from \LL to 
\LS clumps, and a larger decrease into the \SC clumps, with medians
$\Mc \ssim 10^{8.13}, 10^{8.0}, 10^{7.7}\msun$, the latter biased up by the  
selection threshold of $\Mc \sgt 10^7\msun$. 
In terms of clump radius, the \LL clumps tend to be compact, with a median 
radius of $\Rc \ssim 320\pc$ (possibly affected by an effective 
threshold due to the simulation resolution), 
with the \LS and \SC clumps both tending to be more
diffuse, with medians at $\Rc \ssim 400\pc$. 
As a result, the clump virial velocity $\Vc$ tends to be the largest for 
\LL clumps, intermediate for \LS clumps, and smallest for \SC clumps, 
with medians of $\Vc \ssim 50, 33, 22\kms$ for \LL, \LS and \SC respectively.

\begin{figure*} 
\centering
\includegraphics[width=0.33\textwidth]
{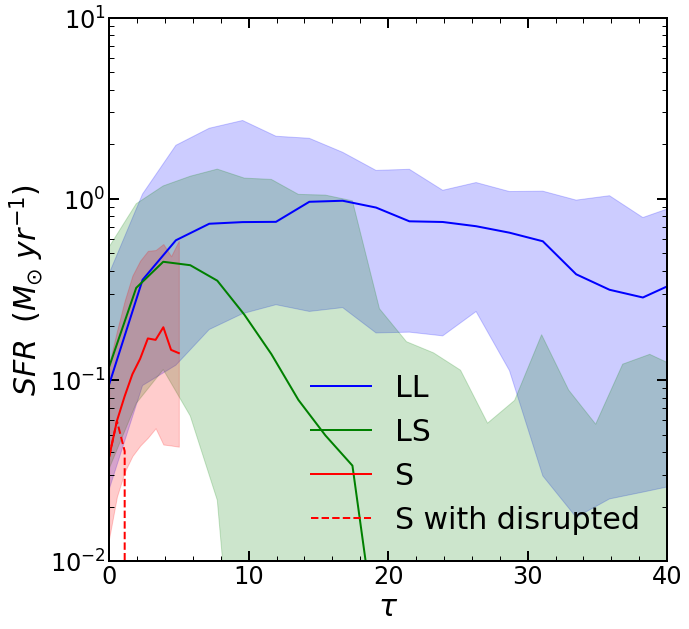} 
\includegraphics[width=0.32\textwidth]
{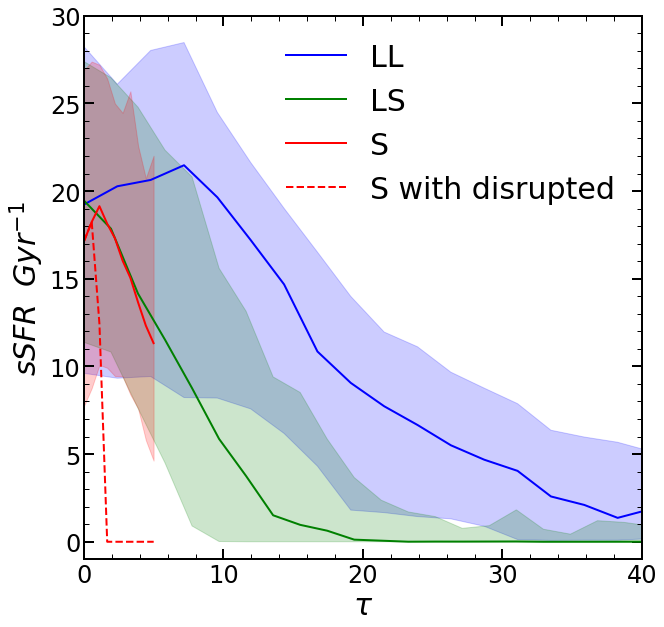} 
\includegraphics[width=0.33\textwidth]
{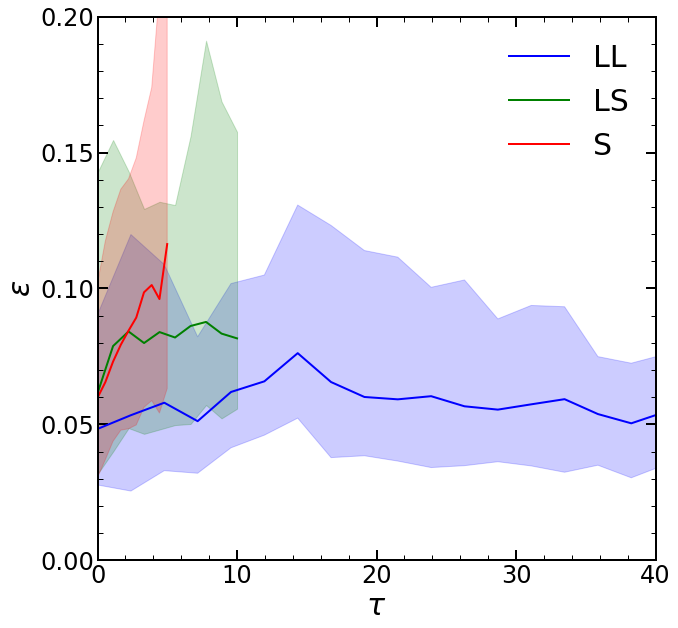} 
\caption{
Evolution of SFR, sSFR and the SFR efficiency per free-fall time $\epsf$
as measured over the whole clump.
The SFR naturally follows the gas mass, so it is low for the \SC clumps,
high for the \LL clumps and roughly constant at $\slsim\!1\msun\yr^{-1}$,
and dropping for the \LS clumps after a peak at $\tau \ssim 5$.
The SFR efficiency is high, especially for the \SC and \LS clumps,
because it refers to the whole clump volume, while the star-forming region is
actually more compact.
}
\label{fig:sfr_t}
\end{figure*}

\smallskip 
The differences and similarities between the three clump types are explored in
more detail by the time evolution of the relevant clump properties, for which
the median and $\pm 34$ percentiles are shown as a function of time $\tau$   
in \figs{mass_t} to \ref{fig:disc_t}.
These detailed properties will serve us in the comparison with the analytic 
model of \se{survival} to \se{thresholds}.


\subsubsection{mass}

\Fig{mass_t} presents the evolution of clump gas and stellar mass.
Most of the \SC clumps disrupt within the first one or two free-fall times, 
so the \SC clumps referred to by the solid curve and shaded area
represent the minority that survive for a little longer. 
The evolution of gas mass $\Mg$ is used for the 
basic classification of the \LC clumps into \LL and \LS clumps via $\taug$, 
which is to be used in the following figures.
Referring to the medians of $\Mg(t)$, 
even the surviving \SC clumps have a median 
gas mass lower by a factor of $2\sdash 3$ than the \LC clumps.
The \LS clumps, after reaching a peak near $\tau$ of a few, 
show a steep decline where half the gas mass is lost
by $\tau \ssim 8$ and practically no gas is left after $\tau \ssim 15$,
while the \LL clumps tend to keep most of their gas till $\tau \ssim 20$, with
some of the \LL clumps gaining gas mass during their evolution.
This relative constancy of gas mass is associated with a roughly constant SFR. 
The stellar mass $\Ms$ for the two \LC types is similar till $\tau \ssim 10$,
growing till $\tau \ssim 5$ and then gradually flattening off.
At later times, the $\Ms$ of the \LL clumps keeps growing roughly linearly
with time, reflecting a constant SFR, while the $\Ms$ of the \LS clumps 
is more constant due to the vaishing SFR (though its median is slowly rising,
perhaps due to the larger masses of the clumps that survive longer, 
or due to accretion).
The gas fraction provides an alternative quantity for the classification, with
similar results to $\Mg$. All the clumps start with $\fg \ssim 0.6 \sdash 0.8$
and the \LS and \LL clumps differ after $\tau \ssim 4$, reaching 
$\fg \seq 0.5$ near $\tau \ssim 8$ and $\sim\!15$ respectively,
with the \LS clumps dropping to negligible gas fractions after 
$\tau \ssim 15$. 

\smallskip   
The median initial total mass of the \LC clumps at $\tau$ of a few
is $\Mc \ssim 10^8\msun$,
while for the \SC clumps it is smaller by a factor of $\sim\!3$. 
For the two types of \LC clumps, $\Mc$ is similar till $\tau \ssim 5$.
After this time, while the median mass of the \LL clumps
is rather constant and even slowly rising, the mass of the \LS clumps is
gradually declining by a factor of $\sim\! 3$ due to gas loss by outflows
to a minimum mass near $\tau \ssim 20$, 
before it is slowly rising at later times.
This could be due to stellar accretion or contamination by background 
disc stars.
From $\mu(t)$,
we learn that the median \LL and \LS clumps are of $\sim\!20\%$ and 
$\sim\!10\%$ of the Toomre mass, respectively.

\subsubsection{size and binding}

\Fig{size_t} highlights a basic continuous trend of $\Vc$ as a function of
clump type, with peak median values in the initial phase 
$\tau \seq 0\sdash10$ of $\Vc \ssim 50,\, 33$ and $20\kms$ 
for \LL, \LS and \SC clumps, respectively. 
This will directly affect the survivability parameter $\Sp$,
and will turn out to be the main distinguishing feature between the \SC and \LC
clumps, as predicted in \se{types_SC_LC}.
The difference in $\Vc^2 \seq G\Mc/\Rc$ may be due to a difference in $\Mc$
or $\Rc$ or both.
\Fig{size_t} shows that during the early phase the median radius is
$\Rc \ssim 400\pc$ for the \SC and \LS clumps, while it is $\Rc \ssim 300\pc$
for the \LL clumps, with some clumps smaller than $200\pc$ (where the radius
is probably overestimated due to resolution).
This implies that the lower $\Vc$ for \SC clumps is predominantly 
due to the lower $\Mc$.
As for the \LS versus \LL clumps,
since the peak of $\Mc$ in the early phase differs by less than $0.2$ dex,
the difference by a factor of almost two in the median $\Vc$ stems mostly 
from the similar difference in $\Rc$.
This may be the basic difference between the \LS and \LL clumps, as discussed
in \se{types_LS_LL}. 
This difference in $\Rc$ is seen in the concentration $c$ as derived
from the Toomre mass at $\tau \slsim 10$ (but less so in $c$ as estimated from
the density contrast).

\smallskip 
\Fig{size_t} also shows a continuous trend in clump free-fall time,
with a value at $\tau \seq 5$ of $\tff \ssim 21, 14$ and $6\Myr$ for
\SC, \LS and \LL clumps, respectively. 
For the \SC clumps, the large $\tff$, reflecting a low density, is both due to
the low $\Mc$ and the high $\Rc$. 
On the other hand, the difference between the two types of \LC clumps is mostly 
due to the difference in $\Rc$ and $c$. 

\smallskip 
As in M17, the clump shape elongation is measured via the axis ratio 
$\ell_1/\ell_3$, the distribution of which for the three clump types at 
$\tau <10$ is shown in \fig{size_t}.
The median \SC clumps are slightly more elongated than the \LC clumps, 
$\ell_1/\ell_3 \ssim 0.3$
versus $\sim\!0.4$, with some of the \LC clumps as round as 
$\ell_1/\ell_3 \ssim 0.8$.
The \LS and \LL clumps are not significantly different in shape.

\subsubsection{SFR}

\Fig{sfr_t} presents the evolution of SFR and sSFR 
and the associated SFR efficiency per free-fall time $\epsf$.
For all types, the early-phase sSFR is high, at the level of 
$\sim\!20\Gyr^{-1}$, characteristic of a starburst in an initial 
stellar-poor clump.
The median SFR in the \SC clumps is lower by a factor of three compared to the
\LC clumps.
Naturally the following decline in sSFR is associated with the decline in
$\Mg$, which occurs in the \SC clumps first, then in the \LS clumps, and later
in the \LL clumps.
\Fig{sfr_t} then shows an apparently surprising result concerning the SFR
efficiency $\epsf \seq \sfr\, \tff\, /\Mg$,
where in the early phase it is $\epsf \ssim 0.08$ and $0.055$ for the 
\LS and \LL clumps, respectively.
This is mostly due to the larger $\tff$ for \LS clumps, 
while $\sfr$ and $\Mg$ are similar.   
This is surprising because the SFR efficiency is argued based on observations 
to be in the same ballpark in all environments and redshifts 
\citep[e.g.][]{kdm12}. 
One way to interpret this is that the SFR in the \LS clumps actually occurs
in regions denser than the average, near the clump centre or in sub-clumps,
but this is not reproduced in the simulated clumps in which sub-clumps are 
not properly resolved.

\smallskip
Another possibility is that the higher $\epsf$ is due to a more efficient
mode of bursty star formation in the \LS clumps, 
e.g., due to shocks induced by clump mergers, intense accretion, 
or strong tidal effects. 
An inspection of Fig.~3 of \citet{kdm12}, which is an accumulation of
star-forming regions in galaxies in different environments and redshifts, 
and focusing on high-redshift galaxies (blue symbols), one can see that the 
starbursts (open symbols), which are likely to represent mergers, 
have $\epsf$ values that are systematically higher than those of the normal 
discs (filled symbols), by a factor of a few.   
Referring to the discussion in \se{types_LS_LL}, these inferred mergers can be
associated with an increase in $\fgrav$ and $\fsn$ compared to the \LL clumps, 
which makes $\Sp$ smaller and thus can explain the early disruption of the 
\LS clumps.
Such an excess of mergers, accretion and tidal effects can be associated with 
the larger radii of the \LS clumps, introducing a larger cross-section for
mergers and capture of gas within the clump Hill sphere, as well as stronger
tidal effects.

\begin{figure*} 
\centering
\includegraphics[width=0.32\textwidth]
{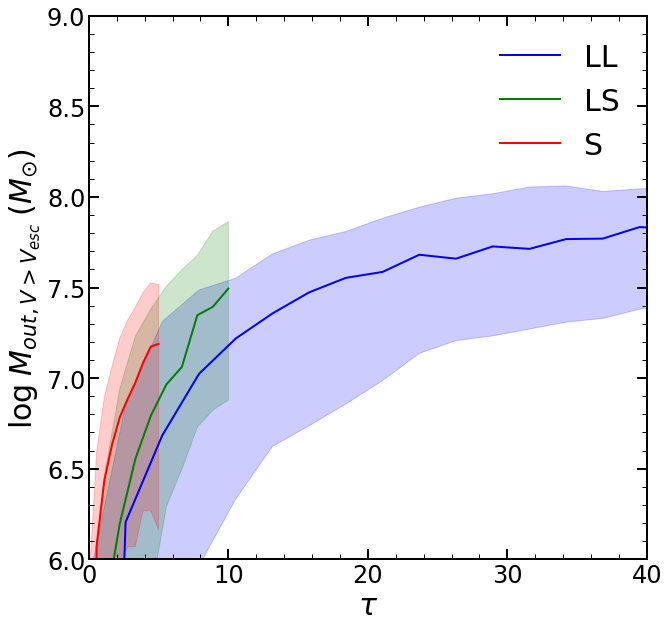} 
\includegraphics[width=0.33\textwidth]
{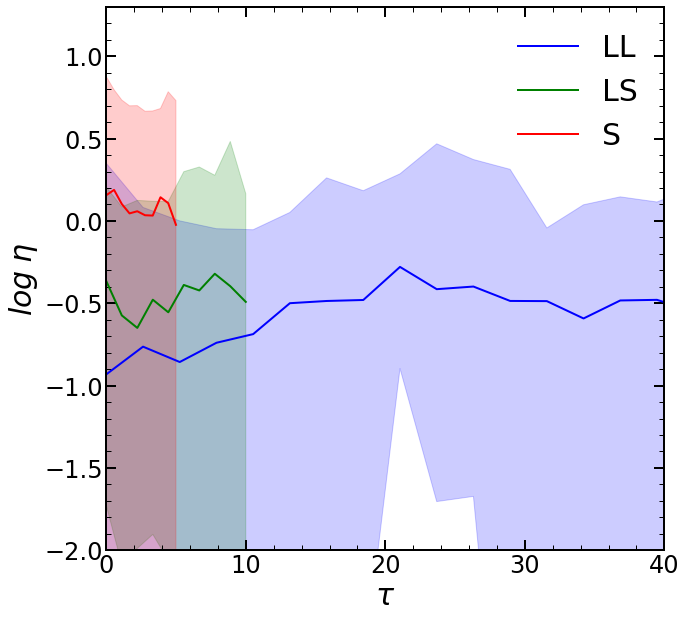} 
\includegraphics[width=0.33\textwidth]
{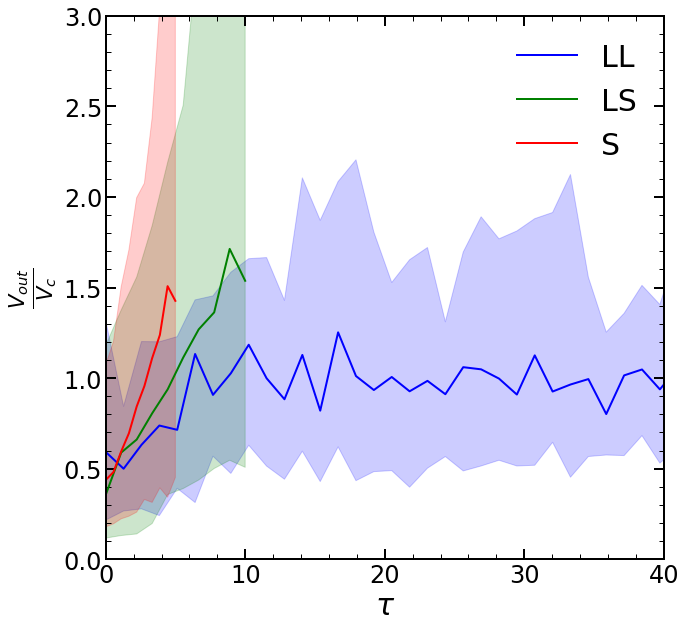} 
\includegraphics[width=0.32\textwidth]
{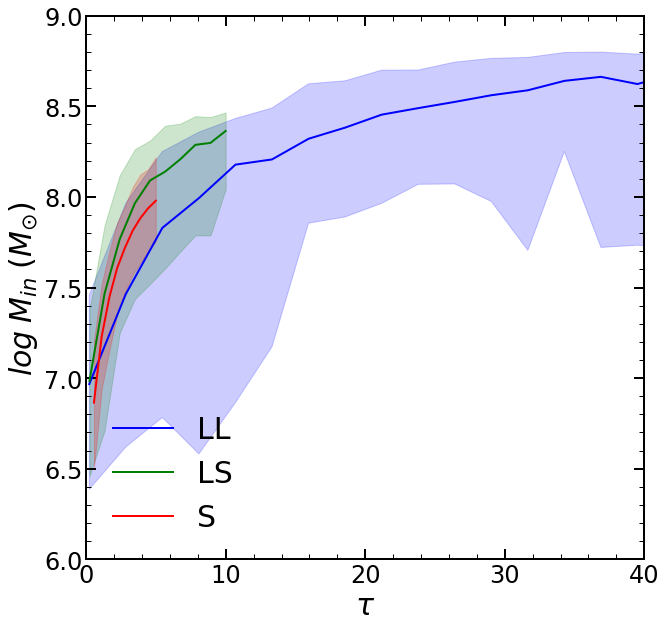} 
\includegraphics[width=0.33\textwidth]
{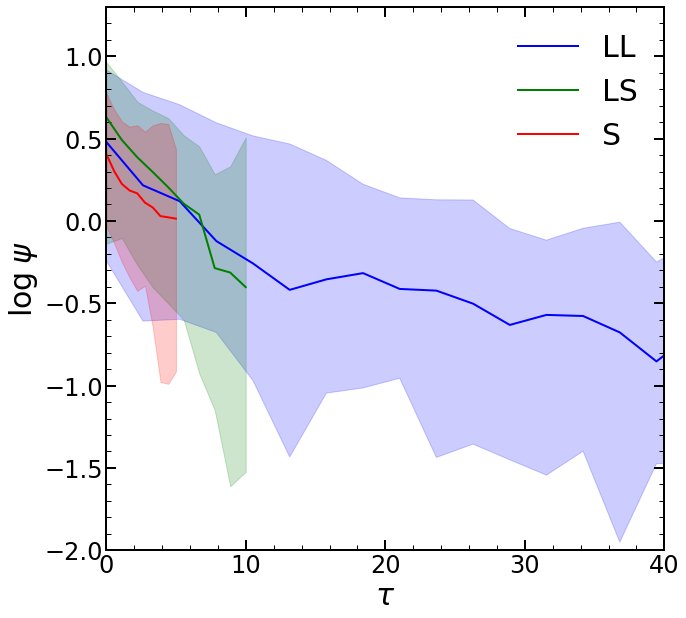} 
\caption{
Evolution of outflow and inflow.
Shown are the cumulative masses in outflow and inflow, as well as the
corresponding parameters $\eta$ and $\psi$ and the outflow velocity $\Vout$.
The outflows are most efficient for the \SC clumps and least efficient for the
\LL clumps, with medians in the early phase of
$\eta \ssim 1.2, 0.3, 0.2$ respectively.
The inflows are more similar between the clump types.
}
\label{fig:out_t}
\end{figure*}

\begin{figure*} 
\centering
\includegraphics[width=0.33\textwidth]
{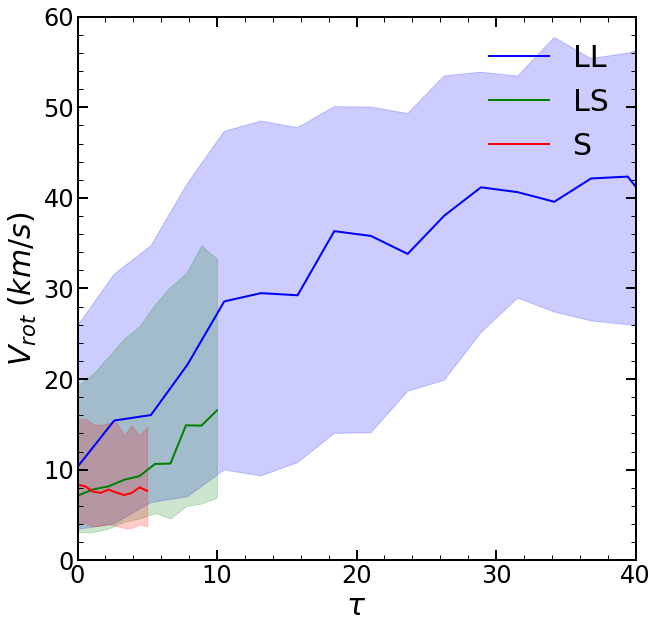} 
\includegraphics[width=0.33\textwidth]
{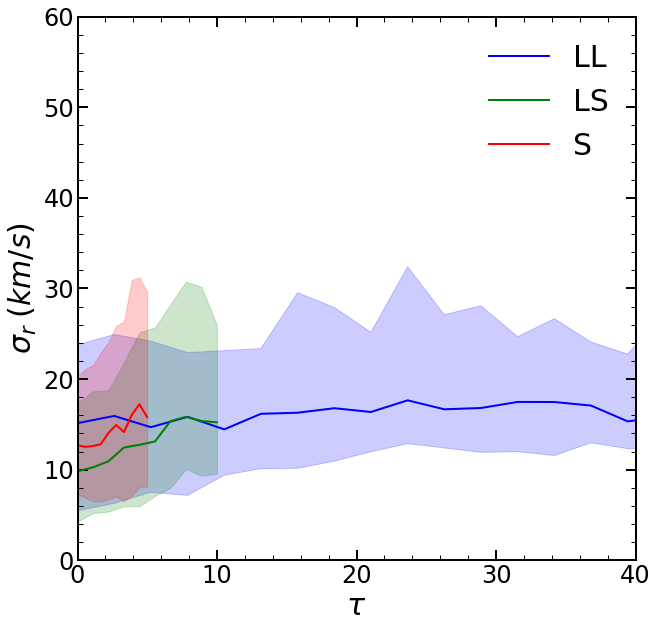} 
\includegraphics[width=0.33\textwidth]
{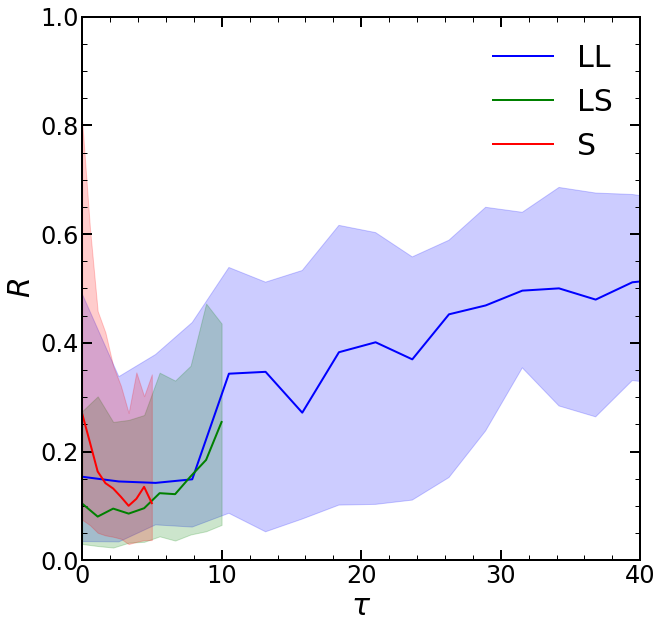} 
\caption{
Evolution of clump internal kinematics, the rotation velocity $\Vrot$,
the radial velocity dispersion $\sigr$,
and the rotation support parameter $R \seq \Vrot^2/\Vc^2$.
The \LL clumps have a larger rotation component, consistent with the higher
contraction factor.
}
\label{fig:kinematics_t}
\end{figure*}

\begin{figure*} 
\centering
\includegraphics[width=0.33\textwidth]
{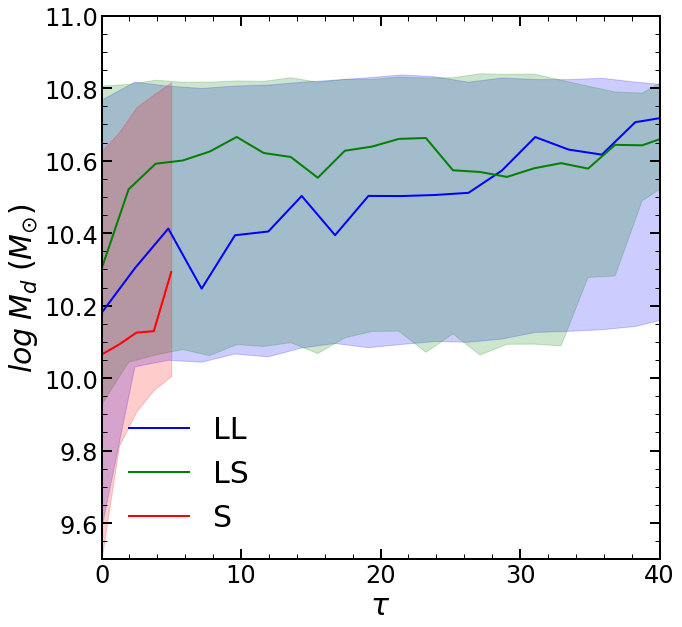} 
\includegraphics[width=0.33\textwidth]
{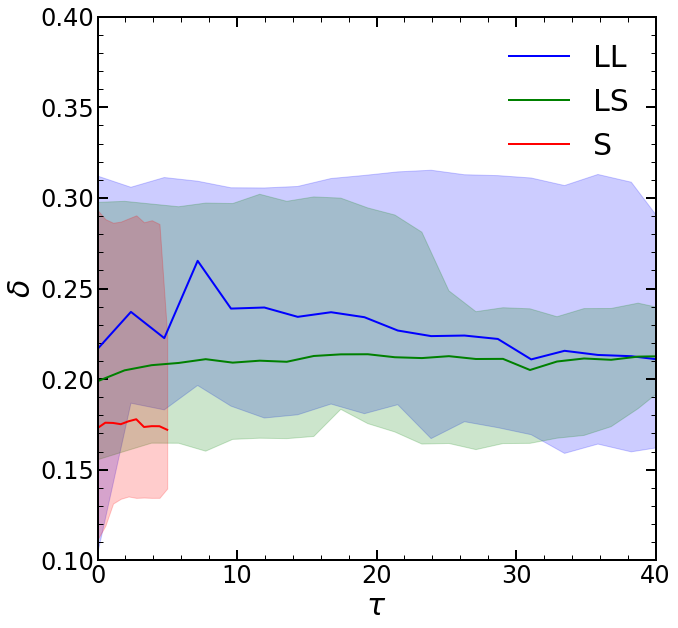} 
\\
\includegraphics[width=0.33\textwidth]
{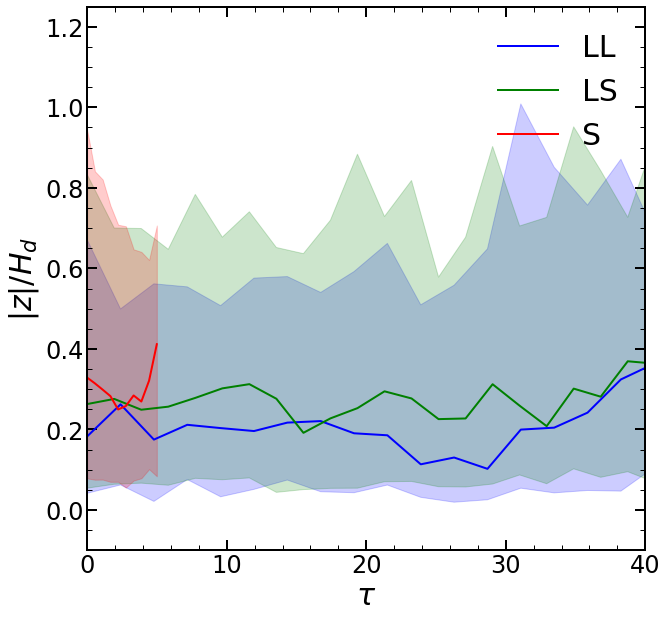} 
\includegraphics[width=0.33\textwidth]
{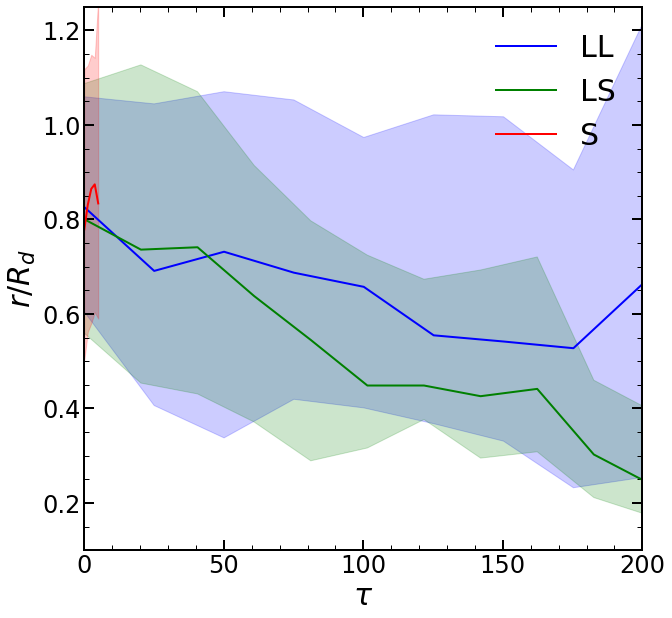} 
\caption{
Evolution of disc mass $\Md$ and mass fraction $\delta$,
and clump position in the disc in the radial and vertical directions.
The disc fraction for galaxies hosting \LL clumps is somewhat higher,
indicating a stronger disc instability with more massive clumps.
The radial migration rate for \LS clumps seems higher in terms of $\tau$,
but this is an artifact of the longer free-fall times for these clumps.
There is no significant difference in the effective $z$ positions of the
different clump types.
}
\label{fig:disc_t}
\end{figure*}

\subsubsection{outflow and inflow}

\Fig{out_t} shows the evolution of gas outflows from and inflows into the
clumps.
During the early phase, the outflowing mass is significantly smaller than
the clump mass, as predicted in \se{Eout}. 
As expected, the outflows in the \SC clumps are the strongest, with a median
larger by a factor of $\sim\!3$ compared to the \LC clumps.
The outflow
is larger by a factor of $\sim\!2$ in the \LS clumps compared to the \LL
clumps, leading to a smaller $\So$ in \equ{S}.  
The outflow mass loading factor varies systematically in a corresponding
way, with a median of $\eta \ssim 1.2, 0.3, 0.2$ for \SC, \LS and \LL clumps
respectively at $\tau \slt 10$.
%
The inflow parameter $\psi$ is defined following eq.~9 of 
\citet[][]{dekel22_mass} (where it is termed $\alpha$), 
\be
\psi = 2\,\Mdotin, \td/\Mc \, .
\ee
Here $\Mdotin$ is measured in analogy to $\Mdotout$, in a shell of radii
$(\Rc,\Rc+100\pc)$, requiring $V_r\slt 0$ but not constraining the absolute 
value of $V$.
The inflowing mass during the early phase is also larger in the \LS clumps
than in the \LL clumps by a factor $\sim\!2$, with the \SC clumps similar to
and slightly smaller than the \LS clumps.
The associated value of $\psi$ is larger for the \LS clumps than the \LL
clumps until $\tau \ssim 0.7$, with the \SC clumps similar to the \LL clumps.
The inflowing mass is consistent with more mergers and intense accretion 
involved in the early phases of the \LS clumps, and an associated higher 
$\fgrav$, leading to a lower $\Sp$. 

\subsubsection{kinematics}

\Fig{kinematics_t} shows that at the early times the clumps are not rotation
supported, with the \LL clumps somewhat more supported by rotation than
the \LS clumps. With $\Vrot$ the mass-weighted rotation velocity, the 
rotation-support parameter is
$R \seq \Vrot^2/\Vc^2 \ssim 0.15$ and $0.1$ for the \LL and \LS clumps
respectively. 
This is consistent with the larger contraction factor $c$ for \LL clumps, 
as predicted in \equ{sigma_Vc} in \se{sigma}, if angular momentum is conserved
during clump collapse.
Both $\Vrot$ and $\sigma_r$ are larger for the \LL clumps, 
though the difference in $\Vrot$ is slightly larger.
We note in passing that the \LL clumps gradually become more rotation supported
at later times.

\subsubsection{disc}

\Fig{disc_t} refers to the clump positions in the host galactic discs.
The initial values of $r/\Rd$ and $\vert z \vert/\Hd$ (at $\tau \ssim 2$)
are similar for the
different clump types, indicating that the differences among the types
are not largely due to the position of formation within the disc.
The two types of \LC clumps migrate radially inwards at a rather slow pace,
in the ball park of the various theoretical predictions 
\citep{dsc09,krum_burkert10,krum18,dekel20_ring}.
%
The non-zero values of $\vert z \vert/\Hd$ and their large scatter
reflect the fact that the clumps oscillate about the disc plane.
While the median position of the \LS clumps is at 
$\vert z \vert \ssim\!0.3\Hd$, 
the \LL clumps get somewhat closer to the disc mid-plane by $\tau \ssim 5$,
and keep a median distance of $\vert z \vert \ssim\!0.2\Hd$ or less till
$\tau \ssim 30$.

The median \LS discs tend to be more massive than the \LL disks,
but the disc mass fraction compared to the total mass including stars and dark
matter, $\delta$, is somewhat larger for the \LL clumps.  
This is consistent with a shorter inward migration time in units of disc
dynamical time for the \LL clumps \citep{dsc09}.
The weak opposite trend in $r(t)/\Rd$ in terms of $\tau$ is probably because
$\tff$ is larger for the \LS clumps.

\begin{figure*} 
\centering
\includegraphics[width=0.45\textwidth]
{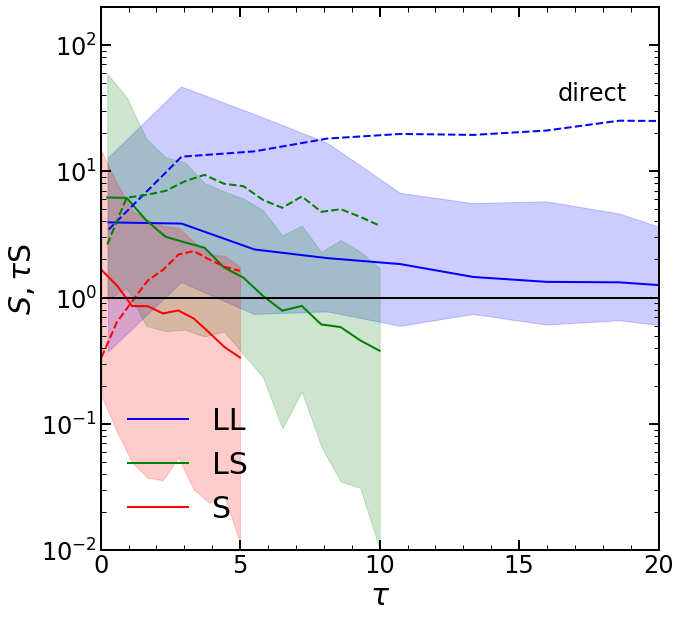} 
\includegraphics[width=0.45\textwidth]
{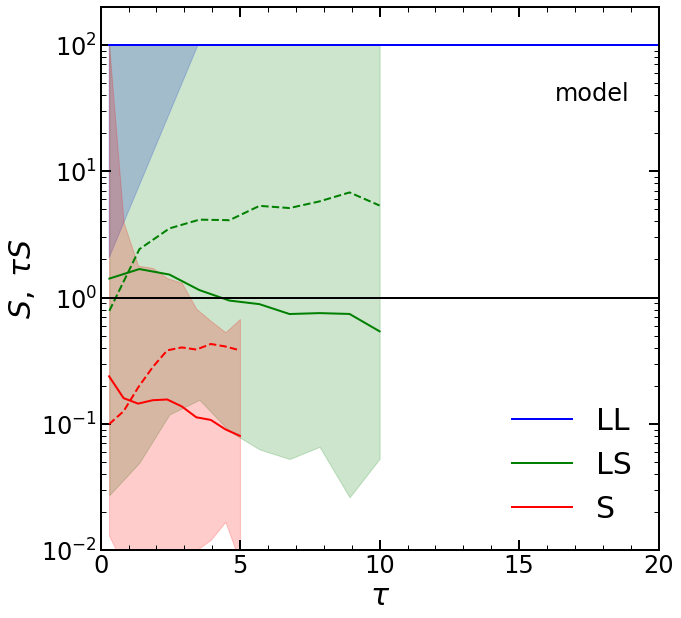} 
\caption{
Time evolution of $\So$ (solid) and $\tau\So$ (dashed), 
the quantities that evaluate clump survivability against gas loss 
and total disruption, respectively, for the three clump types.
{\bf Left:} $\So$ as measured directly from the simulations from $\Mout$, 
\equ{S_sim_f}.
{\bf Right:} $\So$ as measured from the physical parameters assuming energy
balance, \equ{S_mod_f} 
(version 2 with $\gamma\seq 1$).
Values larger than 100 are set to 100.
During the first few dynamical times, the median $\So$ values are significantly
smaller for the \SC clumps compared to the \LC clumps, consistent with the
expected disruption of the \SC clumps compared to the survival of the \LC
clumps.
The median $\So$ for the \LS clumps declines to below unity 
after $\tau \ssim 5 \sdash 10$, while for the \LL clumps it remains above unity
till $\tau \ssim 20$ and beyond.
The median values of $\tau \So$ distinguish between the \SC and \LS clumps,
consistent with the destiny of complete disruption for the \SC clumps
and only gas removal for the \LS clumps, leaving behind bound stellar \LS
clumps.
}
\label{fig:S(t)}
\end{figure*}

\begin{figure*} 
\centering
\includegraphics[width=0.45\textwidth]
{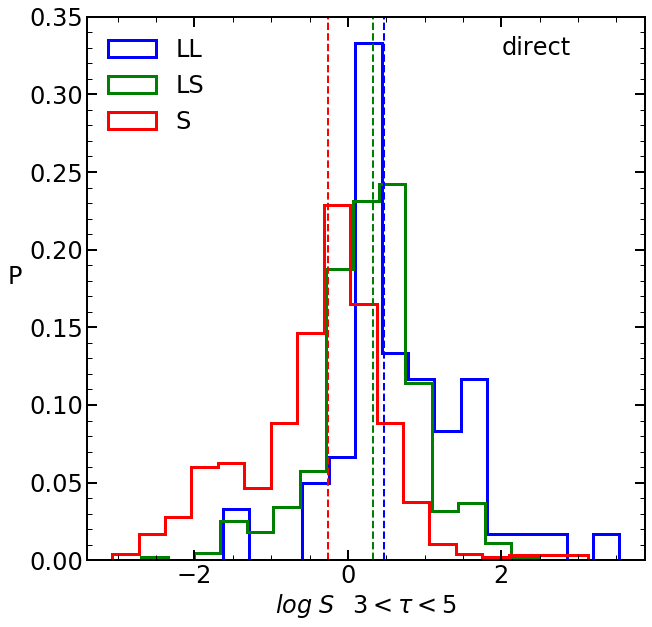} 
\includegraphics[width=0.44\textwidth]
{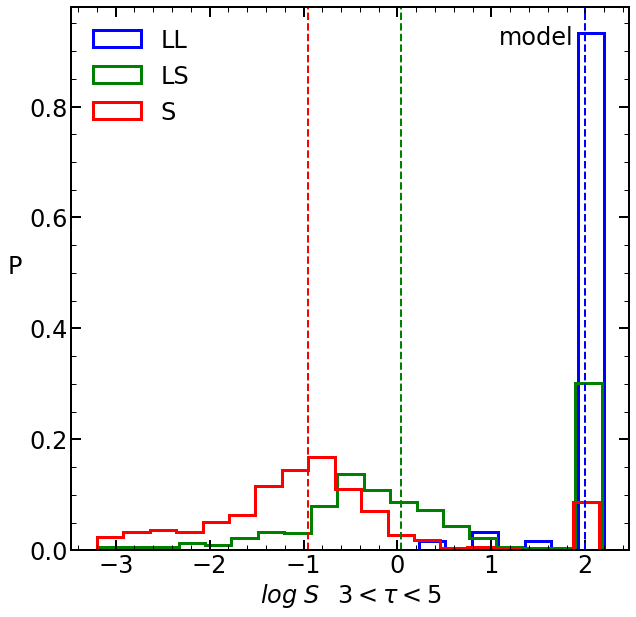} 
\caption{
The probability distributions of the survivability parameter $\So$ for the
three clump types, as in \fig{S(t)} but during $\tau \seq 3\sdash 5$.
{\bf Left:} $\So$ measured directly from the simulations, \equ{S_sim_f}.
Most of the \LC clumps have $\So \sgt 1$ and most of the \SC clumps have 
$\So \slt 1$, with the medians near $3$ and $0.5$ respectively. 
{\bf Right:} $\So$ measured from the physical parameters, \equ{S_mod_f}
(version 2 with $\gamma\seq 1$).
Here all the \LL clumps have $\So \sgg 1$, the \LS clumps have $\So$ 
distributed about unity, and the vast majority of the \SC clumps have 
$\So \sll 1$, with the medians $\gg\!1$, $\sim\!1$ and $\sim\!0.1$ respectively.
}
\label{fig:S_hist}
\end{figure*}

\section{Theory versus simulations}
\label{sec:sim_model}

Using the simulations,
we compute the survivability parameter $\So$ as a function of time since  
clump formation for the simulated clumps of the three types. 
We compute $\So$ in two alternative ways, which we term $\Sdir$ and
$\Smod$.
First, directly from the simulations, we determine $\So$ 
from the measured $\Mout(t)$, $\Mc$ and the correction factor $\f$,
using the definition of $\So$ in \equ{S_sim_f}.
Alternatively, we compute $\So$ from \equ{S_mod_f}, based on our model
assuming energy conservation and the different sources of energy gain and loss,
using the measured $\Vc$ and the factors  
$\fsn$ (\equnp{fsn}), $\fgrav$ (\equnp{fgrav}) and $\fdis$ (\equnp{fdis}).

\smallskip
\Fig{S(t)} shows the evolution of $\So$ and of $\tau\So$ for the three clump
types as computed by the two methods.
Recall that $\So$ is supposed to measure the clump survivability against losing
its gas mass by supernova feedback, thus distinguishing between \SC and \LC 
clumps by $\So$ values below and above unity, respectively.
The product $\tau \So$ is expected to measure the clump survivability against 
total disruption versus partial mass loss followed by a bound stellar system,
thus distinguishing between \SC and \LS clumps,
by $\tau\So$ values below and above unity, respectively.
Inspecting the medians, we see that during the first few dynamical times 
$\So$ is significantly smaller in the \SC clumps compared to the \LC clumps, 
as expected.
The median of $\So$ for the \LS clumps, which starts well above unity at early
times, declines to below unity after $\tau \ssim 5 \sdash 10$, 
while for the \LL clumps it remains above unity till $\tau \ssim 20$ or 
further, as anticipated.
The median values of $\tau\So$ indeed distinguish between the 
\SC clumps that will disrupt completely and the \LS clumps that will keep their
stars bound, as expected.

\smallskip  
Viewing the survivability parameter from a different angle, 
\fig{S_hist} shows the probability distributions of $\So$ for the three clump
types at the early phase $\tau \seq 3\sdash 5$. 
The values of $\So$ as measured directly from the simulations yield that
most of the \LC clumps have $\So \sgt 1$ and most of the \SC clumps have
$\So \slt 1$, with the medians near $3$ and $0.5$ respectively.
The calculation of $\So$ based on the model assuming conservation of energy
yields that all the \LL clumps have $\So \sgg 1$, the \LS clumps have $\So$
distributed about unity, and the vast majority of the \SC clumps have
$\So \sll 1$, with the medians near $\gg\!1$, $1$ and $0.1$ respectively.
We conclude that, in general, $\So$ serves its purpose as a predictor
of clump types.

\smallskip 
Comparing the measured $\So$ and $\tau\So$ by the two methods, 
we note that the medians of these quantities indeed distinguish between the 
clump types in both cases.
We note however that in the direct calculation using \equ{S_sim_f} 
the median values of $\tau\So$ for the \SC clumps are slightly above unity.
This may reflect the uncertainty in measuring $\Mout$.
In particular, our assumption that $\Vout^2 \ssimeq 2\Vc^2$
that enters \equ{Eout0} may be inaccurate.
If we parametrize $\Vout^2 = \fesc\, 2 \Vc^2$, 
we obtain $\Smod \sprop \fesc$. 
A value of $\fesc \ssim 2$, or a range of values for the different clump types,
as in \fig{out_t}, 
may reduce the apparent differences between $\Sdir$ and $\Smod$.
Such refinements of the model are deferred to future work.

\section{Conclusion}
\label{sec:conc}

Energetics considerations lead to a physical criterion for the survival versus 
disruption of the giant clumps that dominate high-redshift galactic discs.
Depending on the disc and clump properties,
the model predicts populations of short-lived clumps (\SCnos) that lose their 
gas in a few free-fall times during formation 
and long-lived clumps (\LCnos) that keep their gas for longer periods. 
The latter are of two types, those that lose most of their gas to outflows 
in $\slsim\! 10$ free-fall times but keep gas-deficient, long-lived,
bound stellar clumps (\LSnos), 
and those that keep most of their baryons for tens of free-fall times and
remain star forming roughly at a constant rate (\LLnos). 

\smallskip
Our model introduces a survivability parameter $\So$, which can predict
the level of survivability against gas loss
at a given time based on the clump and disc
properties and physical parameters that characterize the SFR and
feedback, distinguishing between \SC and \LC clumps.
The quantity $\tau\So$, where $\tau \seq t/\tff$, 
then predicts the survivability against total 
disruption by rapid gas loss, distinguishing \SC from \LS clumps.
These quantities are defined as
\be
\So = \frac{0.5\,\Mc}{\Mout(t)} \, , \quad \tau S \simeq
\frac{0.5\,\Mc}{\Mdotout \tff} \, .
\ee
Thus, the distinction between the three clump types based on the survivability
parameter $S$ during the first few clump free-fall times is as summarized in 
\tab{types}, namely 
\SC clumps have $S\slt 1$ and $\tau S\slt 1$,  
\LS clumps have $S\slsim 1$ but $\tau S \sgsim 1$,
and \LL clumps have $S \sgt 1$ and $\tau S \sgt 1$. 
The model successfully reproduces the distribution of clump properties
in cosmological simulations, and provides a physical basis for the existence 
of the three clump types in these simulations.

\smallskip 
The model considers the balance between the energies supplied by supernova
feedback and by gravitational interactions against the energy required for 
overcoming the clump binding energy, the energy carried away by outflows,
and the dissipative losses of turbulence. 
The supernova input is based on a generalization of the analysis by 
\citet{ds86}.
The clumps are assumed to form by Toomre instability within VDI discs 
\citep{dsc09}.  
If they manage to contract sufficiently, the clumps are assumed to  
reach virial and Jeans equilibrium \citep{ceverino12}.

\begin{table}
\centering
\begin{tabular}{@{}lcccc}
\multicolumn{5}{c}{{\bf Three clump types}} \\
\hline
 Type & Gas loss & Stars  & $S$ & $\tau S$ \\
\hline
\hline
 Short-lived \SC  & $\sim\!\tff$  & none
& $S\slt 1$ & $\tau S \slt 1$ \\
\hline
Long-lived stars \LS & $\sim\!10\tff$  & bound
& $S\slsim 1$ & $\tau S \sgsim 1$ \\
\hline
Long-lived \LL  & $\gg\!\tff$  & bound & $S\sgt 1$ & $\tau S \sgt 1$ \\
\end{tabular}
\caption{
The three clump types are distinguished by the survivability parameter during
the first few clump free-fall times. $S \slt 1$ predicts significant gas loss,
while $\tau S \slt 1$ predicts rapid gas loss that leads to total disruption.
}
\label{tab:types}
\end{table}

\smallskip
We obtain $S \ssimeq (\Sp^{-1}\!-\!1)^{-1}$, namely the critical value of $\Sp$
corresponding to $S\seq 1$ is $\Sp = 0.5$.
For a simple interpretation of the main results,
considering only the balance between the deposited supernova energy and the 
clump binding energy, we approximate near the first free-fall time
\be
\Sp \simeq \frac{\Evir}{\Esn} \simeq \frac{\Vcf}{e_{51}} \, .
\ee
The linear dependence on the clump circular velocity $\Vc$ results from 
the fact that $\Evir \prop \Vc^2$ while $\Esn \prop \Vc$ (via its dependence on
the velocity dispersion in the clump ISM).
The survivability is thus a strong function of the 
{\it clump circular velocity} (which is correlated with the clump mass) 
and the {\it feedback strength} as formulated via $e_{51}$.
In turn, a Toomre analysis connects the clump circular velocity to the disc
properties via 
\be
\Vc \prop \fg \, \Vd \, .
\ee
This introduces strong {\it gas-fraction} dependence and {\it disc mass} 
dependence for the clump survival via $S$.

\smallskip 
If the strength of ejective feedback is at a moderate level,
we find that \LC clumps are likely to exist, with a binding energy 
that is characterized by a circular velocity $\Vc \ssim 50\kms$, 
at or above a clump threshold mass $\sim\!10^8\msun$.
The \LC clumps tend to form in discs of $\Vd \sgeq 200\kms$, 
corresponding at $z\ssim 2$ to stellar masses above $\sim\!10^{9.3}\msun$ 
and galactic halo masses  above $\sim\!10^{11.3}\msun$, which are typically 
required for long-lived discs against disruption by mergers in an orbital
time \citep[as predicted by][]{dekel20_flip}.
\LC clumps are predicted to favor gas fractions $\sgeq\!0.3$
and more so when the central mass of bulge and dark matter is small.
The \LC clumps prefer to form at high redshifts, peaking at $z\ssim 2$.

\smallskip 
On the other hand, the likelihood of \LC clumps is severely reduced if the 
ejective feedback is stronger, e.g., if the effective energy per supernova 
is well above the standard value due to clustering of supernovae, 
if very strong radiative feedback is included,
if the stellar initial mass function is top-heavy,
or if the star-formation-rate efficiency is higher than commonly assumed on
average.
Simulations with varying levels of ejective feedback indeed permit \LC clumps
with different efficiencies.

\smallskip 
The division of the \LC clumps into two sub-populations is demonstrated in the
simulations. The model may explain the existence of \LS clumps
by a smaller contraction factor during formation and 
stronger external gravitational effects, where clump mergers may increase
the SFR efficiency. This is compared to the more compact \LL clumps that 
retain most of their baryons for tens of free-fall times.

\smallskip 
A word of caution concerns the fact that we have modeled the overall stellar 
feedback based on the supernova feedback component. This is a first crude 
treatment that should be improved by detailed considerations of the other 
feedback mechanisms, such as radiative feedback and stellar winds.
Furthermore, our current treatment ignored the clustering of supernovae,
which is expected to increase the deposited feedback energy in low-mass 
star clusters and to decrease it in massive clusters \citep{gentry17}. 
These should be improved in future studies.

\smallskip 
Our findings emphasize a challenging general {\it feedback puzzle}.
While the model successfully predicts that for a moderate feedback strength
the massive clumps are long-lived, when implemented in simulations
(such as \velathreecom)
this moderate feedback leads to an overestimate of the stellar-to-halo mass
ratio as estimated from observations via abundance matching
\citep{rodriguez17,moster18,behroozi19}.
On the other hand, simulations with stronger feedback
(such as \velasixcom), which better match the
stellar-to-halo mass ratio, practically fail to reproduce long-lived clumps.
If the observed massive clumps are long-lived,
as indicated, for example, by their not-so-young stellar ages and the 
gradients in clump properties that are consistent with VDI-driven clump 
radial migration \citep{dekel22_mass}, 
then one is challenged to come up with more sophisticated  
feedback mechanisms. 
The feedback should be more preventive and less ejective, such that
it efficiently suppresses star formation while it is less destructive in 
terms of ejecting gas from the clumps.


\section*{Acknowledgments}

This work was supported by the Israel Science Foundation Grants
ISF 861/20 (AD) and 3061/21 (NM), and by  
Germany-Israel DIP grant STE1869/2-1 GE625/17-1 (AD).  
OG is supported by a Milner Fellowship.
DC is a Ramon-Cajal Researcher and is supported by the Ministerio de Ciencia, 
Innovaci\'{o}n y Universidades (MICIU/FEDER) under research grant 
PID2021-122603NB-C21.
The cosmological simulations were performed at the National
Energy Research Scientific Computing centre (NERSC), Lawrence Berkeley National
Laboratory, and at NASA Advanced Supercomputing (NAS) at NASA Ames Research
Centre.
Development and analysis have been performed in the computing
cluster at the Hebrtew University.

\section*{DATA AVAILABILITY}

Data and results underlying
this article will be shared on reasonable request to the corresponding author.

\bibliographystyle{mn2e}
\bibliography{clump_s}

\appendix


\label{lastpage}
\end{document}